\documentclass[10pt,prd,nofootinbib,preprint]{revtex4}
\usepackage[a4paper, total={7in, 10in}]{geometry}
\usepackage[latin9]{inputenc}
\usepackage{color}
\usepackage{float}
\usepackage{booktabs}
\usepackage{multirow}
\usepackage{amsmath}
\usepackage{graphicx}
\usepackage{esint}
\usepackage{epsfig}
\usepackage{caption}
\usepackage{subcaption}
\usepackage{hyperref}
\makeatletter

\DeclareFontEncoding{LGR}{}{}

\ProvideTextCommand{\~}{LGR}[1]{\char126#1}

\newcommand{\lyxmathsym}[1]{\ifmmode\begingroup\def\b@ld{bold}
  \text{\ifx\math@version\b@ld\bfseries\fi#1}\endgroup\else#1\fi}

\providecommand{\tabularnewline}{\\}

\@ifundefined{textcolor}{}
{%
 \definecolor{BLACK}{gray}{0}
 \definecolor{WHITE}{gray}{1}
 \definecolor{RED}{rgb}{1,0,0}
 \definecolor{GREEN}{rgb}{0,1,0}
 \definecolor{BLUE}{rgb}{0,0,1}
 \definecolor{CYAN}{cmyk}{1,0,0,0}
 \definecolor{MAGENTA}{cmyk}{0,1,0,0}
 \definecolor{YELLOW}{cmyk}{0,0,1,0}
}

\usepackage{xcolor}\usepackage{epsfig}\usepackage{slashed}\usepackage{appendix}
\usepackage{tabularx}
\usepackage{pdfpages}
\usepackage{epstopdf}
\usepackage{multirow}
\usepackage{siunitx}
\usepackage{soul}

\makeatother

\begin{document}
\begin{center}
{\bf\Large\boldmath
Analysis of $b\rightarrow c\tau\bar{\nu}_\tau$ anomalies using weak effective Hamiltonian with complex couplings and their impact on various physical observables
}\\[5mm]
\par\end{center}

\begin{center}
\setlength{\baselineskip}{0.2in} {Muhammad Arslan$^{1,a}$, Tahira Yasmeen$^{2,b}$, Saba Shafaq$^{2,c}$,
Ishtiaq Ahmed$^{3,d}$ and Muhammad Jamil Aslam $^{1,e}$}\\[5mm] $^{1}$~\textit{Department of Physics, Quaid-i-Azam
University, Islamabad 45320, Pakistan.}\\
 $^{2}$~\textit{Department of Physics, International Islamic University,
Islamabad 44000, Pakistan.}\\
 $^{3}$~\textit{National Center for Physics, Islamabad 44000, Pakistan.}\\[5mm] 
\par\end{center}
$^{a}$~{arslan.hep@gmail.com (corresponding author)}\\
$^{b}$~{tahira709@gmail.com}\\
$^{c}$~{saba.shafaq@iiu.edu.pk}\\
$^{d}$~{ishtiaqmusab@gmail.com}\\
$^{e}$~{muhammadjamil.aslam@gmail.com}\\[5mm]

\textbf{Abstract}\\[5mm] Recently, the experimental measurements of the branching ratios and different polarization asymmetries for the processes occurring through flavor-changing-charged current $(b\rightarrow c\tau\overline{\nu}_{\tau})$ transitions by BABAR, Belle, and LHCb show some sparkling differences with the corresponding Standard Model (SM) predictions. This triggered an interest to look for physics beyond the SM in the context of various new physics (NP) models and using model-independent weak effective Hamiltonian (WEH). Assuming the left-handed neutrinos, we add the dimension-six vector, (pseudo-)scalar, and tensor operators with complex Wilson Coefficients (WCs) to the SM WEH. Together with $60\%$, $30\%$
and $10\%$ constraints coming from the branching ratio of  $B_{c}\to\tau\bar{\nu}_{\tau}$, we reassess the parametric space of these new physics WCs accommodating the current anomalies in the purview of the most recent HFLAV data of $R_{\tau/{\mu,e}}\left(D\right)$, $R_{\tau/{\mu,e}}\left(D^*\right)$ and Belle data of $F_{L}\left(D^*\right)$ and $P_{\tau}\left(D^*\right)$. We find that the allowed parametric region of left-handed scalar couplings strongly depends on the constraints of $B_{c}\rightarrow\tau\bar{\nu}_{\tau}$ branching ratio, and the maximum pull from the SM predictions comes for the $<60\%$ branching ratio limit. Also, the parametric region changes significantly if we extend the analysis by adding LHCb data of $R_{\tau/\mu}\left(J/\psi\right)$ and 
$R_{\tau/\ell}\left(\Lambda_c\right)$. Furthermore, due to the large uncertainties in the measurements of $R_{\tau/\mu}\left(J/\psi\right)$ and $R_{\tau/\ell}\left(X_c\right)$, we derive the sum rules which complement them with $R_{\tau/{\mu,e}}\left(D\right)$ and $R_{\tau/{\mu,e}}\left(D^*\right)$. Using the best-fit points of the new complex WCs along with the latest measurements of $R_{\tau/{\mu,e}}\left(D^{(*)}\right)$, we predict the numerical values of the observable $R_{\tau/\ell}\left(\Lambda_c\right)$, $R_{\tau/\mu}\left(J/\psi\right)$ and $R_{\tau/\ell}\left(X_c\right)$ from the sum rules. The simultaneous dependence of above mentioned physical observables on the NP WCs is established by plotting their correlation with $R_{D}$ and $R_{D^*}$, which are useful to discriminate between various NP scenarios. We find that the most significant impact of NP arises from the WC $C_{L}^{S}=4C^{T}$.
Finally, we study the impact of these NP couplings on various angular and the $CP$ triple product asymmetries, that could be measured in some ongoing and future experiments. The precise measurements of these observables are important to check the SM and extract the possible NP.

\maketitle

\section{Introduction}

The role of high-energy physics accelerators is to directly search for new particles, which could be the intermediate quantum states, also known as the resonances. These quantum states often decay quickly and can be observed by detecting their decay products. The heavy to light semileptonic decays of $b-$ hadrons are important to hunt for the new physics (NP) indirectly. Among them, the decays induced by the Flavor Changing Neutral Current (FCNC) transitions, i.e.,  $b\to (d,s)$ are special to test the Standard Model (SM) precisely and to search for possible deviations from it. Particularly, the rare semi-leptonic $B$ meson decays are useful to investigate the Higgs Yukawa interaction by testing the Lepton Flavor Universality (LFU), which is the ratio having different generations of leptons in the final state. LFU is approximately one in SM, with a small breaking due to lepton masses in the decay phase space. The measured values of the branching ratios and angular distributions of FCNC decays, $B^{\pm}\to K^{(*)\pm}\ell^{+}\ell^{-}$,
$B^{0}\to K^{0}\ell^{+}\ell^{-}$ and $B_{s}^{0}\to\phi\ell^{+}\ell^{-}$,
where $\ell=\mu,\;e$, marked some deviations from the SM 
predictions \cite{LHCb:2014cxe,LHCb:2014auh,LHCb:2015wdu,LHCb:2015svh,LHCb:2016ykl,LHCb:2021zwz,LHCb:2021xxq}.
To accommodate these deviations, there exist a plethora of beyond SM studies,
see e.g., \cite{Celis:2017doq,Buttazzo:2017ixm,Aebischer:2019mlg,Alasfar:2020mne,Isidori:2021tzd,Ciuchini:2022wbq}
and references therein.

Contrary to the CKM-suppressed FCNC decays, which occur at a loop level in the SM, the decays involving the quark level flavor-changing-charged-current (FCCC) transitions $b\rightarrow c\ell\nu_{\ell}$ ($\ell=e,\mu,\tau$) have no such suppression. Interestingly, similar tensions with the SM predictions have been
observed in the FCCC exclusive
decays, $B\rightarrow D^{(*)}\ell\nu_\ell$, $B\rightarrow J/\psi\ell\nu_\ell$,
$\Lambda_{b}\rightarrow\Lambda_{c}\ell\nu_\ell$, giving us an opportunity to mark several physical
observables to test the results of the SM. Theoretically, these exclusive decays are prone to uncertainties coming through
the form factors (FFs) (a non-perturbative input) and errors in the CKM matrix elements. However, the dependence on the CKM elements and the uncertainties from FFs are mitigated if we take the ratio of the branching fractions
$(\mathcal{B})$ in these semi-leptonic decays, i.e.,
 
\begin{equation}
R_{\tau/\mu,e}\left(D^{(*)}\right)=\frac{\mathcal{B}\left(B\rightarrow D^{(*)}\tau\overline{\nu}_{\tau}\right)}{\mathcal{B}\left(B\rightarrow D^{(*)}\ell\bar{\nu}_{\ell}\right)},\quad R_{\tau/\mu}\left(J/\psi\right)=\frac{\mathcal{B}\left(B_{c}^{+}\rightarrow J/\psi\tau^{+}\nu_{\tau}\right)}{\mathcal{B}\left(B_{c}^{+}\rightarrow J/\psi\mu^{+}\nu_{\mu}\right)},\quad R_{\tau/\ell}\left(\Lambda_{c}\right)=\frac{\mathcal{B}\left(\Lambda_{b}\rightarrow\Lambda_{c}\tau\bar{\nu}_{\tau}\right)}{\mathcal{B}\left(\Lambda_{b}\rightarrow\Lambda_{c}\ell\bar{\nu}_{\ell}\right)}.\label{RDRDs}
\end{equation}

 In the SM, using heavy-quark effective theory and the lattice QCD approach the LFU ratio $R_{\tau/\mu,e}\left(D^{(*)}\right)$ are studied  \cite{MILC:2015uhg, Na:2015kha, Aoki:2016frl, Fajfer:2012vx, Bigi:2016mdz, Bernlochner:2017jka, Bigi:2017jbd, Jaiswal:2017rve, Gambino:2019sif, Bordone:2019vic, Martinelli:2021onb}. The corresponding SM results are
 \begin{equation}
R^{\text{SM}}_{\tau/\mu,e}\left(D\right)= 0.298\pm 0.004,\quad\quad\quad R^{\text{SM}}_{\tau/\mu,e}\left(D^*\right) = 0.254\pm 0.005. \label{SM-RDDs}
 \end{equation} 
These theoretical results were scrutinized by the measurements from BABAR,
Belle, and LHCb, where the experimental measurements of both $R_{\tau/{\mu, e}}\left(D^{(*)}\right)$
by BABAR \cite{BaBar:2012obs,BaBar:2013mob}, Belle \cite{Belle:2015qfa,Belle:2019rba,Belle:leptonphoton}
and LHCb \cite{LHCb:2015gmp,LHCb:2017smo,LHCb:2017rln,LHCb:2023zxo,LHCb:2023uiv}
marked significant deviations from the SM predictions. Recently,
Heavy Flavor Averaging Group (HFLAV) updated their earlier results
of $R_{\tau/{\mu, e}}\left(D\right)$ and $R_{\tau/{\mu, e}}\left(D^{*}\right)$ \cite{HFLAV:2022pwe,HFLAV:2022link},
and the current world averages \cite{HFLAV:2023link} exceed the SM
predictions by $2.0\sigma$ and $2.2\sigma$,
respectively. Invoking the correlation of $-0.37$ between these universality ratios, the HFLAV global averages are
\begin{equation}
 R_{\tau/\mu,e}\left(D\right) = 0.357\pm0.029, \quad\quad R_{\tau/\mu,e}\left(D^*\right) = 0.284\pm 0.012, \label{HFLAV-RDDs}
\end{equation}
deviating from the SM predictions
by $3.3\sigma$ \cite{HFLAV:2023link}. However, the SM determination is improved by calculating the next-to-leading order QCD corrections for $B\to D^{(*)}$ form factors \cite{Cui:2023bzr}.

Due to the spins of the final state particles $D^*$ and the lepton in $B\to D^* \ell \bar{\nu}_\ell$, the associated polarization asymmetries are useful to have a consistency check of the NP observed in various LFUs.  Experimentally, the lepton polarization asymmetry $P_\tau\left(D^{*}\right)$ and the longitudinal polarization fraction of $D^{*}$ meson, i.e.,  $F_{L}(D^{*})$ are measured by Belle  \cite{Belle:2017ilt,Belle:2016dyj,Belle:2019ewo}
\begin{equation}
F_{L}\left(D^{*}\right) =  0.60 \pm 0.08\pm 0.04,\quad\quad  P_{\tau}\left(D^{*}\right) = -0.38\pm0.51_{-0.16}^{+0.21}, \label{Exp-PTPD}
\end{equation}
which lies within $1.5\sigma$
of the SM prediction \cite{Alok:2016qyh,Iguro:2020cpg}
\begin{equation}
F^{\text{SM}}_{L}\left(D^{*}\right) =  0.464\pm0.003,\quad\quad  P^{\text{SM}}_{\tau}\left(D^{*}\right) = -0.497\pm 0.007. \label{SM-PTPD}
\end{equation}
The incompatibility of
these measurements with SM predictions are a mainstay in discriminating different NP models. 

In the charmed $B-$meson decay $B_c\to J/\psi \ell \nu_\ell $, occurring through the same $b\to c$ quark level transition,  the LFU ratio $R_{\tau/\mu}\left(J/\psi\right)$ was analyzed by the LHCb \cite{LHCb:2017vlu} collaboration, at $7$ TeV and
$8$ TeV center of mass energies of proton-proton collision with an integrated
luminosity of $3fb^{-1}$.  It is found that the mismatch between the experimentally measured value \cite{LHCb:2017vlu} 
\begin{equation}
 R_{\tau/\mu}\left(J/\psi\right) = 0.71\pm 0.17\pm 0.18, \label{Exp-RJpsi}
\end{equation}
and the SM prediction \cite{Watanabe:2017mip, Harrison:2020nrv} 
\begin{equation}
  R^{\text{SM}}_{\tau/\mu}\left(J/\psi\right) =   0.258 \pm 0.038 \label{SMRJPsi}
\end{equation}
is $1.8\sigma$. Similar to these LFU ratios, one can define the ratio for the inclusive $B-$meson decay $B\to X_c\tau\bar{\nu}_{\tau}$
\begin{equation}
  R_{\tau/\ell}\left(X_c\right) = \frac{\mathcal{B}\left(B\to X_c\tau\bar{\nu}_{\tau}\right)}{\mathcal{B}\left( B\to X_c \ell \bar{\nu}_\ell \right)},
\end{equation}
where the SM result $0.216\pm0.003$ \cite{Kamali:2018bdp} lies within $(~0.2\sigma)$ to the experimental measurements \cite{ALEPH:2000vvi} 
\begin{equation}
     R_{\tau/\ell}\left(X_c\right) = 0.223 \pm 0.030 \label{Exp-BXc}.
\end{equation}
    
We know that LHCb is quite rich in producing the $b-$ hadrons and the ratio of $B : \Lambda_b$ is 2:1, giving us an opportunity to study the Cabibbo suppressed decay modes of $\Lambda^0_b$  \cite{LHCb:2013jih, LHCb:2014scu,LHCb:2014nhe,OHanlon:2016eah, LHCb:2017gjj}.  Analogous to the ratios of $B$ meson decays, the ratio $R_{\tau/\ell}(\Lambda_{c})$ involving the
$\Lambda_{b}$ baryon decay, has also been measured by LHCb collaboration
in $\Lambda_{b}\rightarrow\Lambda_{c}\tau\nu_\tau$ decays \cite{LHCb:2022piu}:
\begin{equation}
    R_{\tau/\ell}\left(\Lambda_{c}\right) =  0.242 \pm 0.026\pm 0.040\pm 0.059, \label{Exp-Lambdac}
\end{equation}
where the first, second, and third uncertainties are due to the statistical, system and external branching fractions measurements, respectively. This experimental value conflicts with the corresponding SM result \cite{Detmold:2015aaa, Bernlochner:2018kxh} $0.324\pm0.004$ by $1.8\sigma$. The underabundance of taus in these $\Lambda_b$ decays compared to the other ratios involving $b \to c \tau \bar{\nu}_\tau$ is a perplexing behavior. 

Considering the left-handed (LH) neutrinos, these incompatibilities between the SM predictions and the experimental measurements were explored  through the NP dimension six operators in $b\rightarrow c\tau\overline{\nu}_{\tau}$ transitions \cite{Blanke:2018yud,Blanke:2019qrx,Huang:2018nnq,Alok:2019uqc, Sahoo:2019hbu,Shi:2019gxi,Bardhan:2019ljo, Fedele:2022iib, Asadi:2019xrc,Murgui:2019czp,Mandal:2020htr,Cheung:2020sbq,Colangelo:2020vhu}. Including right-handed (RH) neutrinos and/or the RH quark currents in the model independent WEH, these anomalies were analyzed in a number of studies, see e.g., \cite{Greljo:2018ogz,Azatov:2018kzb,Heeck:2018ntp,Babu:2018vrl,He:2017bft,Gomez:2019xfw,Alguero:2020ukk,Dutta:2013qaa,Dutta:2017xmj,Dutta:2017wpq,Dutta:2018jxz}. Particularly with the assumption of LH neutrinos, using the experimental measurements of $R_{\tau/{\mu,e}}\left(D\right), R_{\tau/{\mu,e}}\left(D^*\right), R_{\tau/{\mu,e}}\left(J/\Psi\right)$ and $F_{L}\left(D^*\right)$,  Blanke \textit{et al.}  \cite{Blanke:2018yud} have elegantly analyzed the four one-dimensional scenarios with real Wilson coefficients (WCs) $C^{V}_L, C^{S}_R, C^{S}_L$ and $C^{S}_{L}=4C^T$, where the subscripts specify the Lorentz structure of the current corresponding to these WCs in WEH. Focusing on the same experimental observables, they had further extended their analysis to 2D scenarios by considering only $C^{S}_L=4C^{T}$ to be the complex-valued.  Considering the fact that the scenarios deeply influence the branching ratio of  $B_{c}\rightarrow\tau\nu_\tau$ decay, which due to a large uncertainty in the lifetime of $B_c$ meson \cite{Celis:2016azn, Alonso:2016oyd} is not measured yet; therefore,  an upper limit of
$60\%, 30\%$, and $10\%$ on the contribution from $B_{c}\rightarrow\tau\nu_\tau$
to total $B_{c}$ decay width has been imposed in literature \cite{Gershtein:1994jw,Bigi:1995fs,Beneke:1996xe,Chang:2000ac,Kiselev:2000pp,Akeroyd:2017mhr}. This impacts the parametric space for these NP WCs, and it was found in \cite{Blanke:2018yud} that for the $10\%$ limit, the only scenario $C^R_S$  departs the fit from measurements; whereas, in the 2D scenarios, the fit is valid for two out of four benchmark scenarios. Later, using these constraints on the available parametric space of the NP WCs, the values of $P_T\left(D^*\right)$ and $R_{\tau/\ell}(\Lambda_{c})$ are predicted for both cases.

Now, after giving this detailed overview of the theoretical and experimental studies, the work presented here has two important features. In the first step, using the model-independent WEH with LH neutrinos and complex NP Wilson coefficients, we take the most up-to-date HFLAV world average summer 2023 \cite{HFLAV:2023link} values for $R_{\tau/\mu,e}(D)$ and $R_{\tau/{\mu,e}}(D^{*})$, and the measurements of the   $F_{L}\left(D^{*}\right), P_{\tau}\left(D^{*}\right), R\left(J/\psi\right)$ and $R_{\tau/\ell}\left(X_c\right)$  given in Eqs. (\ref{RDRDs} - \ref{Exp-Lambdac}) to revisit the global fit analysis. Though we have complex NP WCs, we employ the same technique developed in \cite{Blanke:2018yud} to scrutinize NP parametric space for real WCs. In this situation, we perform a $\chi^{2}$ analysis by considering
two sets of physical observables. In set $\mathcal{S}_1$ we choose $R_{\tau/\mu,e}(D), R_{\tau/{\mu,e}}(D^{*}), F_{L}\left(D^{*}\right)$ and $ P_{\tau}\left(D^{*}\right)$ , whereas in $\mathcal{S}_2$ we included  $R_{\tau/\mu}\left(J/\Psi\right)$ and $R_{\tau/\ell}(X_{c})$ in the list too. The purpose of these two sets is to scrutinize the parametric space of the new WCs.
For the set $\mathcal{S}_1$, we find that the effect of the NP scalar WC $C^{L}_S$ is prominent compared to the other WCs, and it also shows a strong dependence on the constraints arising due to the branching ratio of $B_c \to \tau \nu$. Expecting the $p$-values to be $\sim 50\%$ for a true solution, we observe a less favorable alignment with the observed anomalies for the set $\mathcal{S}_2$.

It is established that the correlation studies of the various observables associated with these decay modes will be an important tool for indirect detection of NP, the LFU ratios $R_{\tau/{\mu,e}}\left(D\right)$, $R_{\tau/{\mu,e}}\left(D^*\right)$ and  $R_{\tau/{\ell}}\left(\Lambda_{c}\right)$ are correlated in a model-independent way \cite{Blanke:2018yud,Fedele:2022iib}. The relation of $R_{\tau/{\ell}}\left(\Lambda_{c}\right)$ with well measured  $R_{\tau/{\mu,e}}\left(D\right)$, $R_{\tau/{\mu,e}}\left(D^*\right)$ is known as the sum-rule, derived in \cite{Blanke:2018yud} and further updated in \cite{Fedele:2022iib}.

In the second step, we find similar relations/sum- rules for the other LFU ratios $R_{\tau/{\mu,e}}\left(J/\Psi\right)$ and $R_{\tau/\ell}(X_{c})$ by expressing them in terms of  $R_{\tau/\mu,e}(D)$ and $R_{\tau/{\mu,e}}(D^{*})$, and calculate their numerical values including the uncertainties from the SM inputs and experimental measurements of  $R_{\tau/\mu,e}(D)$ and $R_{\tau/{\mu,e}}(D^{*})$.  Apart from this correlation established through these sum rules, we also find their correlation with $R_{D}$ and $R_{D^*}$ by plotting them for particular values of NP WCs in $\mathcal{S}_1$.
Finally, we studied the impact of these NP WCs on the physical observables such as the CP even and odd angular observables in $B\to D^*\left(\to D\pi\right) \tau \bar{\nu}_\tau$ decays, and the different CP asymmetry triple products. 

The study is organized as follows; In Sect. \ref{sec2}, after defining the WEH containing the SM and NP operators, we collect  the formulas
of the observables $R_{\tau/{\mu,e}}\left(D\right)$, $R_{\tau/{\mu,e}}\left(D^{*}\right)$,  $P_{\tau}(D^{*})$, $F_{L}(D^{*})$, $R_{\tau/\mu}(J/\psi)$,
$R_{\tau/\ell}(X_{c})$ and $R_{\tau/\ell}\left(\Lambda_{c}\right)$ in terms of the NP WCs.
Considering the most recent data, in Sect. \ref{sec3}, we analyze the parametric space of our complex NP WCs and discuss how the limits on the branching ratio of $B_c\to \tau \bar{\nu}_\tau$ impact allowed regions of these WCs. In Sect. \ref{sec4}, we workout the sum rules for $R_{\tau/{\mu}}\left(J/\Psi\right)$ and $R_{\tau/\ell}(X_{c})$ in terms of  $R_{\tau/\mu,e}(D)$ and $R_{\tau/{\mu,e}}(D^{*})$ and discuss the correlation amongst the physical observables.
Sect. \ref{sec5} discusses the impact of these constraints on polarized branching ratios and various $CP$ even, odd angular observables and the $CP$ asymmetry triple products in $B\to D^*\tau\bar{\nu}_\tau$ decays. Finally, in Sect. \ref{sec6}, we conclude our findings. This work is supplemented by three appendices, discussing the fitting procedure and the derivation of the above-mentioned $CP$ asymmetries.

\section{Theoretical Framework and Analytical Formulae}\label{sec2}
\subsection{Weak Effective Hamiltonian (WEH)}

In the absence of experimental evidence of deviations
from the SM predictions at tree-level transitions involving light leptons, it is generally believed that the NP is supposed to play a role in the third-generation fermions, i.e., $\tau$. Regarding this, to explore any possible NP in the $b\to c\tau\bar{\nu}_\tau$ transition, we consider the WEH containing new dimension-six vector, scalar, and tensor operators with complex WCs. After integrating out the heavy degrees of freedom, the WEH for $b\to c\tau\bar{\nu}_\tau$ transition can be written as \cite{Blanke:2018yud, Mandal:2020htr,Iguro:2018vqb,Asadi:2018wea,Asadi:2018sym,Ligeti:2016npd,Robinson:2018gza}:
\begin{equation}
H=\frac{4G_{F}V_{cb}}{\sqrt{2}}\left\{\left(C_{L}^{V}\right)_{SM}O_{L}^{V}+C_{L}^{V}O_{L}^{V}+C_{R}^{S}O_{R}^{S}+C_{L}^{S}O_{L}^{S}+C^{T}O^{T}\right\}+h.c.\label{eq1}
\end{equation}
Here, $G_{F}$ is the Fermi coupling constant, and $V_{cb}$ is the CKM
matrix element. The first term is the SM contribution, and its corresponding WC is normalized to unity, i.e.,
$\left(C_{L}^{V}\right)_{SM}=1$. Ignoring the small mass of neutrinos and the RH vector currents for quarks, the operators $O_V^L$, $O_S^R$, $O_S^L$ and $O_T$ read as \cite{Buchmuller:1985jz,Grzadkowski:2010es,Aebischer:2015fzz}:
\begin{eqnarray}
O_{L}^{V} & = & \left(\overline{c}\gamma^{\mu}P_{L}b\right)\left(\overline{\tau}\gamma_{\mu}P_{L}\nu\right)\nonumber \\
O_{R}^{S} & = & \left(\overline{c}P_{R}b\right)\left(\overline{\tau}P_{L}\nu\right),\nonumber \\
O_{L}^{S} & = & \left(\overline{c}P_{L}b\right)\left(\overline{\tau}P_{L}\nu\right),\nonumber \\
O^{T} & = & \left(\overline{c}\sigma^{\mu\nu}P_{L}b\right)\left(\overline{\tau}\sigma_{\mu\nu}P_{L}\nu\right),\label{eq2}
\end{eqnarray}
where $P_{R,L}=\left(1\pm\gamma_{5}\right)/2$ are the projection operators, where plus (minus) sign represents the right(left)-handed chiralities. The corresponding NP WCs are $C_{R,L}^{X}$, where the upper index $X$
represents scalar, vector, and tensor contributions, whereas the lower indices $R,L$ denote
quark-current chiral projections.  As we have only considered the LH neutrinos, therefore, the lepton chirality index is not specified in the subscript of the NP WCs. 

The values of the NP WCs
depend on the renormalization scale of the theory. It is important to mention here that $b\to c\tau\bar{\nu}_\tau$ transitions occur at $m_b$ scale, whereas the NP WCs determine at a scale of heavy NP particles, which may set at  $1$ TeV \cite{Blanke:2018yud} or $2$ TeV \cite{Fedele:2022iib}. The WCs at $m_b$ and $1$ TeV scales are connected by the Renormalizable Group equations (RGEs), and the corresponding relationships are
\cite{Gonzalez-Alonso:2017iyc, Blanke:2018yud}
\begin{align}
C_{L}^{V}(m_{b}) & = C_{L}^{V}(1\text{ TeV}),\nonumber \\
C_{R}^{S}(m_{b}) & = 1.737 C_{R}^{S}(1\text{ TeV}),\nonumber \\
\left(\begin{array}{c}
C_{L}^{S}(m_{b})\\
C^{T}(m_{b})
\end{array}\right) & =\left(\begin{array}{cc}
1.752 & -0.287\\
-0.004 & 0.842
\end{array}\right)\left(\begin{array}{c}
C_{L}^{S}(1\text{ TeV})\\
C^{T}(1\text{ TeV})
\end{array}\right).\label{eq3}
\end{align}

For the WEH given in Eq. (\ref{eq1}), the expressions of
the physical observables under consideration can be written in terms of the NP WCs at a scale
$\mu_b = m_{b}$ as \cite{Watanabe:2017mip,Iguro:2018vqb,Asadi:2018sym,Gomez:2019xfw,Cardozo:2020uol,Fedele:2022iib,Mandal:2020htr,Kamali:2018bdp,Iguro:2022yzr}:

\begin{eqnarray}
R_{\tau/{\mu,e}}(D) & = & R_{\tau/{\mu,e}}^{\text{SM}}(D)\left\{ \left|1+C_{L}^{V}\right|^{2}+1.01\left|C_{R}^{S}+C_{L}^{S}\right|^{2}+0.84\left|C^{T}\right|^{2}\right.\nonumber \\
 &  & \left.+1.49\Re\left[\left(1+C_{L}^{V}\right)\left(C_{R}^{S}+C_{L}^{S}\right)^{*}\right]+1.08\Re\left[\left(1+C_{L}^{V}\right)\left(C^{T}\right)^{*}\right]\right\} ,\label{eqn1}\\
R_{\tau/{\mu,e}}(D^{*}) & = & R_{\tau/{\mu,e}}^{\text{SM}}(D^{*})\left\{ \left|1+C_{L}^{V}\right|^{2}+0.04\left|C_{R}^{S}-C_{L}^{S}\right|^{2}+16\left|C^{T}\right|^2\right.\nonumber \\
 &  & \left.\left.+0.11\Re\left[\left(1+C_{L}^{V}\right)\left(C_{R}^{S}-C_{L}^{S}\right)^{*}\right]-5.17\Re\left[\left(1+C_{L}^{V}\right)\left(C^{T}\right)^{*}\right]\right\} \right.,\label{eqn2}\\
P_{\tau}\left(D^{*}\right) & = & P_{\tau}^{SM}\left(D^{*}\right)\left(\frac{R_{\tau/{\mu,e}}(D^{*})}{R_{\tau/{\mu,e}}^{\text{SM}}(D^{*})}\right)^{-1}\left\{ \left|1+C_{L}^{V}\right|^{2}-0.07\left|C_{R}^{S}-C_{L}^{S}\right|^{2}-1.85\left|C^{T}\right|^{2}\right.\nonumber \\
 &  & \left.\left.-0.23\Re\left[\left(1+C_{L}^{V}\right)\left(C_{R}^{S}-C_{L}^{S}\right)^{*}\right]-3.47\Re\left[\left(1+C_{L}^{V}\right)\left(C^{T}\right)^{*}\right]\right\} \right.,\label{eqn3}\\
F_{L}\left(D^{*}\right) & = & F_{L}^{SM}\left(D^{*}\right)\left(\frac{R_{\tau/{\mu,e}}(D^{*})}{R_{\tau/{\mu,e}}^{\text{SM}}(D^{*})}\right)^{-1}\left\{ \left|1+C_{L}^{V}\right|^{2}+0.08\left|C_{R}^{S}-C_{L}^{S}\right|^{2}+6.9\left|C^{T}\right|^{2}\right.\nonumber \\
 &  & \left.\left.+0.25\Re\left[\left(1+C_{L}^{V}\right)\left(C_{R}^{S}-C_{L}^{S}\right)^{*}\right]-4.3\Re\left[\left(1+C_{L}^{V}\right)\left(C^{T}\right)^{*}\right]\right\} \right.,\label{eqn4}\\
R_{\tau/{\mu}}(J/\psi) & = & R_{\tau/{\mu}}^{\text{SM}}(J/\psi)\left\{ \left|1+C_{L}^{V}\right|^{2}+0.04\left|C_{R}^{S}-C_{L}^{S}\right|^{2}+14.7\left|C^{T}\right|^{2}\right.\nonumber \\
 &  & \left.+0.1\Re\left[\left(1+C_{L}^{V}\right)\left(C_{R}^{S}-C_{L}^{S}\right)^{*}\right]-5.39\Re\left[\left(1+C_{L}^{V}\right)\left(C^{T}\right)^{*}\right]\right\} ,\label{eqn5}
 \end{eqnarray}
 \begin{eqnarray}
R_{\tau/{\ell}}(X_{c}) & = & R_{\tau/{\ell}}^{\text{SM}}(X_{c})\left\{ 1+1.147\left(\left|C_{L}^{V}\right|^{2}+2\Re\left[C_{L}^{V}\right]\right)+0.327\left|C_{R}^{S}+C_{L}^{S}\right|^{2}+0.031\left|C_{R}^{S}-C_{L}^{S}\right|^{2}\right.\nonumber \\
 &  & \left.+12.637\left|C^{T}\right|^{2}+0.493\Re\left[\left(1+C_{L}^{V}\right)\left(C_{R}^{S}+C_{L}^{S}\right)^{*}\right]+0.096\Re\left[\left(1+C_{L}^{V}\right)\left(C_{R}^{S}-C_{L}^{S}\right)^{*}\right]\right\}, \label{eqn6}\\
R_{\tau/{\ell}}\left(\Lambda_{c}\right) & = & R_{\tau/{\ell}}^{\text{SM}}\left(\Lambda_{c}\right)\left\{ \left|1+C_{L}^{V}\right|^{2}+0.32\left(\left|C_{R}^{S}\right|^{2}+\left|C_{L}^{S}\right|^{2}\right)+0.52\Re\left[C_{R}^{S}\left(C_{L}^{S}\right)^{*}\right]+10.4\left|C^{T}\right|^{2}\right.\nonumber \\
 &  &\left. +0.5\Re\left[\left(1+C_{L}^{V}\right)\left(C_{R}^{S}\right)^{*}\right]+0.33\Re\left[\left(1+C_{L}^{V}\right)\left(C_{L}^{S}\right)^{*}\right]-3.11\Re\left[\left(1+C_{L}^{V}\right)\left(C^{T}\right)^{*}\right]\right\}. \label{eqn7}
\end{eqnarray}
Similarly, the expression of branching
ratio of the $B_c \to \tau \bar{\nu}_\tau$ decay read as 
\begin{eqnarray*}
\mathcal{B}\left(B_{c}^{-}\rightarrow\tau^{-}\bar{\nu}_{\tau}\right) & = & \mathcal{B}\left(B_{c}^{-}\rightarrow\tau^{-}\bar{\nu}_{\tau}\right)^{\text{SM}}\left\{ \left|1+C_{L}^{V}+4.3\left(C_{R}^{S}-C_{L}^{S}\right)\right|^{2}\right\} ,
\end{eqnarray*}
where in the SM $\mathcal{B}\left(B_{c}^{-}\rightarrow\tau^{-}\bar{\nu}_{\tau}\right)^{\text{SM}}\approx 0.022$  \cite{Iguro:2022yzr}.

\section{Analysis of the parametric space of New Physics Wilson Coefficients}\label{sec3}
In this section, we will scrutinize the parametric space for the NP complex WCs. We follow the same $\chi^2$ fitting technique that is developed for real WCs in \cite{Blanke:2018yud}, and it is summarized in Appendix \ref{GoF}. In addition to the complex WCs in our case, the other differences from \cite{Blanke:2018yud} are; we have incorporated the most updated results of observables from HFLAV \cite{HFLAV:2023link}, and in Eqs. (\ref{eqn1} - \ref{eqn7}), the latest values of the FFs are used.
Furthermore, based on the number of experimentally measured observables included in the analysis, we will consider two different sets, namely  
\begin{itemize}
    \item In
    Set 1 $(\mathcal{S}_1)$, by choosing  $R_{\tau/\mu,e}(D), R_{\tau/{\mu,e}}(D^{*}), F_{L}\left(D^{*}\right)$ and $ P_{\tau}\left(D^{*}\right)$, we find the best-fit points (BFP) for these complex WCs, along with $1\sigma$ and $2\sigma$ variations.
    \item In Set 2 $(\mathcal{S}_2)$, we will add the two more observables $R_{\tau/\mu}(J/\psi)$ and $ R_{\tau/\ell}\left(X_c\right)$ to $\mathcal{S}_1$ and see if we could further constraint the parametric space of these NP WCs.
\end{itemize}

Due to complex NP WCs, to perform the $\chi^2$ analysis,  the number of parameters, $N_{par}$, equal to $2$ and $N_{obs}$ to be $4$ and $6$ for $\mathcal{S}_1$ and $\mathcal{S}_2$, respectively. Thus, the number of degrees of freedom (dof):  $N_{dof}=N_{obs}-N_{par}$ gives $2$ and $4$ for $\mathcal{S}_1$ and $\mathcal{S}_2$, respectively. The $p-$value quantifies the goodness of fit and it is attributed to the probability of difference between the theoretical (SM) and experimental values. This can be hypothesized as 
\begin{equation}
p=\intop_{\chi^{2}}^{\infty}f\left(z;N_{dof}\right)dz,
\end{equation}
where $f(z;N_{dof})$ is the $\chi^{2}$ probability distribution function. As $N_{par}=0$ for the SM, therefore, the $p$-value is very small $\left(\approx 10^{-3}\right)$ \cite{Blanke:2018yud}. Our main objective is to estimate the BFP, 
$p-$value, $\chi_{\text{SM}}^{2}$, $\text{pull}_{\text{SM}}$ and $1\sigma$, $2\sigma$
intervals for the WCs.

Just to emphasize again, the parametric space of the WCs depends upon the constraints on the branching ratio of a helicity suppressed SM decay mode $B^-_{c}\to \tau^- \bar{\nu}_\tau$, which is not precisely measured yet (see e.g., \cite{Blanke:2018yud} for an elaborated discussion). However, we have some experimental bounds on it, therefore, to impose the cuts on the allowed parametric space, one has to put some upper bounds on the branching ratio of this decay. To begin with, we consider the most conservative bound of $<60\%$ contribution of the $B^-_{c}\to \tau^- \bar{\nu}_\tau$ decay rate to the total decay width of $B_c$ meson, then take the commonly used $<30\%$ and finally a hypothetical future
limit of $<10\%$  \cite{Blanke:2018yud}. 

\begin{figure}[H]
\centering{} \subfloat[]{\includegraphics[width=6.5cm, height=6cm]{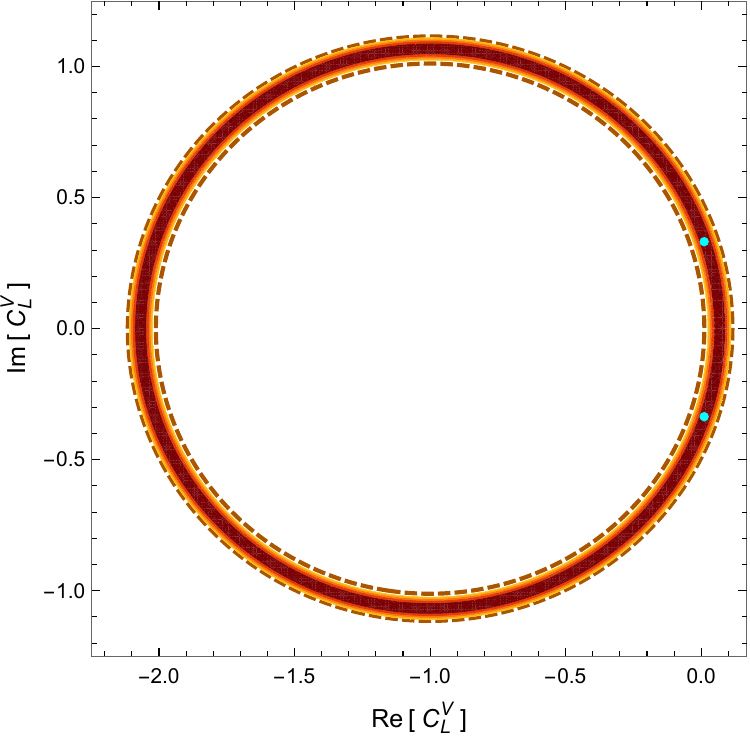}}\quad \subfloat[]{\includegraphics[width=6.5cm, height=6cm]{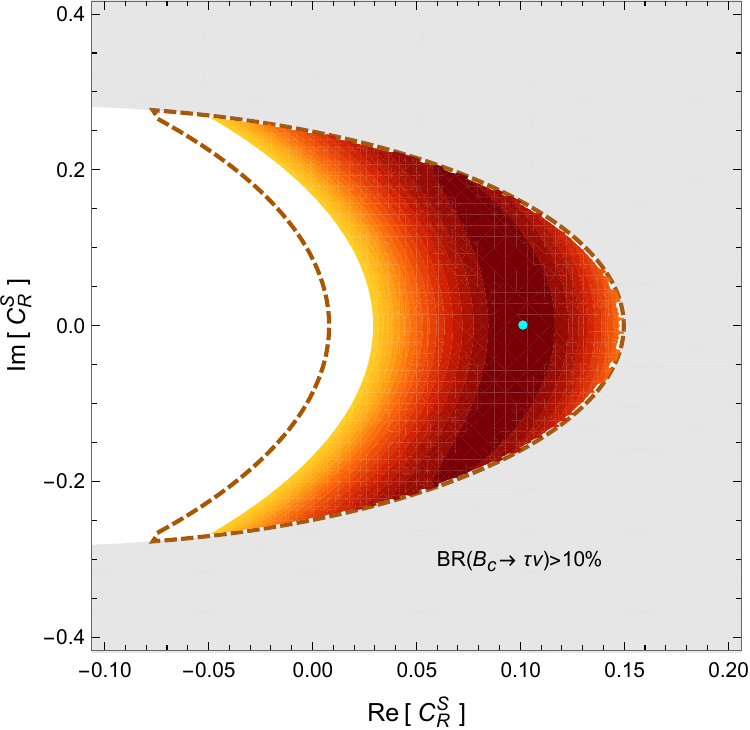}} \includegraphics[width=1cm, height=6cm]{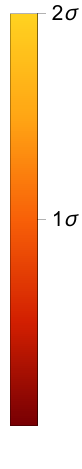}

\centering{} \subfloat[]
{\includegraphics[width=6.5cm, height=6cm]{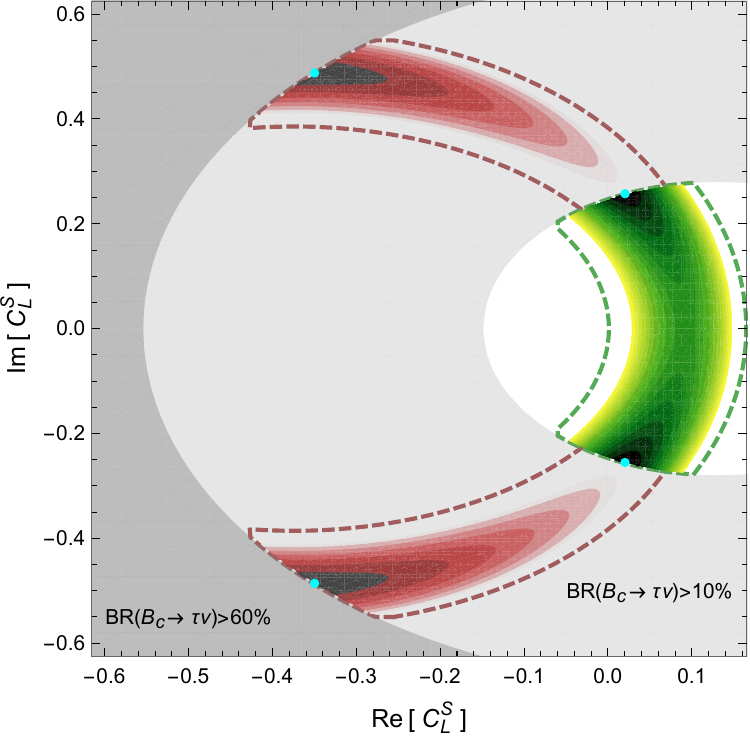}}\quad \subfloat[]{\includegraphics[width=6.5cm, height=6cm]{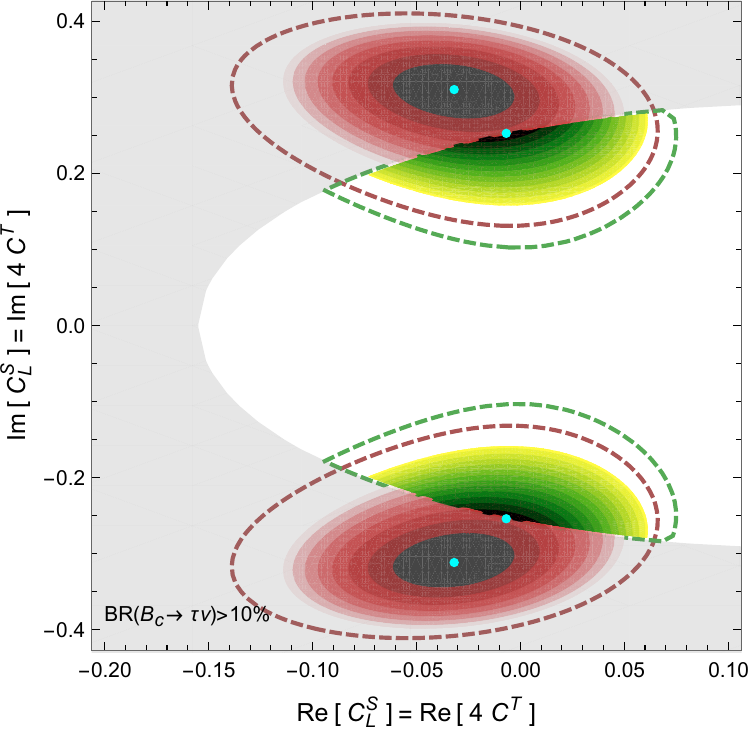}} \includegraphics[width=2cm, height=6cm]{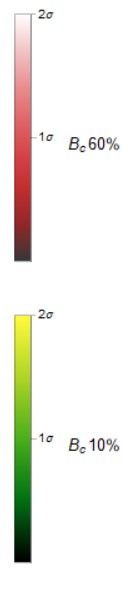}
\caption{\label{Monika fig2}Results of the fits for NP scenarios. The light and dark gray color shows the $10\%$ and $60\%$ branching ratio constraints. The light (dark) color contours represent the $1(2\sigma)$ deviations from the BFP (cyan color). Panels (a), (b), (c), and (d) show the ranges of real and imaginary parts of $C_{L}^{V}$, $C_{R}^{S}$, $C_{L}^{S}$ and 
$C_{L}^{S}=4C^{T}$, respectively restricted by the $\mathcal{S}_1$. The contours drawn with dotted lines show the maximum limit of $2\sigma$ region for the set $\mathcal{S}_2$. In Figs. 1a and 1b, the (orange color coding) is not affected by either of these constraints, whereas in Figs. 1c and 1d, the red and green colors coding are the $60\%$ and $10\%$ constraints, respectively. The cyan dots show the BPF in two sets.}
\end{figure}

In our analysis, we adopted the strategy of switching on one NP WC at a time, and the corresponding results are shown in Fig. \ref{Monika fig2} for both sets $\mathcal{S}_{1}$ (solid colors) and $\mathcal{S}_2$ (dotted lines). We have also included the $10\%$ and $60\%$ constraints of $\mathcal{B}\left(B^-_c\to \tau^- \bar{\nu}_\tau\right)$, drawn by light and dark gray colors, respectively. Any point lying inside the gray shades shows the disallowed parametric space corresponding to a particular limit on the branching ratio. The description of various colors is given in the caption of Fig. \ref{Monika fig2}. 

\begin{itemize}
    \item In Fig. \ref{Monika fig2}a the allowed parametric space of $C^{V}_L$ is plotted. We can see that for $C^{V}_L$, the parametric space is independent of the constraints from $\mathcal{B}\left(B^-_c\to \tau^- \bar{\nu}_\tau\right)$, which is not the case for the real WCs  \cite{Blanke:2018yud}.  The corresponding real and imaginary values of the BFP, $\chi^2_{\text{min}}$, the percentage $p-\text{value}$, $\text{pull}_{\text{SM}}$ and $1\sigma$, $2\sigma$ ranges around BFP are given in Table \ref{Monika tab2}. We can see that going from the $\mathcal{S}_1$ (solid colors) to $\mathcal{S}_2$ (dotted lines) the corresponding BFP value is reduced by $38\%$. Also, the $p-\text{value}$ is reduced by $30\%$ making it less likely to fit the anomalies. Additionally, the $1\sigma$ and $2\sigma$ ranges around the BFP are drawn with light and dark colors for $\mathcal{S}_1$ and $\mathcal{S}_2$, respectively. Here, the dotted line represents the maximum of $2\sigma$ for $\mathcal{S}_2$.
    \item Fig. \ref{Monika fig2}b represents the allowed regions for $C^{S}_R$ for sets $\mathcal{S}_1$ and $\mathcal{S}_2$ with the color description being the same as in Fig. \ref{Monika fig2}a. One can see that the $1\sigma$ and $2\sigma$ ranges lie beyond the $10\%$ constraints of the branching ratio. From Table \ref{Monika tab2}, one can read that for the BFP the imaginary component of $C^{S}_R$ is zero to a given accuracy, and the corresponding $p-\text{value}$ is small too. However, for the corresponding $1\sigma$ and $2\sigma$ intervals, this is not the case, as the real and imaginary components are of equal weight here. 
     \item Compared to $C^{V}_L$ and $C^{S}_R$, the complex WC $C_{L}^{S}$ is notably influenced by the branching ratio constraints for both sets $\mathcal{S}_1$ and $\mathcal{S}_2$. This dependence is expected because the SM helicity suppression of $B_c\to \tau^{-}\bar{\nu}_\tau$ is lifted due to these left-handed scalar/pseudo-scalar type operators. In the case of $\mathcal{S}_1$, for a $60\%$ bound on the branching ratio, its $p-\text{value}$ is $64\%$. However, for the commonly used bound of
$30\%$ on it, the $p-\text{value}$ decreases by quarter $16\%$, and in the hypothetical
future limit $10\%$, this value reduces to just
$2.7\%$, disfavoring the pseudo-scalar explanation of the data. Contrary to $\mathcal{S}_1$, in the set $\mathcal{S}_2$, the $p-\text{value}$ is already reduced by a factor of half even for the $60\%$ bound on the branching ratio. The reason is the large uncertainty in the measurement of $R_{\tau/\mu}\left(J/\psi\right)$. In Fig. \ref{Monika fig2}c, we have plotted the corresponding BFP and the $1\sigma$ and $2\sigma$ ranges. We can see that the large parametric space is available when we consider the $60\%$ limit, and this reduces when we apply the $10\%$ bound, justifying our findings presented in Table \ref{Monika tab2}.

\item Furthermore, our fourth scenario, which is $C_{L}^{S}=4C^{T}$ shows
moderate sensitivity only when the stringent future limit of $<10\%$ on the branching ratio
is used. It reduces the $p-\text{value}$ from $26\%$ to $10\%$, when we changed the $\mathcal{B}\left(B^-_c\to \tau^-\bar{\nu}_\tau\right)$ from $60\%$ to $10\%$, showing an adequate dependence on these branching ratio constraints. This is also clear from Fig. \ref{Monika fig2}d. 
\end{itemize}

\begin{table}[H]
\centering{}%
\begin{tabular}{|c|c|c|c|c|c|c|c|}
\toprule 
\multicolumn{8}{c}{Set $\mathcal{S}_1\left(\mathcal{S}_2\right)$ ( $\chi_{\text{SM}}^{2}=17.22\left(20.53\right)$, $p-\text{value}=1.75\left(3.38\right)\times10^{-3}$
)}
\tabularnewline
\midrule 
\hline\hline
WC & BR & BFP & $\chi_{\text{min}}^{2}$ & $p-\text{value}$ $\%$ & $\text{pull}_{\text{SM}}$ & $1\sigma$-range & $2\sigma$-range\tabularnewline
\midrule
\midrule 
\hline
\multirow{3}{*}{$C_{L}^{V}$} & \multirow{2}{*}{-} & $(0.01,\pm0.33)$ & $2.76$ & $25.2$ & $3.8$ & $[-2.09,0.09],[-1.09,1.09]$ & $[-2.11,0.11],[-1.11,1.11]$\tabularnewline
 &  & $(-0.03,\pm0.45)$ & $6.31$ & $17.7$ & $3.77$ & $[-2.1,0.1],[-1.1,1.1]$ & $[-2.12,0.12],[-1.12,1.12]$\tabularnewline
\cmidrule{2-8} \cmidrule{3-8} \cmidrule{4-8} \cmidrule{5-8} \cmidrule{6-8} \cmidrule{7-8} \cmidrule{8-8}
 & & $ 38\%$ & $3.56$ & $-30\%$ & $-0.03$ & $1\%$ & $1\%$\tabularnewline
\midrule 
\hline
\multirow{3}{*}{$C_{R}^{S}$} & \multirow{2}{*}{-} & $(0.1,0)$ & $5.67$ & $5.9$ & $3.4$ & $[0,0.14],[-0.25,0.25]$ & $[-0.05,0.03],[-0.27,0.27]$\tabularnewline
 &  & $(0.1,0)$ & $9.18$ & $5.7$ & $3.37$ & $[-0.04,0.04],[-0.27,0.27]$ & $[-0.08,0.01],[-0.28,0.28]$\tabularnewline
\cmidrule{2-8} \cmidrule{3-8} \cmidrule{4-8} \cmidrule{5-8} \cmidrule{6-8} \cmidrule{7-8} \cmidrule{8-8} 
 &  & $0\%$ & $3.51$ & $-3\%$ & $-0.03$ & $14\%$ & $4\%$\tabularnewline
\midrule 
\hline
\multirow{9}{*}{$C_{L}^{S}$} & \multirow{2}{*}{$<60\%$} & $(-0.35,\pm0.49)$ & $0.89$ & $64.1$ & $4.04$ & $[-0.39,-0.15],[-0.53,0.53]$ & $[-0.42,0.02],[-0.55,0.55]$\tabularnewline
 &  & $(-0.42,\pm0.48)$ & $4.36$ & $35.9$ & $4.02$ & $[-0.41,-0.03],[-0.54,0.54]$ & $[-0.44,0.12],[-0.55,0.55]$\tabularnewline
\cmidrule{2-8} \cmidrule{3-8} \cmidrule{4-8} \cmidrule{5-8} \cmidrule{6-8} \cmidrule{7-8} \cmidrule{8-8} 
 &  & $12\%$ & $3.47$ & $-44\%$ & $-0.02$ & $13\%$ & $10\%$\tabularnewline
\cmidrule{2-8} \cmidrule{3-8} \cmidrule{4-8} \cmidrule{5-8} \cmidrule{6-8} \cmidrule{7-8} \cmidrule{8-8} 
 & \multirow{2}{*}{$<30\%$} & $(-0.13,\pm0.41)$ & $3.65$ & $16.1$ & $3.68$ & $[-0.17,-0.02],[-0.43,0.43]$ & $[-0.2,0.11],[-0.45,0.45]$\tabularnewline
 &  & $(-0.13,\pm0.41)$ & $7.07$ & $13.2$ & $3.67$ & $[-0.19,0.06],[-0.44,0.44]$ & $[-0.21,0.15],[-0.45,0.45]$\tabularnewline
\cmidrule{2-8} \cmidrule{3-8} \cmidrule{4-8} \cmidrule{5-8} \cmidrule{6-8} \cmidrule{7-8} \cmidrule{8-8} 
 &  & $0\%$ & $3.42$ & $-18\%$ & $-0.01$ & $12\%$ & $5\%$\tabularnewline
\cmidrule{2-8} \cmidrule{3-8} \cmidrule{4-8} \cmidrule{5-8} \cmidrule{6-8} \cmidrule{7-8} \cmidrule{8-8} 
 & \multirow{2}{*}{$<10\%$} & $(0.02,\pm0.26)$ & $7.26$ & $2.7$ & $3.16$ & $[-0.02,0.1],[-0.27,0.27]$ & $[-0.05,0.15],[-0.28,0.28]$\tabularnewline
 &  & $(0.02,\pm0.26)$ & $10.65$ & $3.1$ & $3.14$ & $[-0.04,0.13],[-0.27,0.27]$ & $[-0.07,0.16],[-0.28,0.28]$\tabularnewline
\cmidrule{2-8} \cmidrule{3-8} \cmidrule{4-8} \cmidrule{5-8} \cmidrule{6-8} \cmidrule{7-8} \cmidrule{8-8} 
 &  & $0\%$ & $3.38$ & $17\%$ & $-0.02$ & $9\%$ & $5\%$\tabularnewline
\midrule 
\hline
\multirow{6}{*}{$C_{L}^{S}=4C^{T}$} & \multirow{2}{*}{$\begin{array}{c}
<60\%\\
\&30\%
\end{array}$} & $(-0.03,\pm0.31)$ & $2.69$ & $26.1$ & $3.81$ & $[-0.08,0.02],[-0.37,0.37]$ & $[-0.11,0.05],[-0.4,0.4]$\tabularnewline
 &  & $(-0.04,\pm0.31)$ & $5.98$ & $20.1$ & $3.81$ & $[-0.11,0.04],[-0.38,0.38]$ & $[-0.14,0.07],[-0.41,0.41]$\tabularnewline
\cmidrule{2-8} \cmidrule{3-8} \cmidrule{4-8} \cmidrule{5-8} \cmidrule{6-8} \cmidrule{7-8} \cmidrule{8-8} 
 &  & $3\%$ & $3.29$ & $-23\%$ & $0.$ & $7\%$ & $7\%$\tabularnewline
\cmidrule{2-8} \cmidrule{3-8} \cmidrule{4-8} \cmidrule{5-8} \cmidrule{6-8} \cmidrule{7-8} \cmidrule{8-8} 
 & \multirow{2}{*}{$<10\%$} & $(-0.01,\pm0.25)$ & $4.51$ & $10.5$ & $3.56$ & $[-0.05,0.03],[-0.27,0.27]$ & $[-0.07,0.06],[-0.28,0.28]$\tabularnewline
 &  & $(-0.01,\pm0.25)$ & $7.74$ & $10.2$ & $3.58$ & $[-0.07,0.05],[-0.28,0.28]$ & $[-0.09,0.07],[-0.28,0.28]$\tabularnewline
\cmidrule{2-8} \cmidrule{3-8} \cmidrule{4-8} \cmidrule{5-8} \cmidrule{6-8} \cmidrule{7-8} \cmidrule{8-8} 
 &  & $0\%$ & $3.23$ & $-3\%$ & $0.02$ & $8\%$ & $7\%$\tabularnewline
\bottomrule
\hline\hline
\end{tabular}\caption{\label{Monika tab2} The results of the fit for complex WCs including
BFP $(\text{Real},\; \text{Imaginary})$, $\chi_{\text{min}}^{2}$,\; $p-\text{value}$ $\%$,\;  $\text{pull}_{\text{SM}}$,\;
$1\sigma$ and $2\sigma$-ranges of $(\text{Real},\;\text{Imaginary})$ parts of the corresponding WCs. The numbers are given after putting the bounds on $\mathcal{B}\left(B^-_{c}\to\tau^-\bar{\nu}_\tau\right)<60\%, 30\%$
 and $<10\%$ for both sets of observables i.e., $\mathcal{S}_1$ and $\mathcal{S}_2$. In each sub-row of WCs, the first, second, and third rows represent
data for sets $\mathcal{S}_1, \mathcal{S}_2$, and the difference between the values obtained from the two sets, respectively.}
\end{table}

\subsection{Observable circumstance at BFP}

In this section, we will evaluate the impact of the BFP of these NP WCs on the given observables and calculate the difference between the predicted and experimental values for both sets $\mathcal{S}_1$ and $\mathcal{S}_2$. The discrepancies between the experimental and predicted observables can be defined in the units of $\sigma$ as:
\begin{equation}
dO_{i}=\frac{O_{i}^{\text{NP}}-O_{i}^{\text{exp.}}}{\sigma^{O_{i}^{\text{exp.}}}}.
\end{equation}

The corresponding results are given in Table \ref{Monika tab2split}. We compared these results with the corresponding central values of the experimental measurements. 

It can be seen that the difference between the predicted and measured value of $R_{\tau/{\mu,e}}\left(D\right)$ differs by $\leq 1\sigma$ for both $\mathcal{S}_1$ and $\mathcal{S}_2$. Except, for $C^{S}_R$, and $10\%,\; 30\%$ bounds on the branching ratio for $C_L^{S}$, the predicted value is smaller than the corresponding measured central value. We can see that in the case of $C_{L}^S$, there is a concurrence to data for both sets when we used the $60\%$ bound on the branching ratio, whereas for $C_L^{S}=4C^T$ it is the case for both $60\%$ and $30\%$ limits.  Similarly, for both sets of observables, the best-fit result of $R_{\tau/{\mu,e}}\left(D^\ast\right)$ is almost $2\sigma$ smaller than its experimental value for NP WC $C_R^{S}$ and for $C^{S}_L$ with $10\%$ branching ratio bound. Just like $R_{\tau/{\mu,e}}\left(D\right)$ the NP WC $C_L^{S}=4C^T$ align the results for $60\%$ and $30\%$ limits of $\mathcal{B}\left(B^-_c\to \tau^-\bar{\nu}_\tau\right)$.
 
 The $\tau$ polarization asymmetry, $P_{\tau}\left(D^*\right)$, deviates by $8\%$ to $10\%$  from measurements for all the NP WCs. The situation is not that favorable for $F_{L}\left(D^*\right)$. In this case, we can see that the agreement between the predicted and measured values lies only within $-(1.2 - 1.6)\sigma$, except for the most conservative $60\%$ bound of branching ratio in $C_{L}^S$ where the predicted result falls short by $0.8\sigma$. Likewise, the best-fit value of $R_{\tau/\mu}\left(J/\psi\right)$ is small compared to its experimental measurement, and the suppression is almost $2\sigma$ for all the NP couplings.  In contrast to the observables discussed above, the values of $R_{\tau/\ell}\left(X_c\right)$ and $R_{\tau/\ell}\left(\Lambda_c\right)$ at the BFPs are above their experimental measurements. In the case of $R_{\tau/\ell}\left(X_c\right)$ this value lies within $1\sigma$, whereas, for $R_{\tau/\ell}\left(\Lambda_c\right)$ the deviations are by $(1.4 -1.7)\sigma$ for both $\mathcal{S}_1$ and $\mathcal{S}_2$.

To summarize, we can see that different observables show different dependence on the complex nature of these NP WCs, showing their discriminatory power. The significant deviations for $R_{\tau/\ell}\left(\Lambda_c\right)$, and experimentally improved measurement of $P_{\tau}\left(D^*\right)$ will disfavor some of these complex scenarios. Also, among the physical observables $F_{L}\left(D^*\right),\; R_{\tau/\ell}\left(J/\psi\right)$, $R_{\tau/\ell}\left(X_c\right)$ and $R_{\tau/\ell}\left(\Lambda_c\right)$ are sensitive to the new vector-like coupling, and hence their refined measurements would make it possible to explore new vector-like particles in different beyond SM scenarios.

\begin{table}
\fontsize{10}{10}\selectfont
\begin{tabular}{|c|c|c|r@{\extracolsep{0pt}.}l|c|c|c|c|c|c|}
\hline 
WC & Br & BFP & \multicolumn{2}{c|}{$R_{\tau/\mu,e}\left(D\right)$} & $R_{\tau/\mu,e}\left(D^{*}\right)$ & $P_{\tau}\left(D^{*}\right)$ & $F_{L}\left(D^{*}\right)$ & $R_{\tau/\mu}\left(J/\psi\right)$ & $R_{\tau/\ell}\left(X_{c}\right)$ & $R_{\tau/\ell}\left(\Lambda_{c}\right)$\tabularnewline
\hline 
\hline 
\multirow{3}{*}{$C_{V}$} & \multirow{3}{*}{-} & $\mathcal{S}1:$$(0.01,\pm0.33)$ & \multicolumn{2}{c|}{$\begin{array}{c}
0.340\\
-0.6\sigma
\end{array}$} & $\begin{array}{c}
0.290\\
+0.5\sigma
\end{array}$ & $\begin{array}{c}
-0.497\\
-0.2\sigma
\end{array}$ & $\begin{array}{c}
0.464\\
-1.5\sigma
\end{array}$ & $\begin{array}{c}
0.274\\
-1.8\sigma
\end{array}$ & $\begin{array}{c}
0.251\\
+0.9\sigma
\end{array}$ & $\begin{array}{c}
0.37\\
+1.7\sigma
\end{array}$\tabularnewline
\cline{3-11} \cline{4-11} \cline{6-11} \cline{7-11} \cline{8-11} \cline{9-11} \cline{10-11} \cline{11-11} 
 &  & $\mathcal{S}2:$$(0.03,\pm0.45)$ & \multicolumn{2}{c|}{$\begin{array}{c}
0.338\\
-0.7\sigma
\end{array}$} & $\begin{array}{c}
0.288\\
+0.3\sigma
\end{array}$ & $\begin{array}{c}
-0.497\\
-0.2\sigma
\end{array}$ & $\begin{array}{c}
0.464\\
-1.5\sigma
\end{array}$ & $\begin{array}{c}
0.294\\
-1.7\sigma
\end{array}$ & $\begin{array}{c}
0.249\\
+0.9\sigma
\end{array}$ & $\begin{array}{c}
0.367\\
+1.6\sigma
\end{array}$\tabularnewline
\cline{3-11} \cline{4-11} \cline{6-11} \cline{7-11} \cline{8-11} \cline{9-11} \cline{10-11} \cline{11-11} 
 &  & Diff. & \multicolumn{2}{c|}{$\begin{array}{c}
-0.6\%\\
-0.1
\end{array}$} & $\begin{array}{c}
-0.6\%\\
-0.2
\end{array}$ & $\begin{array}{c}
0\%\\
0
\end{array}$ & $\begin{array}{c}
0\%\\
0
\end{array}$ & $\begin{array}{c}
-0.6\%\\
0.1
\end{array}$ & $\begin{array}{c}
-0.7\%\\
0
\end{array}$ & $\begin{array}{c}
-0.6\%\\
-0.1
\end{array}$\tabularnewline
\hline 
\multirow{3}{*}{$C_{R}^{S}$} & \multirow{3}{*}{} & $\mathcal{S}1:$$(0.1,0)$ & \multicolumn{2}{c|}{$\begin{array}{c}
0.385\\
+1\sigma
\end{array}$} & $\begin{array}{c}
0.259\\
-2.1\sigma
\end{array}$ & $\begin{array}{c}
-0.466\\
-0.2\sigma
\end{array}$ & $\begin{array}{c}
0.476\\
-1.4\sigma
\end{array}$ & $\begin{array}{c}
0.244\\
-1.9\sigma
\end{array}$ & $\begin{array}{c}
0.241\\
+0.6\sigma
\end{array}$ & $\begin{array}{c}
0.356\\
+1.5\sigma
\end{array}$\tabularnewline
\cline{3-11} \cline{4-11} \cline{6-11} \cline{7-11} \cline{8-11} \cline{9-11} \cline{10-11} \cline{11-11} 
 &  & $\mathcal{S}2:$$(0.1,0)$ & \multicolumn{2}{c|}{$\begin{array}{c}
0.382\\
+0.9\sigma
\end{array}$} & $\begin{array}{c}
0.259\\
-2.1\sigma
\end{array}$ & $\begin{array}{c}
-0.467\\
-0.2\sigma
\end{array}$ & $\begin{array}{c}
0.475\\
-1.4\sigma
\end{array}$ & $\begin{array}{c}
0.263\\
-1.8\sigma
\end{array}$ & $\begin{array}{c}
0.24\\
+0.6\sigma
\end{array}$ & $\begin{array}{c}
0.354\\
+1.5\sigma
\end{array}$\tabularnewline
\cline{3-11} \cline{4-11} \cline{6-11} \cline{7-11} \cline{8-11} \cline{9-11} \cline{10-11} \cline{11-11} 
 &  & Diff. & \multicolumn{2}{c|}{$\begin{array}{c}
-0.8\%\\
-0.1
\end{array}$} & $\begin{array}{c}
-0.1\%\\
0
\end{array}$ & $\begin{array}{c}
0.2\%\\
0
\end{array}$ & $\begin{array}{c}
-0.1\%\\
0
\end{array}$ & $\begin{array}{c}
-0.1\%\\
-0.1
\end{array}$ & $\begin{array}{c}
-0.4\%\\
0
\end{array}$ & $\begin{array}{c}
-0.3\%\\
0
\end{array}$\tabularnewline
\hline 
\multirow{9}{*}{$C_{L}^{S}$} & \multirow{3}{*}{$<60\%$} & $\mathcal{S}1:$$(-0.44,\pm0.49)$ & \multicolumn{2}{c|}{$\begin{array}{c}
0.355\\
-0.1\sigma
\end{array}$} & $\begin{array}{c}
0.287\\
0.2\sigma
\end{array}$ & $\begin{array}{c}
-0.32\\
+0.1\sigma
\end{array}$ & $\begin{array}{c}
0.53\\
-0.8\sigma
\end{array}$ & $\begin{array}{c}
0.269\\
-1.8\sigma
\end{array}$ & $\begin{array}{c}
0.252\\
+1\sigma
\end{array}$ & $\begin{array}{c}
0.377\\
+1.8\sigma
\end{array}$\tabularnewline
\cline{3-11} \cline{4-11} \cline{6-11} \cline{7-11} \cline{8-11} \cline{9-11} \cline{10-11} \cline{11-11} 
 &  & $\mathcal{S}2:$$(-0.42,\pm0.48)$ & \multicolumn{2}{c|}{$\begin{array}{c}
0.351\\
-0.2\sigma
\end{array}$} & $\begin{array}{c}
0.285\\
0.1\sigma
\end{array}$ & $\begin{array}{c}
-0.326\\
+0.1\sigma
\end{array}$ & $\begin{array}{c}
0.528\\
-0.8\sigma
\end{array}$ & $\begin{array}{c}
0.289\\
-1.7\sigma
\end{array}$ & $\begin{array}{c}
0.25\\
+0.9\sigma
\end{array}$ & $\begin{array}{c}
0.374\\
+1.7\sigma
\end{array}$\tabularnewline
\cline{3-11} \cline{4-11} \cline{6-11} \cline{7-11} \cline{8-11} \cline{9-11} \cline{10-11} \cline{11-11} 
 &  & Diff & \multicolumn{2}{c|}{$\begin{array}{c}
-1.1\%\\
-0.1
\end{array}$} & $\begin{array}{c}
-0.5\%\\
-0.1
\end{array}$ & $\begin{array}{c}
2.1\%\\
0
\end{array}$ & $\begin{array}{c}
-0.5\%\\
0
\end{array}$ & $\begin{array}{c}
-0.5\%\\
-0.1
\end{array}$ & $\begin{array}{c}
0.8\%\\
-0.1
\end{array}$ & $\begin{array}{c}
-0.7\%\\
-0.1
\end{array}$\tabularnewline
\cline{2-11} \cline{3-11} \cline{4-11} \cline{6-11} \cline{7-11} \cline{8-11} \cline{9-11} \cline{10-11} \cline{11-11} 
 & \multirow{3}{*}{$<30\%$} & $\mathcal{S}1:$$(-0.13,\pm0.41)$ & \multicolumn{2}{c|}{$\begin{array}{c}
0.369\\
+0.4\sigma
\end{array}$} & $\begin{array}{c}
0.265\\
-1.6\sigma
\end{array}$ & $\begin{array}{c}
-0.432\\
-0.1\sigma
\end{array}$ & $\begin{array}{c}
0.488\\
-1.3\sigma
\end{array}$ & $\begin{array}{c}
0.25\\
-1.9\sigma
\end{array}$ & $\begin{array}{c}
0.24\\
+0.6\sigma
\end{array}$ & $\begin{array}{c}
0.358\\
+1.5\sigma
\end{array}$\tabularnewline
\cline{3-11} \cline{4-11} \cline{6-11} \cline{7-11} \cline{8-11} \cline{9-11} \cline{10-11} \cline{11-11} 
 &  & $\mathcal{S}2:$$(-0.13,\pm0.41)$ & \multicolumn{2}{c|}{$\begin{array}{c}
0.365\\
+0.3\sigma
\end{array}$} & $\begin{array}{c}
0.265\\
-1.5\sigma
\end{array}$ & $\begin{array}{c}
-0.431\\
-0.1\sigma
\end{array}$ & $\begin{array}{c}
0.488\\
-1.2\sigma
\end{array}$ & $\begin{array}{c}
0.269\\
-1.8\sigma
\end{array}$ & $\begin{array}{c}
0.24\\
+0.6\sigma
\end{array}$ & $\begin{array}{c}
0.357\\
+1.5\sigma
\end{array}$\tabularnewline
\cline{3-11} \cline{4-11} \cline{6-11} \cline{7-11} \cline{8-11} \cline{9-11} \cline{10-11} \cline{11-11} 
 &  & Diff. & \multicolumn{2}{c|}{$\begin{array}{c}
-0.9\%\\
-0.1
\end{array}$} & $\begin{array}{c}
0\%\\
-0.1
\end{array}$ & $\begin{array}{c}
-0.2\%\\
0
\end{array}$ & $\begin{array}{c}
0.1\%\\
-0.1
\end{array}$ & $\begin{array}{c}
0\%\\
-0.1
\end{array}$ & $\begin{array}{c}
-0.3\%\\
0
\end{array}$ & $\begin{array}{c}
-0.2\%\\
0
\end{array}$\tabularnewline
\cline{2-11} \cline{3-11} \cline{4-11} \cline{6-11} \cline{7-11} \cline{8-11} \cline{9-11} \cline{10-11} \cline{11-11} 
 & \multirow{3}{*}{$<10\%$} & $\mathcal{S}1:$$(0.02,\pm0.26)$ & \multicolumn{2}{c|}{$\begin{array}{c}
0.375\\
+0.6\sigma
\end{array}$} & $\begin{array}{c}
0.255\\
-2.4\sigma
\end{array}$ & $\begin{array}{c}
-0.492\\
-0.2\sigma
\end{array}$ & $\begin{array}{c}
0.465\\
-1.5\sigma
\end{array}$ & $\begin{array}{c}
0.241\\
-1.9\sigma
\end{array}$ & $\begin{array}{c}
0.235\\
+0.4\sigma
\end{array}$ & $\begin{array}{c}
0.349\\
+1.4\sigma
\end{array}$\tabularnewline
\cline{3-11} \cline{4-11} \cline{6-11} \cline{7-11} \cline{8-11} \cline{9-11} \cline{10-11} \cline{11-11} 
 &  & $\mathcal{S}2:$$(0.02,\pm0.26)$ & \multicolumn{2}{c|}{$\begin{array}{c}
0.373\\
+0.6\sigma
\end{array}$} & $\begin{array}{c}
0.255\\
-2.4\sigma
\end{array}$ & $\begin{array}{c}
-0.491\\
-0.2\sigma
\end{array}$ & $\begin{array}{c}
0.466\\
-1.5\sigma
\end{array}$ & $\begin{array}{c}
0.26\\
-1.8\sigma
\end{array}$ & $\begin{array}{c}
0.234\\
+0.4\sigma
\end{array}$ & $\begin{array}{c}
0.348\\
+1.4\sigma
\end{array}$\tabularnewline
\cline{3-11} \cline{4-11} \cline{6-11} \cline{7-11} \cline{8-11} \cline{9-11} \cline{10-11} \cline{11-11} 
 &  & Diff. & \multicolumn{2}{c|}{$\begin{array}{c}
-0.6\%\\
0
\end{array}$} & $\begin{array}{c}
0\%\\
0
\end{array}$ & $\begin{array}{c}
-0.1\%\\
0
\end{array}$ & $\begin{array}{c}
0\%\\
0
\end{array}$ & $\begin{array}{c}
0\%\\
-0.1
\end{array}$ & $\begin{array}{c}
-0.2\%\\
-0.1
\end{array}$ & $\begin{array}{c}
-0.2\%\\
0
\end{array}$\tabularnewline
\hline 
\multirow{6}{*}{$\begin{array}{c}
C_{L}^{S}\\
=4C^{T}
\end{array}$} & \multirow{3}{*}{$\begin{array}{c}
<60\%\\
\&\\
<30\%
\end{array}$} & $\mathcal{S}1:$$(-0.03,\pm0.31)$ & \multicolumn{2}{c|}{$\begin{array}{c}
0.356\\
-0\sigma
\end{array}$} & $\begin{array}{c}
0.284\\
-0\sigma
\end{array}$ & $\begin{array}{c}
-0.437\\
-0.1\sigma
\end{array}$ & $\begin{array}{c}
0.454\\
-1.6\sigma
\end{array}$ & $\begin{array}{c}
0.267\\
-1.8\sigma
\end{array}$ & $\begin{array}{c}
0.244\\
+0.7\sigma
\end{array}$ & $\begin{array}{c}
0.367\\
+1.7\sigma
\end{array}$\tabularnewline
\cline{3-11} \cline{4-11} \cline{6-11} \cline{7-11} \cline{8-11} \cline{9-11} \cline{10-11} \cline{11-11} 
 &  & $\mathcal{S}2:$$(-0.04,\pm0.31)$ & \multicolumn{2}{c|}{$\begin{array}{c}
0.35\\
-0.2\sigma
\end{array}$} & $\begin{array}{c}
0.258\\
+0.1\sigma
\end{array}$ & $\begin{array}{c}
-0.437\\
-0.1\sigma
\end{array}$ & $\begin{array}{c}
0.454\\
-1.6\sigma
\end{array}$ & $\begin{array}{c}
0.287\\
-1.7\sigma
\end{array}$ & $\begin{array}{c}
0.243\\
+0.7\sigma
\end{array}$ & $\begin{array}{c}
0.366\\
+1.6\sigma
\end{array}$\tabularnewline
\cline{3-11} \cline{4-11} \cline{6-11} \cline{7-11} \cline{8-11} \cline{9-11} \cline{10-11} \cline{11-11} 
 &  & Diff. & \multicolumn{2}{c|}{$\begin{array}{c}
-1.6\%\\
-0.2
\end{array}$} & $\begin{array}{c}
0.3\%\\
-0.1
\end{array}$ & $\begin{array}{c}
-0.1\%\\
0
\end{array}$ & $\begin{array}{c}
0.1\%\\
0
\end{array}$ & $\begin{array}{c}
0.3\%\\
-0.1
\end{array}$ & $\begin{array}{c}
-0.6\%\\
0
\end{array}$ & $\begin{array}{c}
-0.3\%\\
-0.1
\end{array}$\tabularnewline
\cline{2-11} \cline{3-11} \cline{4-11} \cline{6-11} \cline{7-11} \cline{8-11} \cline{9-11} \cline{10-11} \cline{11-11} 
 & \multirow{3}{*}{$<10\%$} & $\mathcal{S}1:$$(-0.01,\pm0.25)$ & \multicolumn{2}{c|}{$\begin{array}{c}
0.348\\
-0.3\sigma
\end{array}$} & $\begin{array}{c}
0.269\\
-1.2\sigma
\end{array}$ & $\begin{array}{c}
-0.462\\
-0.2\sigma
\end{array}$ & $\begin{array}{c}
0.456\\
-1.6\sigma
\end{array}$ & $\begin{array}{c}
0.254\\
-1.8\sigma
\end{array}$ & $\begin{array}{c}
0.237\\
+0.5\sigma
\end{array}$ & $\begin{array}{c}
0.352\\
+1.5\sigma
\end{array}$\tabularnewline
\cline{3-11} \cline{4-11} \cline{6-11} \cline{7-11} \cline{8-11} \cline{9-11} \cline{10-11} \cline{11-11} 
 &  & $\mathcal{S}2:$$(-0.01,\pm0.25)$ & \multicolumn{2}{c|}{$\begin{array}{c}
0.344\\
-0.5\sigma
\end{array}$} & $\begin{array}{c}
0.27\\
-1.1\sigma
\end{array}$ & $\begin{array}{c}
-0.461\\
-0.1\sigma
\end{array}$ & $\begin{array}{c}
0.457\\
-1.6\sigma
\end{array}$ & $\begin{array}{c}
0.273\\
-1.8\sigma
\end{array}$ & $\begin{array}{c}
0.236\\
+0.4\sigma
\end{array}$ & $\begin{array}{c}
0.352\\
+1.4\sigma
\end{array}$\tabularnewline
\cline{3-11} \cline{4-11} \cline{6-11} \cline{7-11} \cline{8-11} \cline{9-11} \cline{10-11} \cline{11-11} 
 &  & Diff. & \multicolumn{2}{c|}{$\begin{array}{c}
-1.2\%\\
-0.2
\end{array}$} & $\begin{array}{c}
0.4\%\\
-0.1
\end{array}$ & $\begin{array}{c}
-0.2\%\\
-0.1
\end{array}$ & $\begin{array}{c}
0.1\%\\
0
\end{array}$ & $\begin{array}{c}
0.4\%\\
0
\end{array}$ & $\begin{array}{c}
-0.4\%\\
-0.1
\end{array}$ & $\begin{array}{c}
-0.1\%\\
-0.1
\end{array}$\tabularnewline
\hline 
\end{tabular}
\caption{\label{Monika tab2split} The best-fit points (BFP) of the WCs for $\mathcal{S}_1$ and $\mathcal{S}_2$ and the corresponding predictions of various observables along with the deviations from the experimental values for different bounds on the branching ratio. These are expressed in multiples of $\sigma^{O_{i}^{exp.}}$, where Diff. signifies the difference between the BF predicted and experimental results in percent, and the corresponding $\sigma$.}
\end{table}

\section{Correlating different Physical observables}\label{sec4}

In this section, we investigate the correlation between observables from the numerical expressions presented in Section \ref{sec2}. As these decays occur through the same FCCC quark level transition $\left(b\to c \tau \bar{\nu}_\tau\right)$, therefore, it is possible to express $R_{\tau/\ell}\left(\Lambda_{c}\right)$ in terms of $R_{\tau/\mu,e}\left(D,D^*\right)$. This is known as the sum rule, which can be derived from Eqs. (\ref{eqn1}, \ref{eqn2}) and Eq. (\ref{eqn7}) as \cite{Fedele:2022iib}:
\begin{equation}
\frac{R_{\tau/\ell}\left(\Lambda_{c}\right)}{R_{\tau/\ell}^{\text{SM}}\left(\Lambda_{c}\right)}=0.275\frac{R_{\tau/{\mu,e}}\left(D\right)}{R^{\text{SM}}_{\tau/{\mu,e}}\left(D\right)}+0.725\frac{R_{\tau/{\mu,e}}\left(D^{*}\right)}{R^{\text{SM}}_{\tau/{\mu,e}}\left(D^{*}\right)}+x_{1},\label{sum-n1}
\end{equation}
where small remainder $x_{1}$ can be approximated in terms of WCs at a scale
$m_{b}$ as:
\begin{equation}
x_{1}=\Re\left[\left(1+C_{L}^{V}\right)\left(0.011\left(C_{R}^{S}\right)^{*}+0.341\left(C^{T}\right)^{*}\right)\right]+0.013\left(\left|C_{R}^{S}\right|^{2}+\left|C_{L}^{S}\right|^{2}\right)+0.023\Re\left[C_{L}^{S}\left(C_{R}^{S}\right)^{*}\right]-1.431\left|C^{T}\right|^{2}\label{rem1}.
\end{equation}
This is an updated version of the sum rule reported earlier \cite{Blanke:2018yud}.
In Eq. (\ref{sum-n1}), we can see that in $R_{\tau/\ell}\left(\Lambda_c\right)$, the relative weight of the $R_{\tau/{\mu,e}}\left(D^*\right)/R^{\text{SM}}_{\tau/{\mu,e}}\left(D^*\right)$ is $72\%$, and hence with better control over the errors in its measurements and the SM predictions, will help us to predict $R_{\tau/\ell}\left(\Lambda_c\right)$ to good accuracy.

In Eq. (\ref{Exp-RJpsi}), we can see that the the discrepancy between the measured and SM predicted value of $R_{\tau/\mu}\left(J/\psi\right)$ is $1.8\sigma$, therefore, it will be interesting to see if we can write similar sum rules for $R_{\tau/\mu}\left(J/\psi\right)$ and $R_{\tau/\ell}\left(X_{c}\right)$. The corresponding results derived from Eqs. (\ref{eqn1}, \ref{eqn2}, \ref{eqn5}), and (\ref{eqn6}) are
\begin{align}
\frac{R_{\tau/\mu}\left(J/\psi\right)}{R^{\text{SM}}_{\tau/\mu}\left(J/\psi\right)} & =0.006\frac{R_{\tau/{\mu,e}}\left(D\right)}{R^{\text{SM}}_{\tau/{\mu,e}}\left(D\right)}+0.994\frac{R_{\tau/\mu}\left(D^{*}\right)}{R^{\text{SM}}_{\tau/{\mu,e}}\left(D^{*}\right)}+x_{2},\label{sum2}\\
\frac{R_{\tau/\ell}\left(X_{c}\right)}{R^{\text{SM}}\left(X_{c}\right)} & =0.347\frac{R_{\tau/{\mu,e}}\left(D\right)}{R^{\text{SM}}_{\tau/{\mu,e}}\left(D\right)}+0.653\frac{R_{\tau/{\mu,e}}\left(D^{*}\right)}{R^{\text{SM}}_{\tau/{\mu,e}}\left(D^{*}\right)}+x_{3},\label{sum3}
\end{align}
where the remainder $x_{2}$ and $x_{3}$ can be written as
\begin{eqnarray}
x_{2} & = &-\Re\left[\left(1+C_{L}^{V}\right)\left(0.019C_{R}^{S\;*}+0.259C^{T\;*}\right)\right]-0.006\left(\left|C_{R}^{S}\right|^{2}+\left|C_{L}^{S}\right|^{2}\right)-0.013\Re\left[C_{L}^{S}C_{R}^{S\;*}\right]-1.205\left|C^{T}\right|^{2},\label{rem2}\\
x_{3} & = &  -\Re\left[\left(1+C_{L}^{V}\right)\left(0.048C_{L}^{S\;*}-3.001C^{T\;*}\right)\right]+0.147\left(\left|C_{L}^{V}\right|^{2}+2\Re\left[C_{L}^{V}\right]\right)-0.019\left(\left|C_{R}^{S}\right|^{2}+\left|C_{L}^{S}\right|^{2}\right)\notag\\
& &-0.057\Re\left[C_{L}^{S}C_{R}^{S\;*}\right]+1.899\left|C^{T}\right|^{2}\label{rem3}.
\end{eqnarray}
In Eq. (\ref{sum2}), the LFU ratio $R_{\tau/\mu}\left(J/\psi\right)$ normalized with the corresponding SM prediction, has negligible dependence on the $R_{\tau/{\mu,e}}\left(D\right)/R^{\text{SM}}_{\tau/{\mu,e}}\left(D\right)$, therefore, the refined measurement of the $R_{\tau/{\mu,e}}\left(D^*\right)$ will help us to get good control over $R_{\tau/\mu}\left(J/\psi\right)$. Also, we can see that if $R_{\tau/{\mu,e}}\left(D\right)$ and $R_{\tau/{\mu,e}}\left(D^{*}\right)$ are enhanced
over their SM values, it follows that $R_{\tau/\ell}\left(X_{c}\right)$ must also experience an enhancement, which is restricted by the small difference between measured and SM predicted values. 

By evolving the BFPs of Table \ref{Monika tab2} for $\mathcal{S}_{1,2}$, we find that remainders
in Eqs. (\ref{rem1}, \ref{rem2}) and (\ref{rem3}) are approximately: 
$x_{1}<10^{-3}$, $x_{2}<10^{-3}$, and $x_{3}<10^{-2}$ for all the
NP WCs, which ensure the validity of these sum rules. Being model-independent, these sum rules remain valid in any NP model, indicating that future measurements of $R_{\tau/\ell}\left(\Lambda_{c}\right)$,
$R_{\tau/\ell}\left(X_{c}\right)$, and $R_{\tau/\ell}\left(J/\psi\right)$ can serve as essential
crosschecks for the measurements of $R_{\tau/{\mu,e}}\left(D\right)$ and $R_{\tau/{\mu,e}}\left(D^{*}\right)$.

Using the values from Eqs. (\ref{RDRDs},\ref{SM-RDDs}) in Eq. (\ref{sum-n1}), we can predict
\begin{align*}
R_{\tau/\ell}\left(\Lambda_{c}\right) & =R^{\text{SM}}_{\tau/\ell}\left(\Lambda_{c}\right)\left(1.14\pm0.047\right) =0.369\pm0.015\pm0.005,
\end{align*}
as given in ref. \cite{Fedele:2022iib}. Similarly, for the other two sum rules [c.f. Eqs. (\ref{sum2}, \ref{sum3})], we have
\begin{equation}
R_{\tau/\mu}\left(J/\psi\right)  = R^{\text{SM}}_{\tau/\mu}\left(J/\psi\right) \left(1.119\pm0.052\right) =0.289\pm0.013\pm0.043, \label{RJPsiSRule}
\end{equation}
and
\begin{equation}
R_{\tau/\ell}\left(X_{c}\right)  = R^{\text{SM}}_{\tau/\ell}\left(X_{c}\right)\left(1.146\pm0.048\right) =0.248\pm0.01\pm0.003.\label{RXcSRule}
\end{equation}
In these results, the first error comes from the experimental measurements in LFU ratios of $D$ and $D^*$
and second is due to the uncertainties in the SM predictions of the corresponding ratios. In the case of $R_{\tau/\ell}\left(J/\psi\right)$, we can see the result is slightly larger than the SM predictions but still smaller than the experimental values. However, there are the large errors in the experimental results, and we hope with better measurements in future will help us to make these results more conclusive.

\subsection{Correlation between observables for particular NP WCs}

\begin{figure}[H]
\centering 
\begin{subfigure}[b]{0.48\textwidth}
\centering
\includegraphics[width=6cm, height=5cm]{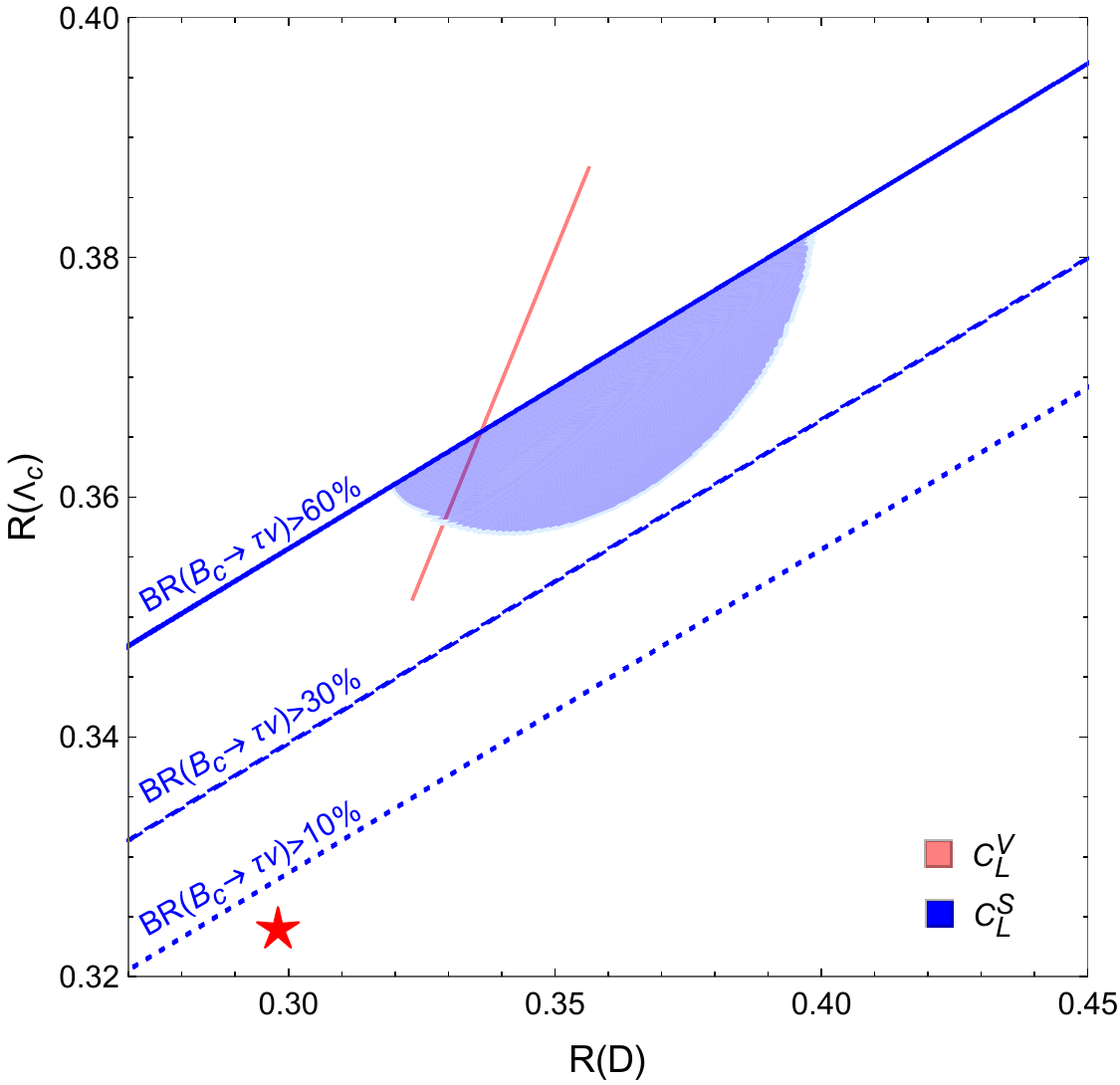}
\caption{}
\end{subfigure}
\begin{subfigure}[b]{0.48\textwidth}
\centering
\includegraphics[width=6cm, height=5cm]{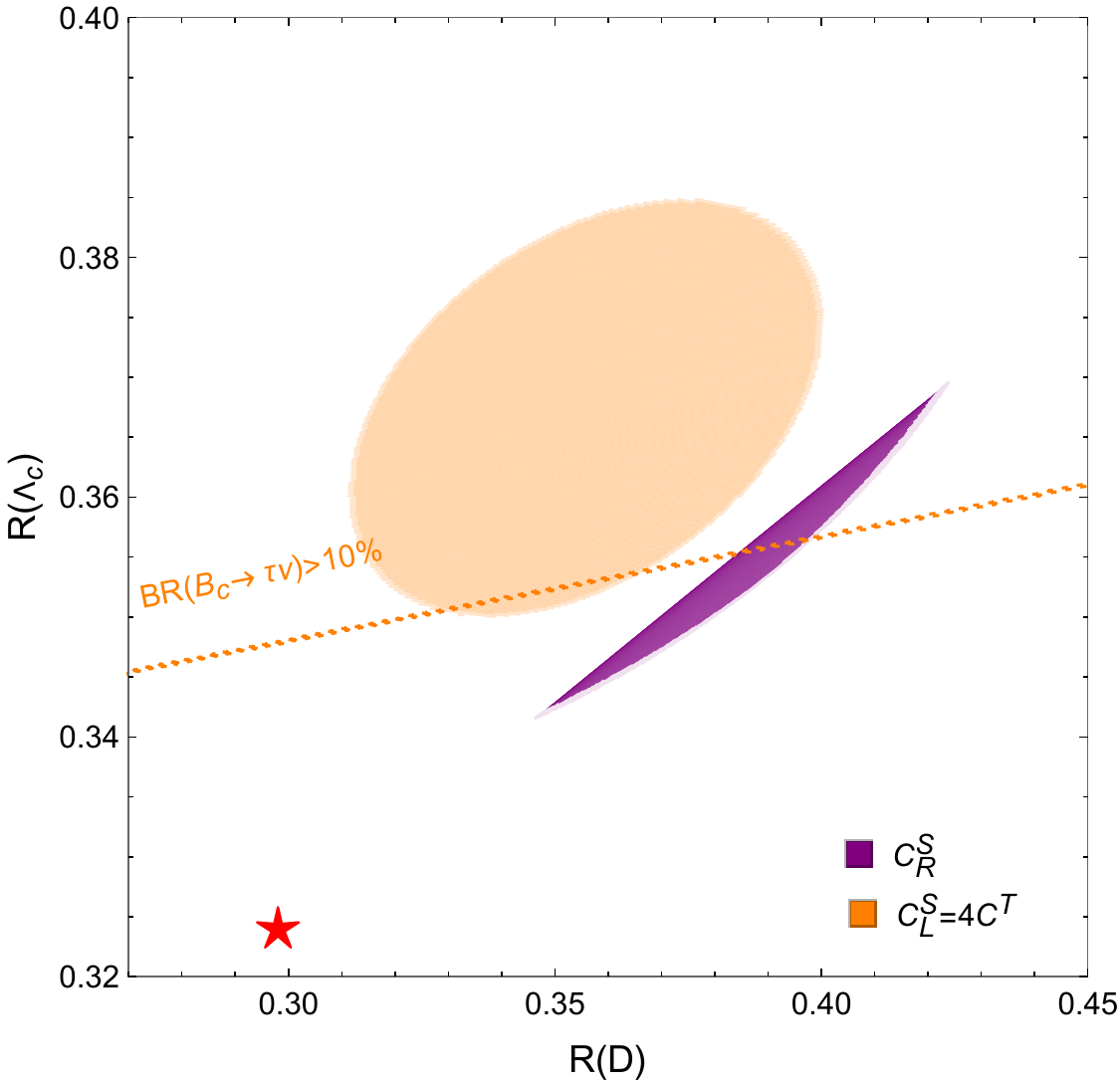}
\caption{}
\end{subfigure}
\begin{subfigure}[b]{0.48\textwidth}
\centering
\includegraphics[width=6cm, height=5cm]{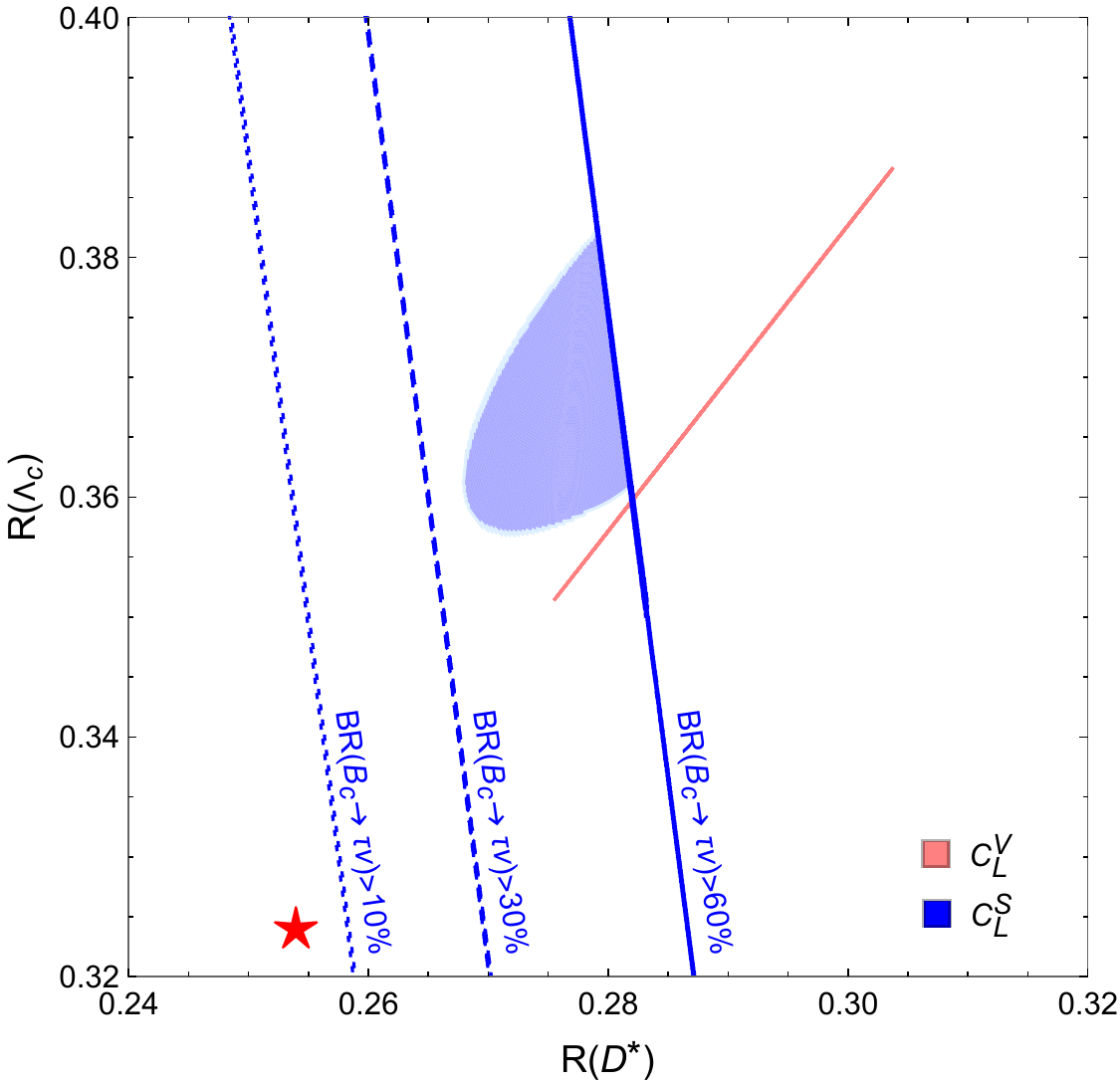}
\caption{}
\end{subfigure}
\begin{subfigure}[b]{0.48\textwidth}
\centering
\includegraphics[width=6cm, height=5cm]{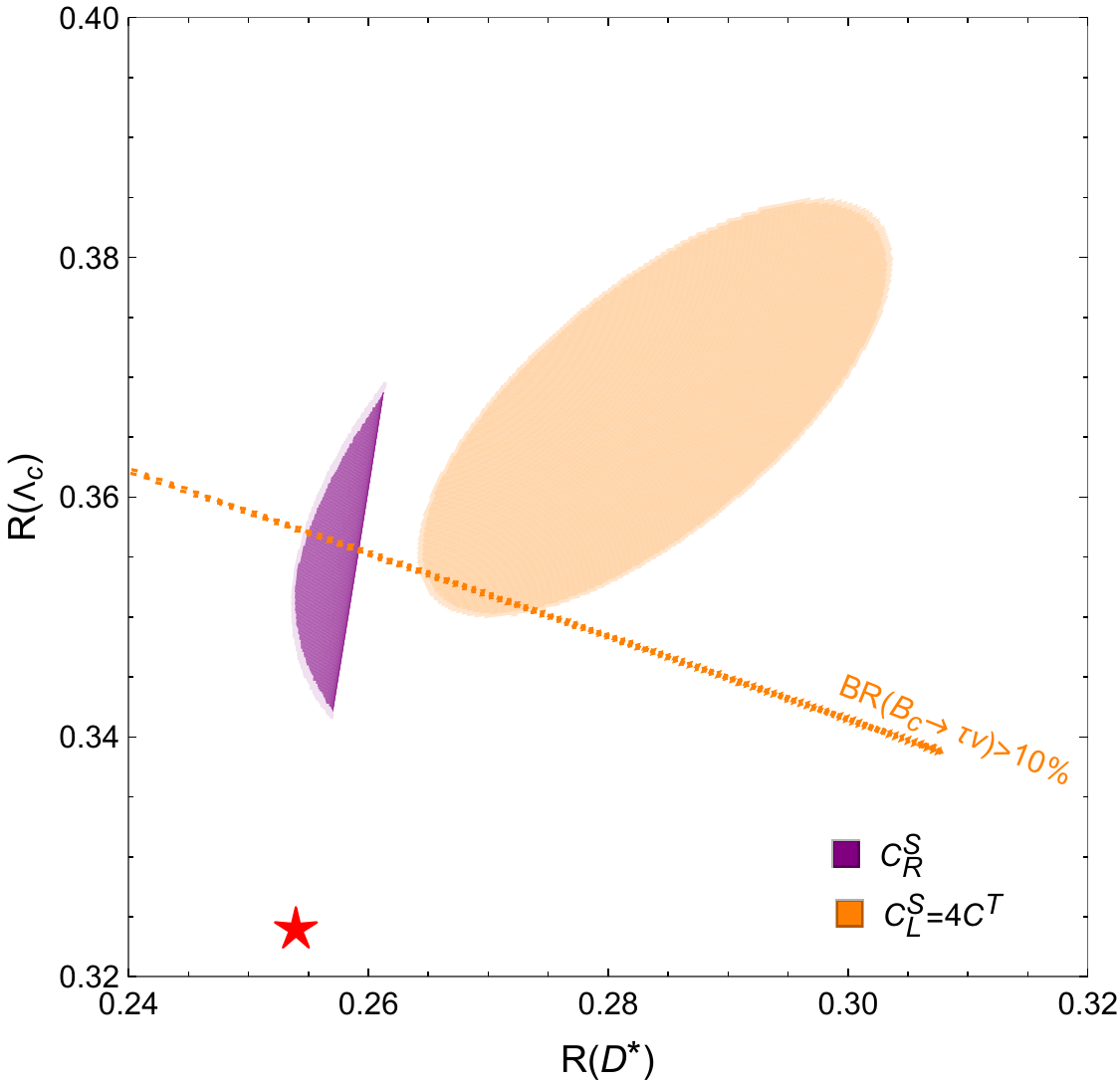}
\caption{}
\end{subfigure}
\caption{\label{fig3 4obs} Preferred $1\sigma$ regions for the four 2-dimensional
complex scenarios for set $\mathcal{S}_1$ in the $R_{\tau/{\mu,e}}\left(D\right)-R_{\tau/\ell}\left(\Lambda_{c}\right)\text{plane (above) and }R_{\tau/{\mu,e}}\left(D^{*}\right)-R_{\tau/\ell}\left(\Lambda_{c}\right)$
plane (below) for the $BR\left(B_{c}\to\tau\bar{\nu}_\tau\right)<60\%$. The regions
of the plot in the left panel correspond to the $C_{L}^{V}$ (pink)
and $C_{L}^{S}$ (Blue), while the plots on the right panel correspond
to $C_{R}^{S}$ (Purple) $C_{L}^{S}=4C^{T}$(Orange). The solid, dashed,
dotted lines refer to a constraint on $\mathcal{B}\left(B^-_{c}\to\tau^-\nu_\tau\right)<60\%,30\%$  and
$10\%$, respectively. The red stars represent SM predictions. In all the figures, we removed the subscripts from the LFU ratios to have better visibility.}
\end{figure}

In this section, we analyze
the correlations among the above-mentioned observables, except the already established $R_{D}$ and $R_{D^*}$, by taking specific NP WCs, and  Fig. \ref{fig3 4obs} shows
the preferred $1\sigma$ regions for the four two-dimensional complex
scenarios under consideration. Particularly, by considering the different bounds on the $\mathcal{B}(B^-_{c}\rightarrow\tau^-\bar{\nu}_\tau)$, these scenarios are depicted in the $R_{\tau/{\mu,e}}\left(D\right)-R_{\tau/\ell}\left(\Lambda_{c}\right)$
and $R_{\tau/{\mu,e}}\left(D^{*}\right)-R_{\tau/\ell}\left(\Lambda_{c}\right)$ planes for $\mathcal{S}_1$.
For the NP WC $C_{L}^{V}$,
we found a direct correlation, resulting in the region shrinking
to a line in $R_{\tau/\ell}\left(\Lambda_{c}\right)$ with respect to both $R_{\tau/{\mu,e}}\left(D\right)$
and $R_{\tau/{\mu,e}}\left(D^{*}\right)$. Notably, these correlations are accompanied
by higher $p-\text{values}$, indicating stronger statistical support for
the relationship among them; however, for our choice of observables in $S_{1}$ include observables that depend only on the absolute value of $C_{L}^{V}$. Furthermore, a high degree of positive correlation
was identified for the WC $C_{R}^{S}$ in $R_{\tau/\ell}\left(\Lambda_{c}\right)$
with respect to both $R_{\tau/{\mu,e}}\left(D,D^*\right)$ but in this scenario BFP is real to a given accuracy.
Similarly, for the other two WCs, $C_{L}^{S}$ and $C_{L}^{S}=4C^{T}$,
we also found moderate positive correlations between these LFU ratios. Notably WC $C_{L}^{S}=4C^{T}$ is consistent with the experimental values of $R_{\tau/{\mu,e}}\left(D\right)$ and $R_{\tau/{\mu,e}}\left(D^{*}\right)$. This agreement between theory and experiment provides support for the validity of the proposed WC relationship in explaining the observed phenomena in the context of flavor-changing charged current transitions.

\begin{figure}[H]
\centering 
\begin{subfigure}[b]{0.48\textwidth}
\centering
\includegraphics[width=6cm, height=5cm]{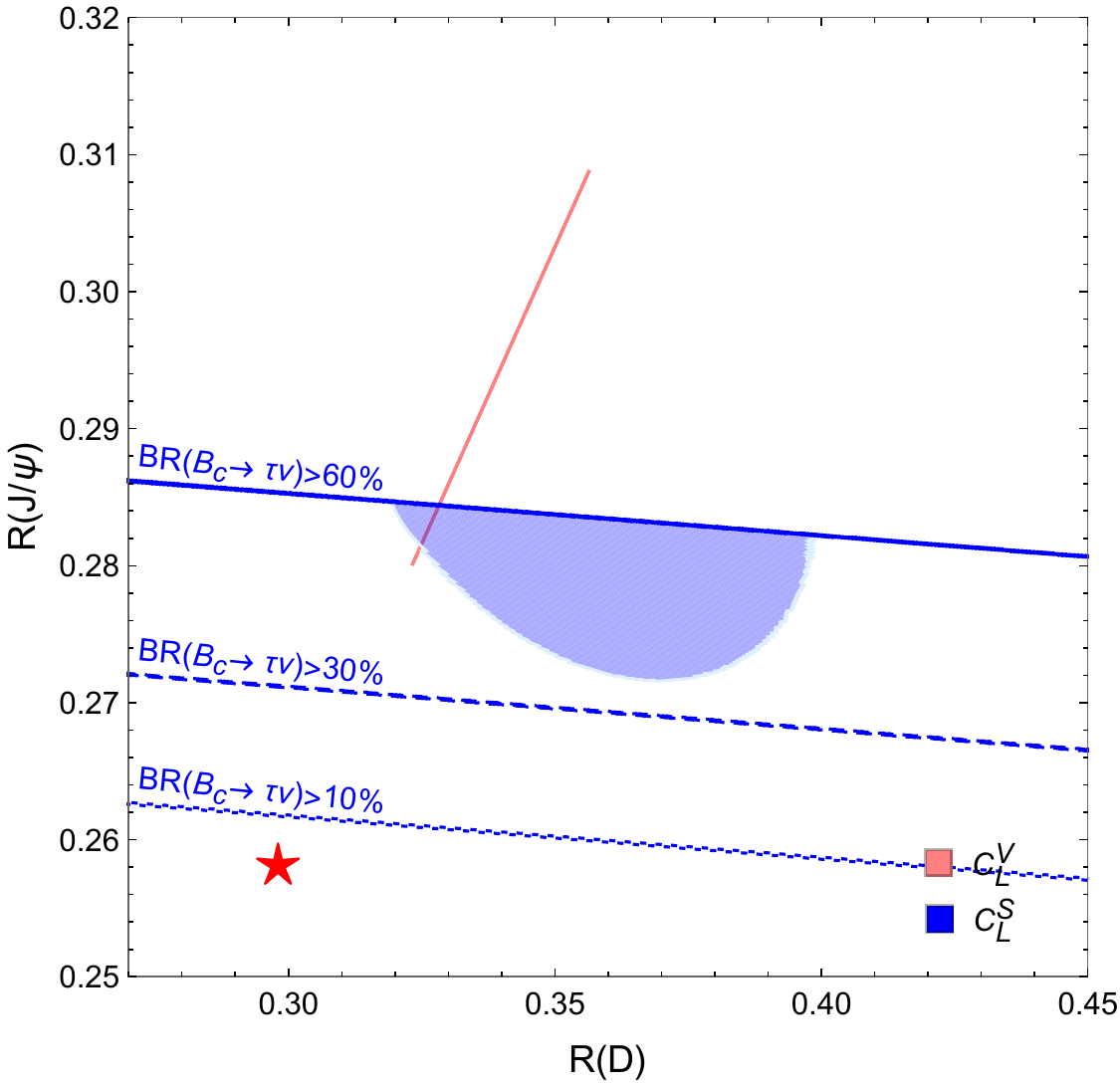}
\caption{}
\end{subfigure}
\begin{subfigure}[b]{0.48\textwidth}
\centering
\includegraphics[width=6cm, height=5cm]{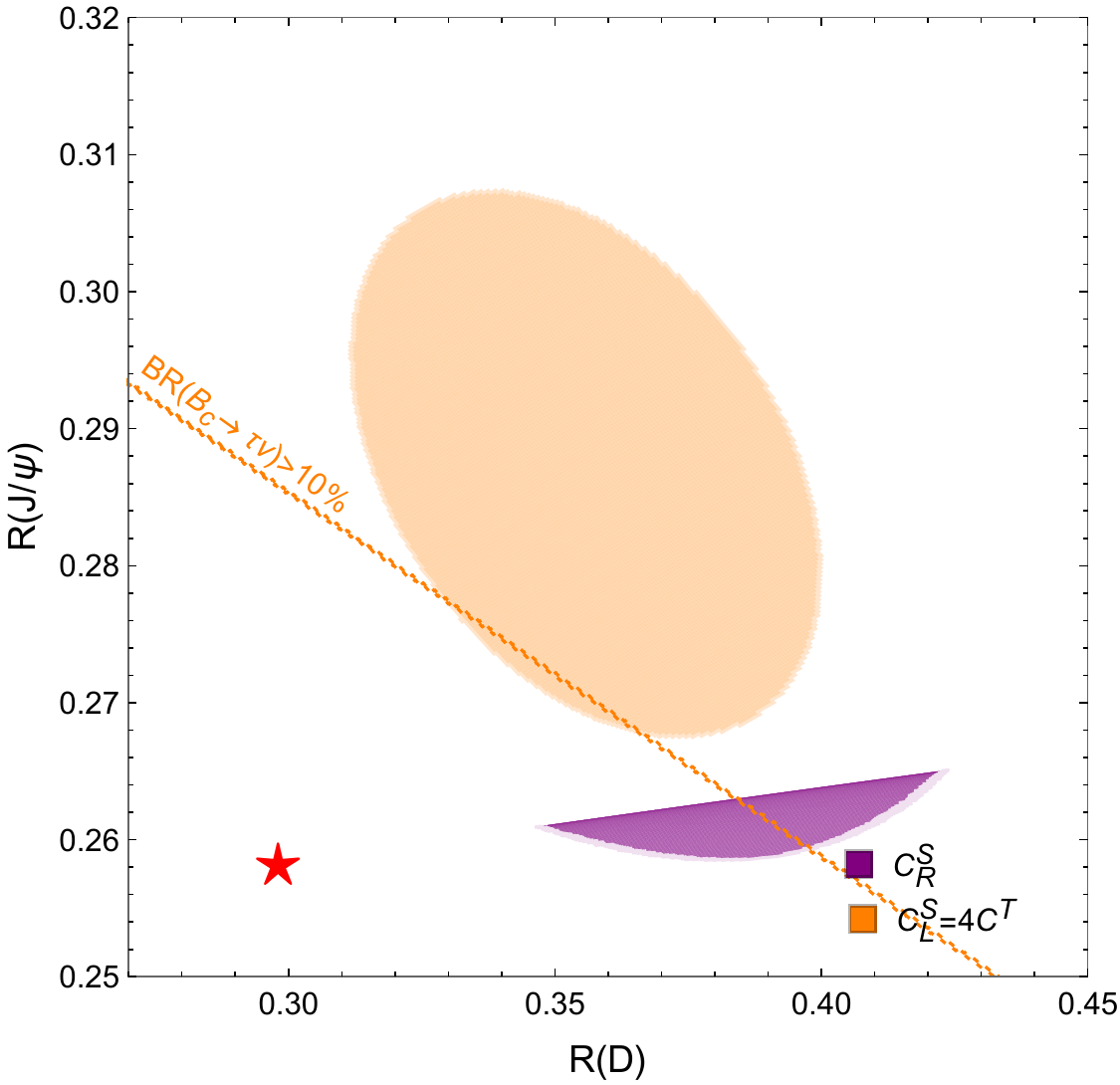}
\caption{}
\end{subfigure}
\begin{subfigure}[b]{0.48\textwidth}
\centering
\includegraphics[width=6cm, height=5cm]{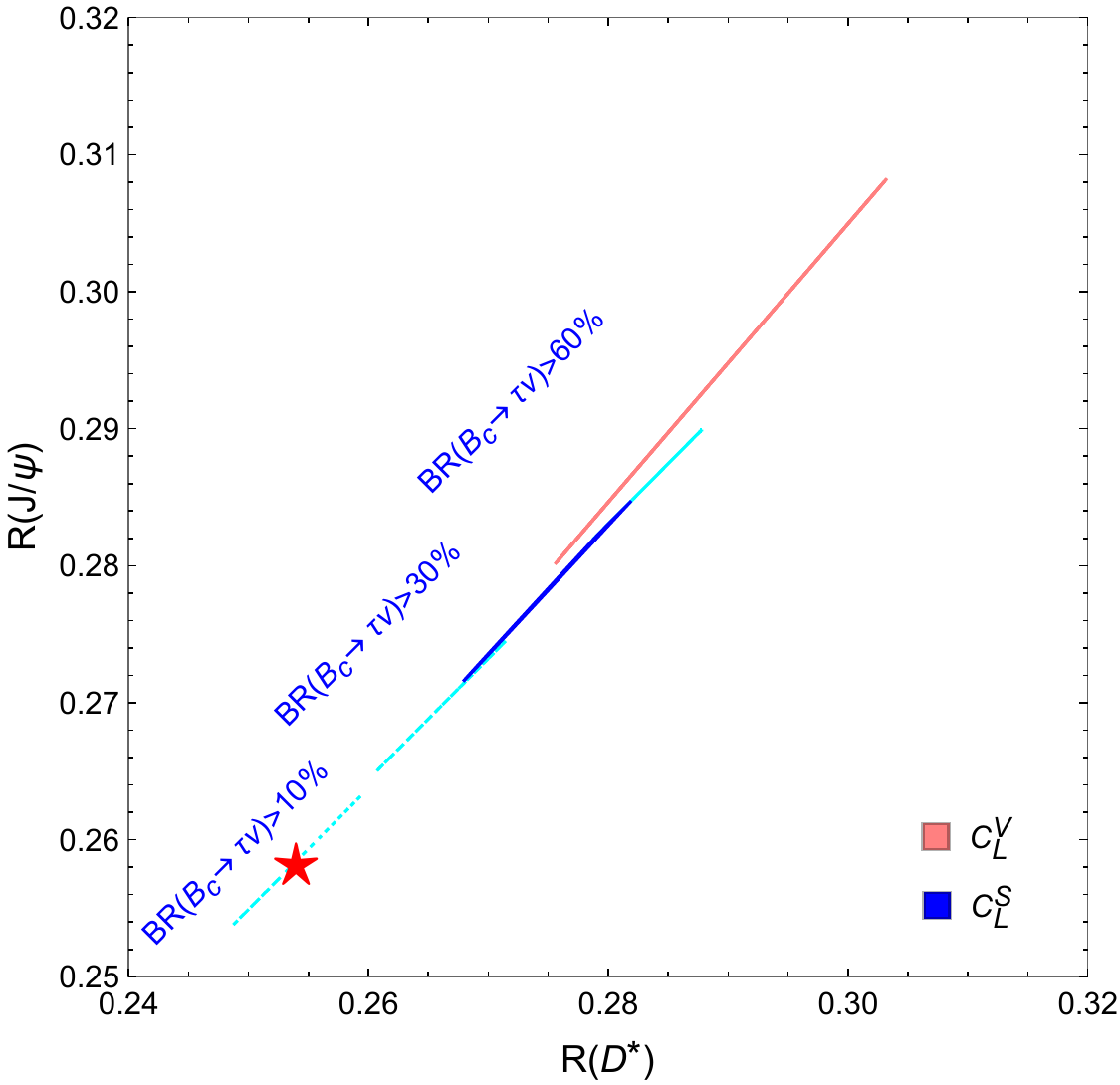}
\caption{}
\end{subfigure}
\begin{subfigure}[b]{0.48\textwidth}
\centering
\includegraphics[width=6cm, height=5cm]{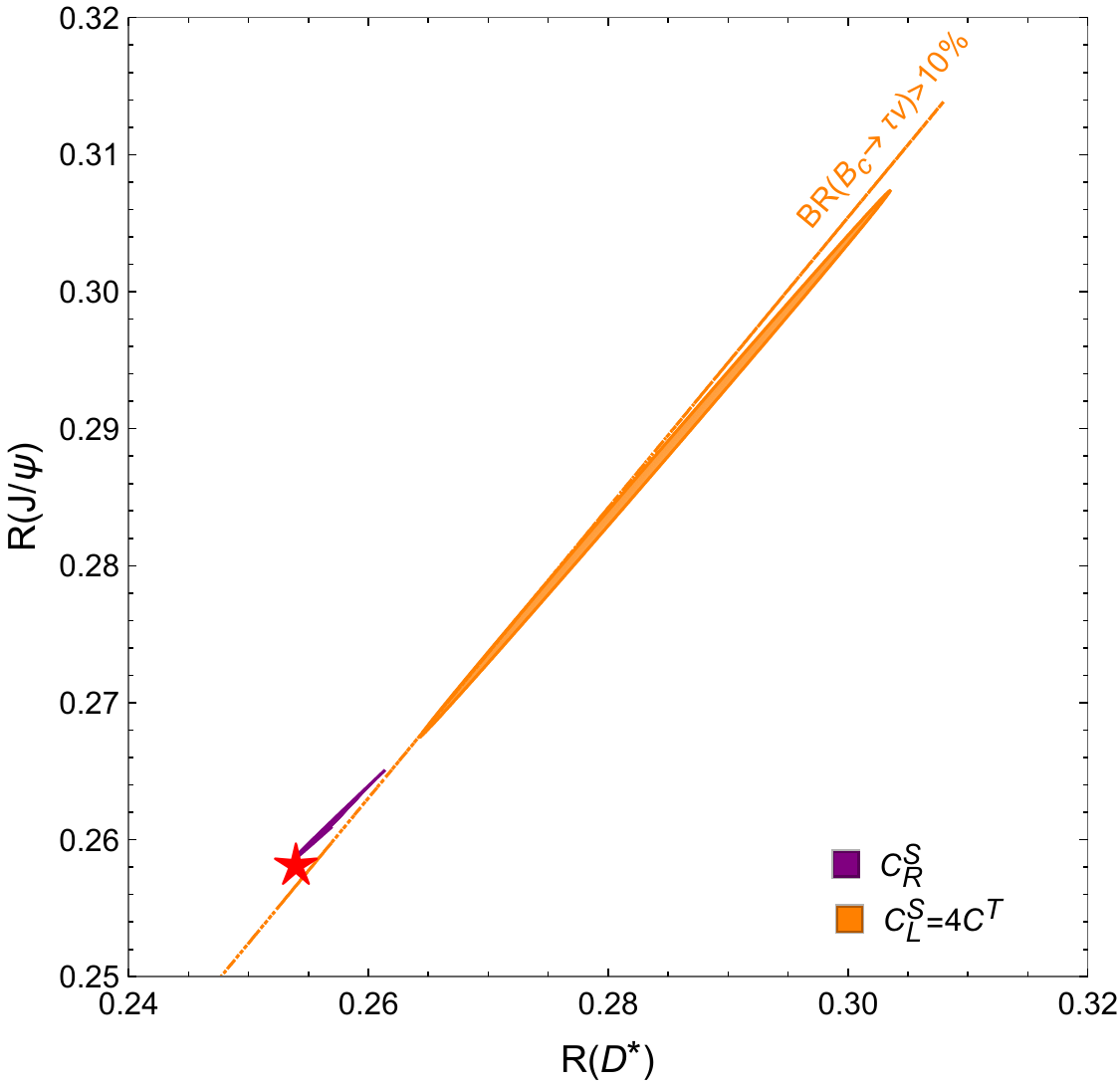}
\caption{}
\end{subfigure}
\caption{\label{fig3 4obs2} The preferred $1\sigma$ regions for the four 2- dimensional
complex scenarios for set $\mathcal{S}_1$ in the $R_{\tau/{\mu,e}}\left(D\right)-R_{\tau/\mu}\left(J/\psi\right)\text{plane (above) and }R_{\tau/{\mu,e}}\left(D^{*}\right)-R_{\tau/{\mu}}\left(J/\psi\right)$
plane (below) for the $\mathcal{B}(B^-_{c}\to\tau^-\bar{\nu}_\tau)<60\%$. The color coding is the same as in Fig. \ref{fig3 4obs}. The red stars represent SM predictions in this case.}
\end{figure}

Similarly, for $\mathcal{S}_1$, Fig. \ref{fig3 4obs2} shows the preferred $1\sigma$ regions for
the four two-dimensional complex scenarios in the $R_{\tau/\mu}\left(J/\psi\right)-R_{\tau/{\mu,e}}\left(D\right)$
and $R_{\tau/{\mu}}\left(J/\psi\right)-R_{\tau/{\mu,e}}\left(D^{*}\right)$ planes for $\mathcal{B}(B^-_{c}\to\tau^-\bar{\nu}_\tau)<60\%$ bounds. 
We observed
a direct correlation for the WC $C_{L}^{V}$, a modest positive correlation for the WCs $C_{R}^{S}$ scenarios. For $C_{L}^{S}$ and $C_{L}^{S}=4C^{T}$ scenarios, we found a reasonable negative correlation
between $R_{\tau/\mu}\left(J/\psi\right)- R_{\tau/{\mu,e}}\left(D\right)$. Thus, our findings indicate that
this correlation pattern
can significantly vary for our specific WC scenario. Turning our attention to the correlation between $R_{\tau/\mu}\left(J/\psi\right)$
and $R_{\tau/{\mu,e}}\left(D^{*}\right)$, we discovered an direct correlation
across all WC scenarios. These findings highlight the significance
of the interplay between $R_{\tau/\mu}\left(J/\psi\right)$ and
$R_{\tau/{\mu,e}}\left(D^{*}\right)$ to gain valuable insights into the underlying
NP.

\begin{figure}[H]
\centering
\begin{subfigure}[b]{0.48\textwidth}
         \centering
     \includegraphics[width=6cm, height=5cm]{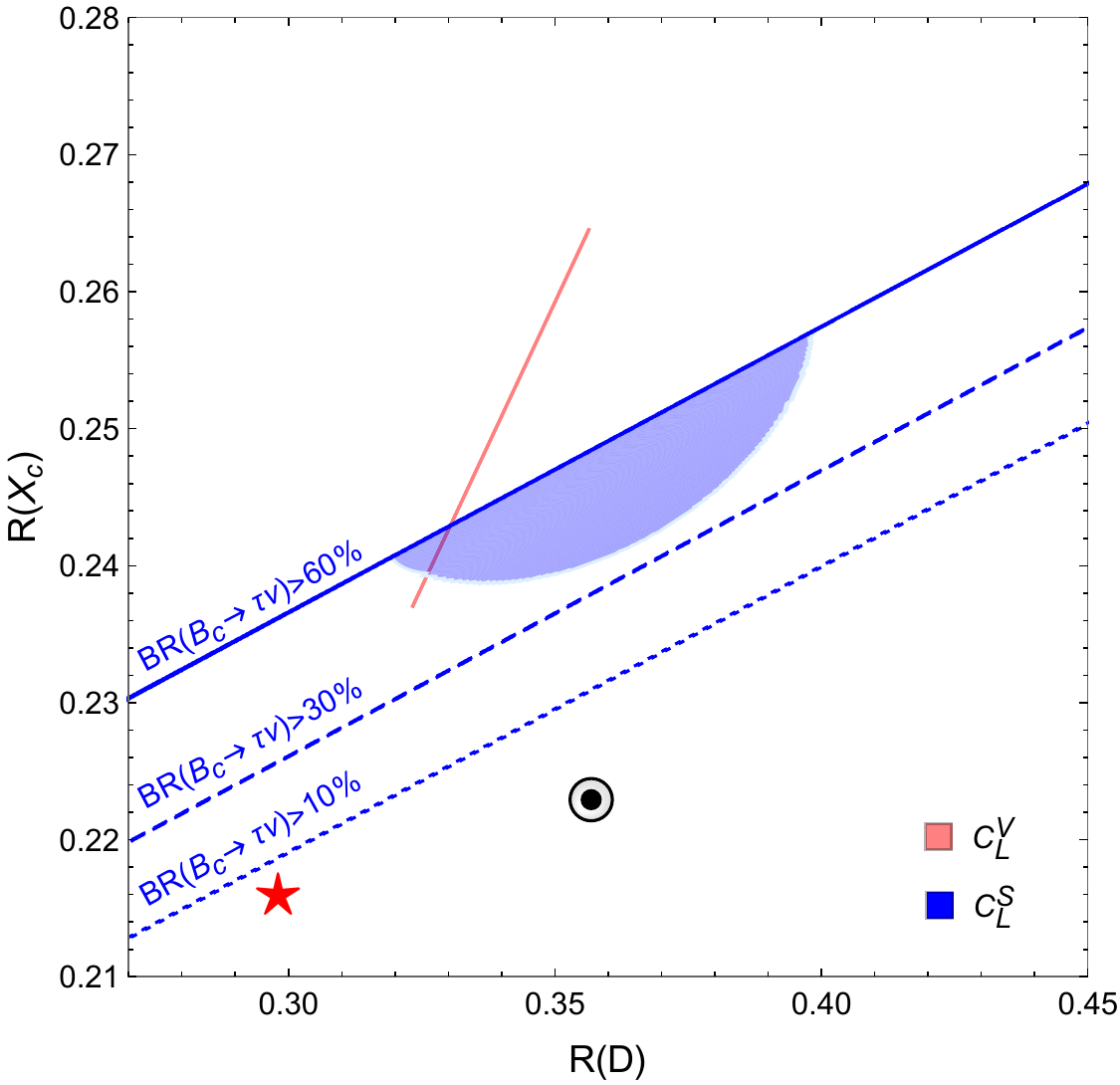}
\caption{}
\end{subfigure}
\begin{subfigure}[b]{0.48\textwidth}
\centering 
\includegraphics[width=6cm, height=5cm]{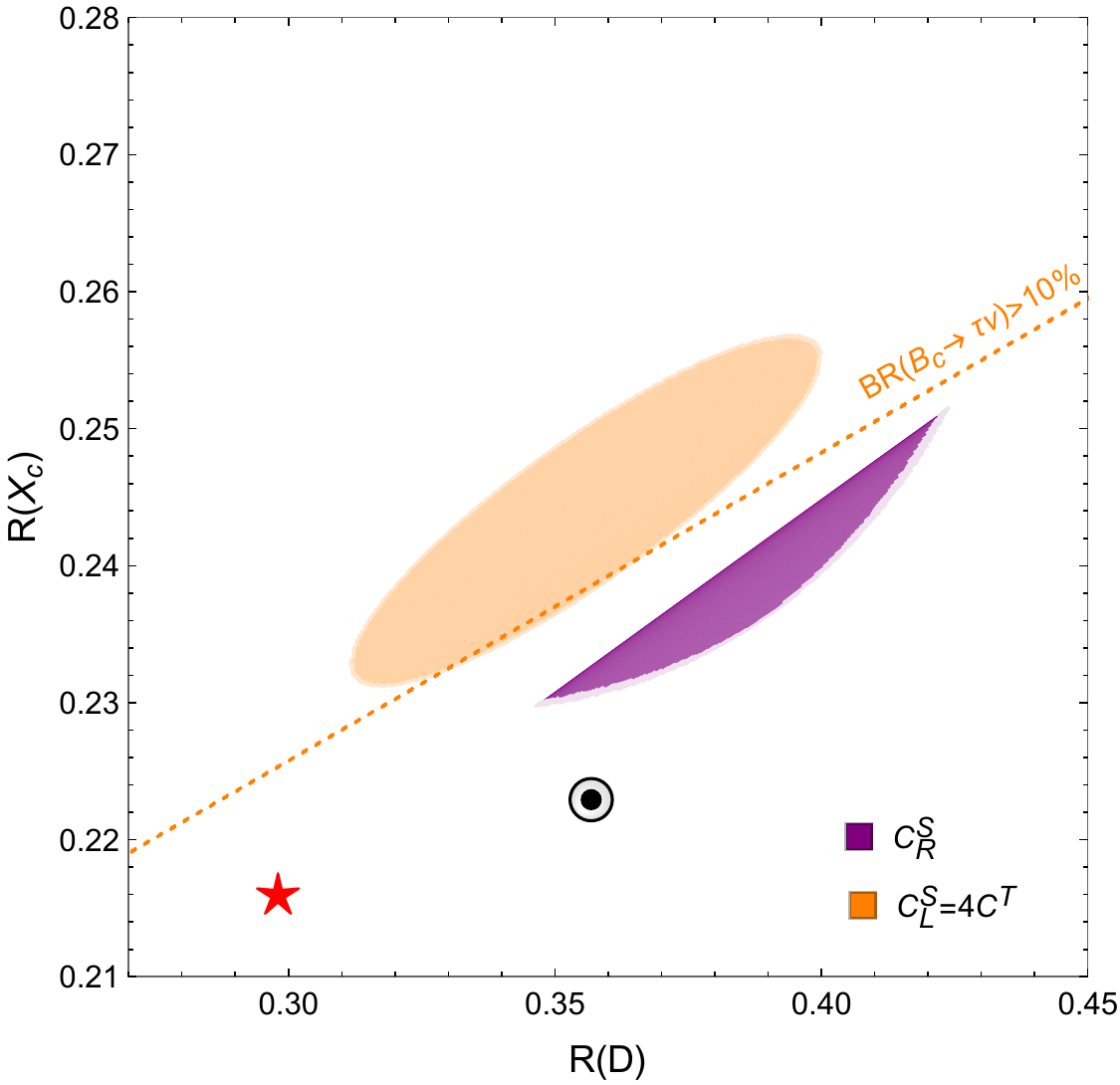}
\caption{}
\end{subfigure}
\begin{subfigure}[b]{0.48\textwidth}
\centering 
\includegraphics[width=6cm, height=5cm]{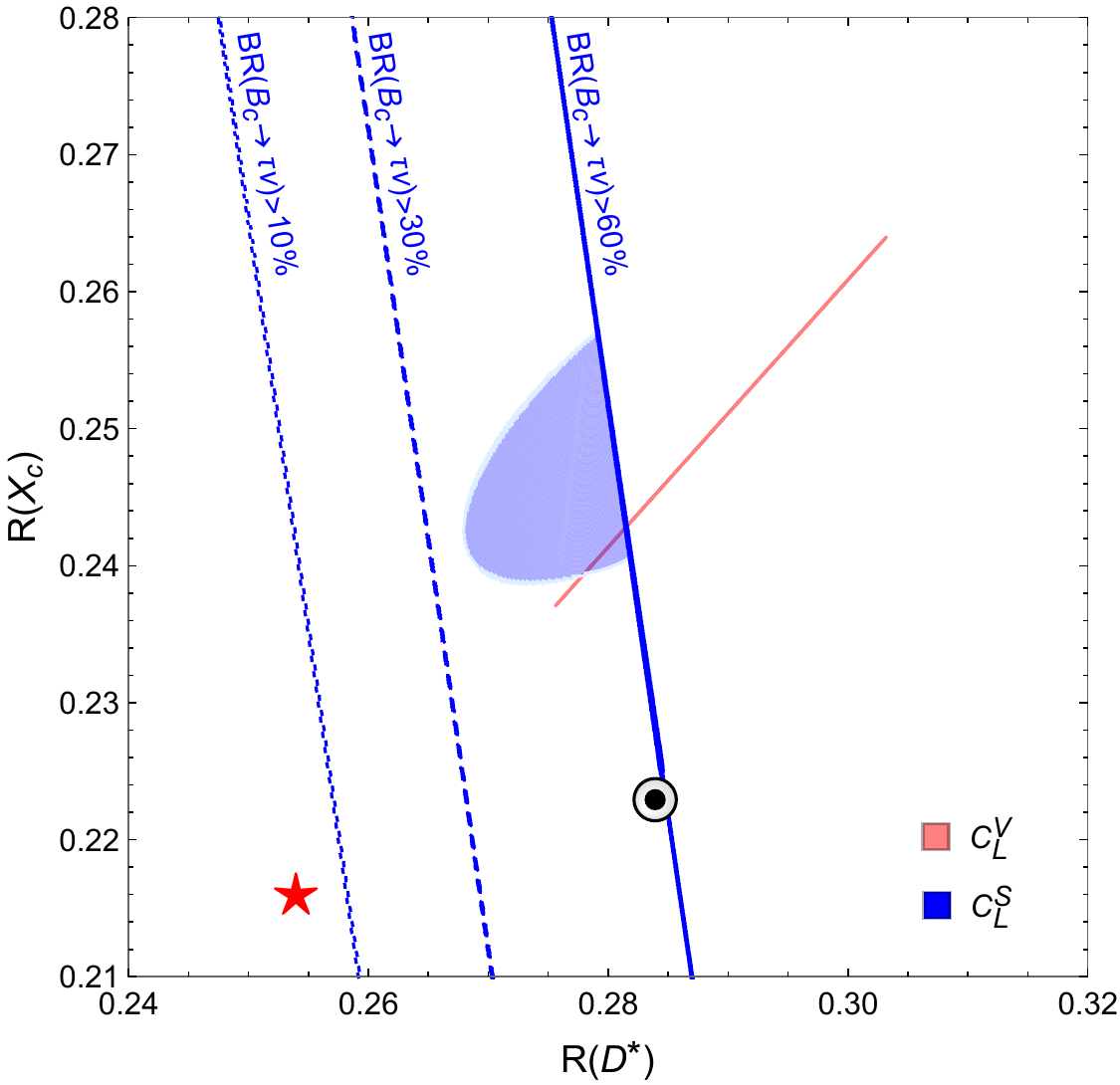}
\caption{}
\end{subfigure}
\begin{subfigure}[b]{0.48\textwidth}
\centering 
\includegraphics[width=6cm, height=5cm]{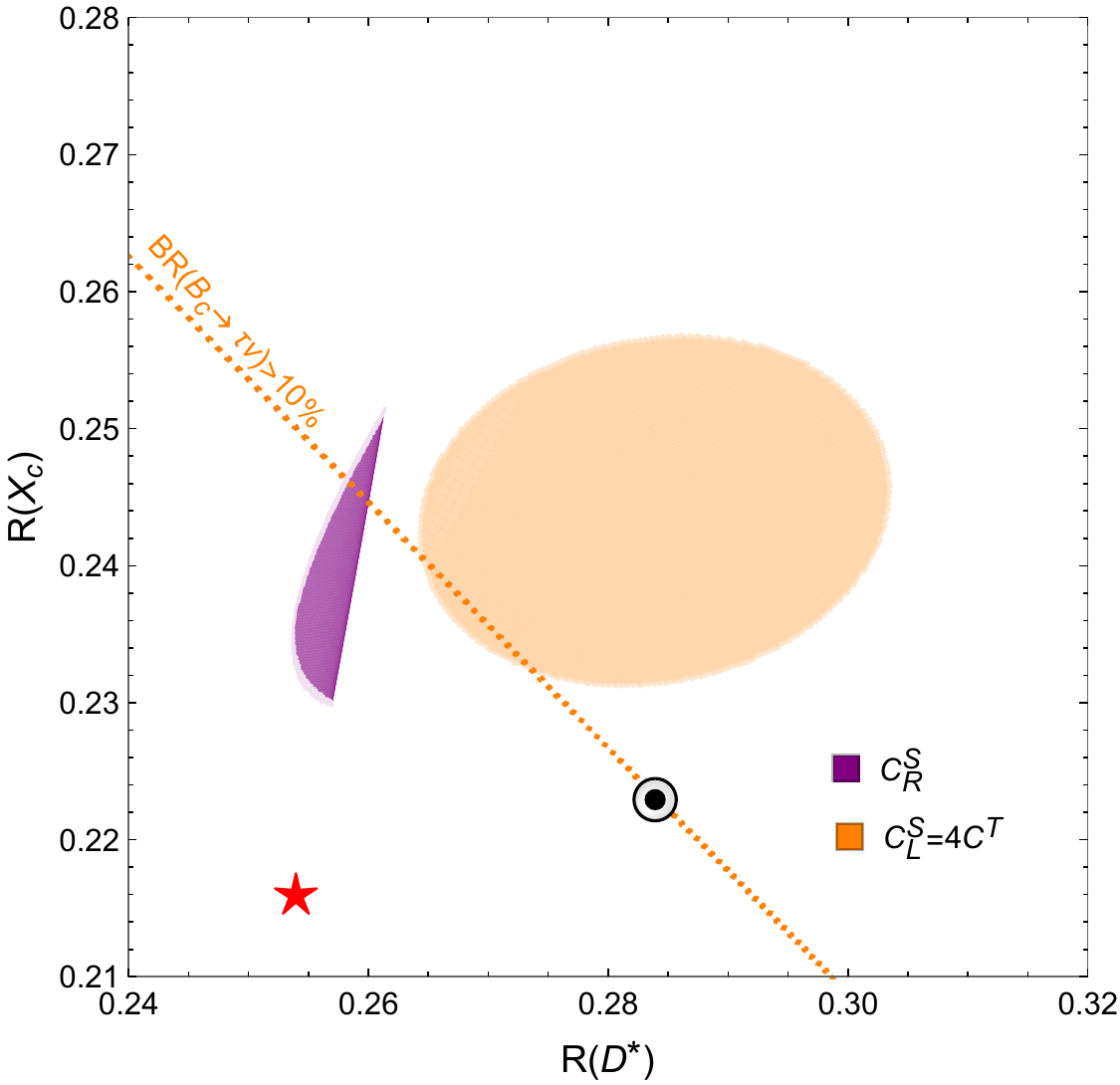}
\caption{}
\end{subfigure}
\caption{\label{fig3 4obs3}The preferred $1\sigma$ regions for the four 2- dimensional
complex scenarios for set $\mathcal{S}_1$ in the $R_{\tau/{\mu,e}}\left(D\right)-R_{\tau/\ell}\left(X_{c}\right)\text{plane (above) and }R_{\tau/{\mu,e}}\left(D^{*}\right)-R_{\tau/\ell}\left(X_{c}\right)$
plane (below) for the $\mathcal{B}(B^-_{c}\to\tau^-\bar{\nu}_\tau)<60\%$. The color coding is the same as in Fig. \ref{fig3 4obs}. The red stars represent SM predictions. The black radio button represents experimental measurements.}
\end{figure}

Finally, Fig. \ref{fig3 4obs3} shows the same for $R_{\tau/\ell}\left(X_{c}\right)-R_{\tau/{\mu,e}}\left(D,D^*\right)$ plane.
The correlation patterns between $R_{\tau/\ell}\left(X_{c}\right)$ and $R_{\tau/{\mu,e}}\left(D\right)$
or $R_{\tau/{\mu,e}}\left(D^{*}\right)$ show similar trend as for the 
$R_{\tau/\ell}\left(\Lambda_{c}\right)$. However, one minor difference arises
in the case of the WC scenario $C_{L}^{S}=4C^{T}$. In this particular
scenario, we observe a moderate positive correlation between $R_{\tau/\ell}\left(X_{c}\right)$
and $R_{\tau/{\mu,e}}\left(D^{*}\right)$, compared to the one between
$R_{\tau/\ell}\left(X_{c}\right)$ and $R_{\tau/{\mu,e}}\left(D\right)$. An additional noteworthy observation is that
all the WCs scenarios closely approximate the experimental values
of the observable, $R_{\tau/\ell}\left(X_{c}\right)$, $R_{\tau/{\mu,e}}\left(D\right)$ and
$R_{\tau/{\mu,e}}\left(D^{*}\right)$, indicating that our results align well with the experimental measurements, hence making our findings reliable.

\section{Impact of complex WCs on various physical observables}\label{sec5}

In $B\to D^*\tau \bar{\nu}_{\tau}$ decay, the $D^*$ meson is unstable, decaying further to $D\pi$. The geometry of the planes in which $B\to D^*\tau \bar{\nu}_{\tau}$ and then $D^*\to  D\pi$  decays are happening helps us to express the decay rate in terms of various angular observables (see. e.g.,  \cite{Mandal:2020htr,Becirevic:2019tpx,Becirevic:2016hea}). In this section, we will summarize the expressions of various physical observables and discuss the impact of our constrained NP WCs on them. Here, we also study the $CP$ violation triple product asymmetries, which hopefully be accessible in some ongoing and future flavor physics experiments. 
\subsection{Analytical expressions}
The angular analysis of a four-body $B\to D^{*}\left(\to D\pi\right)\tau\bar{\nu}$,
gives us additional observables. The differential decay rate distribution
of the transition process $B\rightarrow D^{*}\tau\bar{\nu}_\tau$
with $D^{*}\to D\pi$ on the mass shell has the form \cite{Alok:2016qyh,Mandal:2020htr,Becirevic:2019tpx,Becirevic:2016hea}:
\begin{eqnarray}
\frac{d^{4}\Gamma\left(B\to D^{*}\left(\to D\pi\right)\tau\bar{\nu}_\tau\right)}{dq^{2}d\cos\theta_{\tau}d\cos\theta_{D}d\phi} & \equiv & I\left(q^{2,},\theta_{\tau},\theta_{D},\phi\right)\nonumber \\
& =&\frac{9}{32\pi}\left\{ I_{1}^{s}\sin^{2}\theta_{D}+I_{1}^{c}\cos^{2}\theta_{D}+\left(I_{2}^{s}\sin^{2}\theta_{D}+I_{2}^{c}\cos^{2}\theta_{D}\right)\cos2\theta_{\tau}\right.\nonumber \\
&& \left.+\left(I_{3}\cos2\phi+I_{9}\sin2\phi\right)\sin^{2}\theta_{D}\sin^{2}\theta_{\tau}+\left(I_{4}\cos\phi+I_{8}\sin\phi\right)\sin2\theta_{D}\sin2\theta_{\tau}\right.\nonumber \\
&& \left.+\left(I_{5}\cos\phi+I_{7}\sin\phi\right)\sin2\theta_{D}\sin\theta_{\tau}+\left(I_{6}^{s}\sin^{2}\theta_{D}+I_{6}^{c}\cos^{2}\theta_{D}\right)\cos\theta_{\tau}\right\} \label{four-foldedrate} ,
\end{eqnarray}
where, the lepton-pair invariant mass squared $q^{2}=\left(p_{\tau}+p_{\bar{\nu}_\tau}\right)^{2}$,
and kinematical variables $\theta_{\tau}\text{,}\theta_{D}\text{ and }\phi$
defined as follows. Taking $D^{*}$ momentum in the $B$ rest frame
along $z-$axis, $D$ and $D^{*}$momentum lie in the $xz-$plane
with positive $x-$component, $\theta_{\tau}\text{ and }\theta_{D}$
are the polar angles between $\tau$ and final $D$ meson in the $\tau\bar{\nu}$
and $D\pi$ rest frames, respectively. 
The angular coefficients $I_{i}$'s are function of $q^{2}$ that
encode short- and long-distance physics contributions and these are given in \cite{Mandal:2020htr}.

Integrating Eq. (\ref{four-foldedrate}) over different angles, the differential decay rate of $B\to D^*\left(\to D\pi\right)\tau \bar{\nu}_\tau$ can be written as \cite{Mandal:2020htr,Becirevic:2019tpx,Becirevic:2016hea}:
\begin{equation}
\frac{d\Gamma}{dq^{2}}=\frac{1}{4}\left(3I_{1}^{c}+6I_{1}^{s}-I_{2}^{c}-2I_{2}^{s}\right).
\end{equation}

The decay rate fractions $R_{A,B}$, $D^{*}$ polarization
fraction $R_{L,T}$ and forward-backward asymmetry $A_{FB}$ are given
in terms of these angular coefficients $I's$ as \cite{Becirevic:2019tpx}
\begin{equation}
R_{A,B}\left(q^{2}\right)=\frac{d\Gamma_{A}/dq^{2}}{d\Gamma_{B}/dq^{2}},\quad\quad\quad\quad R_{L,T}\left(q^{2}\right)=\frac{d\Gamma_{L}/dq^{2}}{d\Gamma_{T}/dq^{2}},\quad\quad\quad\quad A_{FB}\left(q^{2}\right)=\frac{3}{8}\frac{I^c_{6}+2I^s_{6}}{d\Gamma/dq^{2}},
\end{equation}
where
\begin{align}
\frac{d\Gamma_{A}}{dq^{2}} & =\frac{1}{4}\left(I^c_{1}+2I^s_{1}-3I^c_{2}-6I^s_{2}\right),\quad\quad\quad\frac{d\Gamma_{B}}{dq^{2}}=\frac{d\Gamma}{dq^{2}}-\frac{d\Gamma_{A}}{dq^{2}}=\frac{1}{2}\left(I^c_{1}+2I^s_{1}+I^c_{2}+2I^s_{2}\right),\nonumber \\
\frac{d\Gamma_{L}}{dq^{2}} & =\frac{1}{4}\left(3I^c_{1}-I^c_{2}\right),\quad\quad\quad\quad\quad\quad\quad\quad\quad\frac{d\Gamma_{T}}{dq^{2}}=\frac{1}{4}\left(3I^s_{1}-I^s_{2}\right).
\end{align}
Here, $\Gamma_{A}$ and $\Gamma_{B}$ represent partial decay
rates with respect to $\theta_{\ell}$, i.e, angle between the lepton and neutrino in $B$ rest frame, $\Gamma_{L}$ and $\Gamma_{T}$ represent the rate corresponding to the longitudinal and transverse $D^{*}$-meson polarization, respectively. Apart from these decay fractions, the other interesting observables are $A_{3}$, $A_{4}$,
$A_{5}$, $A_{6s}$, $A_{7}$, $A_{8}$ and $A_{9}$, and these can be written
as \cite{Becirevic:2019tpx}:
\begin{equation}
A_{3}=\frac{1}{2\pi}\frac{I_{3}}{d\Gamma/dq^{2}},\quad\quad\quad A_{4}=-\frac{2}{\pi}\frac{I_{4}}{d\Gamma/dq^{2}},\quad\quad\quad A_{5}=-\frac{3}{4}\frac{I_{5}}{d\Gamma/dq^{2}},\quad\quad\quad A_{6s}=-\frac{27}{8}\frac{I^s_{6}}{d\Gamma/dq^{2}}.
\end{equation}

\begin{equation}
A_{7}=-\frac{3}{4}\frac{I_{7}}{d\Gamma/dq^{2}},\quad\quad\quad\quad A_{8}=\frac{2}{\pi}\frac{I_{8}}{d\Gamma/dq^{2}},\quad\quad\quad\quad A_{9}=\frac{1}{2\pi}\frac{I_{9}}{d\Gamma/dq^{2}}.\label{ang 4}
\end{equation}
Out of these seven observables $A_{7}, A_{8}$ and $A_{9}$ are CP-odd, whereas
all the others are CP even. After using the values of form factors and other input parameters, and integrating over $q^2$, the expressions of the $I_i's$ in terms of the NP WCs are given in Appendix \ref{AO}.

\subsection{$CP-$ violating triple product asymmetries}

The inclusion of NP operators and the corresponding WCs in $b\to c \tau \bar{\nu}_{\tau}$ decays may add phase which leads  to the $CP-$ asymmetries in the corresponding exclusive decays. This additional new phase gives a marked difference between the decay amplitude and its $CP$ conjugate. However, the possibility of introducing a strong phase due to the interference between the higher $D^*$ resonances was ruled out by the $F_{L}\left(D^*\right)$ measurements. In this situation, there are some possibilities to investigate the violation of $CP-$ symmetry through triple product asymmetries
(TPA). This has been discussed in detail in \cite{Kumbhakar:2020jdz, Aloni:2018ipm,Duraisamy:2014sna,Duraisamy:2013pia,Alok:2011gv,Bhattacharya:2019olg,Bhattacharya:2020lfm}, and to make the study self-sufficient, these details are given in the Appendix \ref{CPTP}. The three different transverse asymmetries $\left(A_{C}^{(i=1,2,3)}\right)$ are \cite{Duraisamy:2013pia}:
\begin{equation}
A_{C}^{\left(1\right)}\left(q^{2}\right)=\frac{4V_{4}^{T}}{3\left(A_{L}+A_{T}\right)},\quad\quad\quad A_{C}^{\left(2\right)}\left(q^{2}\right)=\frac{V_{2}^{0T}}{A_{L}+A_{T}},\quad\quad\quad A_{C}^{\left(3\right)}\left(q^{2}\right)=\frac{V_{1}^{0T}}{A_{L}+A_{T}},\label{asy}
\end{equation}
and for the conjugate mode

\begin{equation}
\bar{A}_{C}^{\left(1\right)}\left(q^{2}\right)=\frac{4\bar{V}_{4}^{T}}{3\left(\bar{A}_{L}+\bar{A}_{T}\right)},\quad\quad\quad\bar{A}_{C}^{\left(2\right)}\left(q^{2}\right)=\frac{-\bar{V}_{2}^{0T}}{\bar{A}_{L}+\bar{A}_{T}},\quad\quad\quad\bar{A}_{C}^{\left(3\right)}\left(q^{2}\right)=\frac{\bar{V}_{1}^{0T}}{\bar{A}_{L}+\bar{A}_{T}},\label{asy-1}
\end{equation}
and due to CP even transformation, we are not interested in that.
The $CP$ violating triple products are given as:
\begin{equation}
A_{T}^{\left(1\right)}\left(q^{2}\right)=\frac{4V_{5}^{T}}{3\left(A_{L}+A_{T}\right)},\quad\quad\quad A_{T}^{\left(2\right)}\left(q^{2}\right)=\frac{V_{3}^{0T}}{A_{L}+A_{T}},\quad\quad\quad A_{T}^{\left(3\right)}\left(q^{2}\right)=\frac{V_{4}^{0T}}{A_{L}+A_{T}}.\label{CP}
\end{equation}
where, the angular coefficients are denoted as $V's$, while the longitudinal
and transverse amplitudes are represented as $A_{L}$ and $A_{T}$,
respectively. In the case of the CP-conjugate decay,
the descriptions outlined in Eq. (\ref{CP}) adopt the subsequent
formulations 
\begin{equation}
\bar{A}_{T}^{\left(1\right)}\left(q^{2}\right)=\frac{-4\bar{V}_{5}^{T}}{3\left(\bar{A}_{L}+\bar{A}_{T}\right)}\quad\quad\quad\bar{A}_{T}^{\left(2\right)}\left(q^{2}\right)=\frac{\bar{V}_{3}^{0T}}{\bar{A}_{L}+\bar{A}_{T}},\quad\quad\quad\bar{A}_{T}^{\left(3\right)}\left(q^{2}\right)=\frac{-\bar{V}_{4}^{0T}}{\bar{A}_{L}+\bar{A}_{T}}.\label{CP-1}
\end{equation}
By employing Eqs. (\ref{CP}) and (\ref{CP-1}), these TPAs are defined as follows \cite{Kumbhakar:2020jdz}:
\begin{align}
\left\langle A_{T}^{\left(1\right)}\left(q^{2}\right)\right\rangle  & =\frac{1}{2}\left(A_{T}^{\left(1\right)}\left(q^{2}\right)+\bar{A}_{T}^{\left(1\right)}\left(q^{2}\right)\right),\\
\left\langle A_{T}^{\left(2\right)}\left(q^{2}\right)\right\rangle  & =\frac{1}{2}\left(A_{T}^{\left(2\right)}\left(q^{2}\right)-\bar{A}_{T}^{\left(2\right)}\left(q^{2}\right)\right),\\
\left\langle A_{T}^{\left(3\right)}\left(q^{2}\right)\right\rangle  & =\frac{1}{2}\left(A_{T}^{\left(3\right)}\left(q^{2}\right)+\bar{A}_{T}^{\left(3\right)}\left(q^{2}\right)\right).\label{24}
\end{align}
To get a comprehensive understanding of these TPAs, we have expressed the longitudinal, transverse, and mixed amplitudes $V's$ at the central values of the form factors. In addition, to assess the sensitivity of NP, the  analytical
expressions of these asymmetries in terms of the complex NP WCs are summarized in Appendix \ref{CPTP}. 
\subsection{Numerical Analysis }
\begin{figure}[H]
\centering 
\begin{subfigure}[b]{0.48\textwidth}
\centering 
\includegraphics[width=7cm, height=4cm]{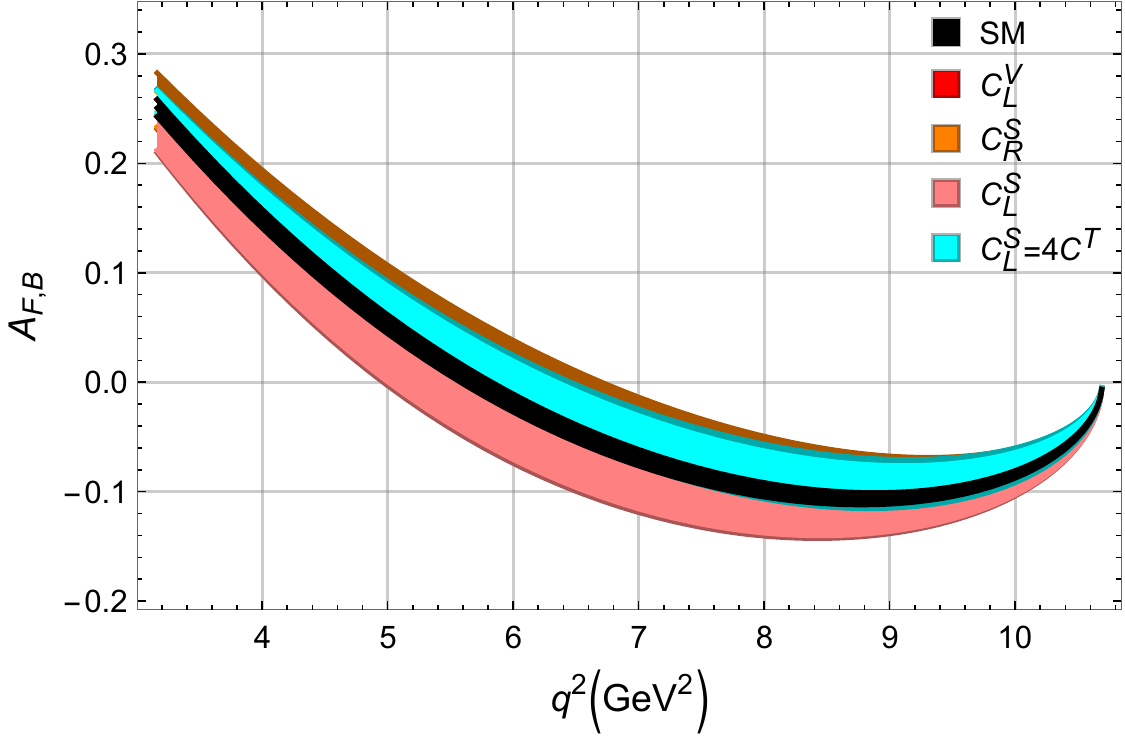} 
\caption{}
\end{subfigure}
\begin{subfigure}[b]{0.48\textwidth}
\centering 
\includegraphics[width=7cm, height=4cm]{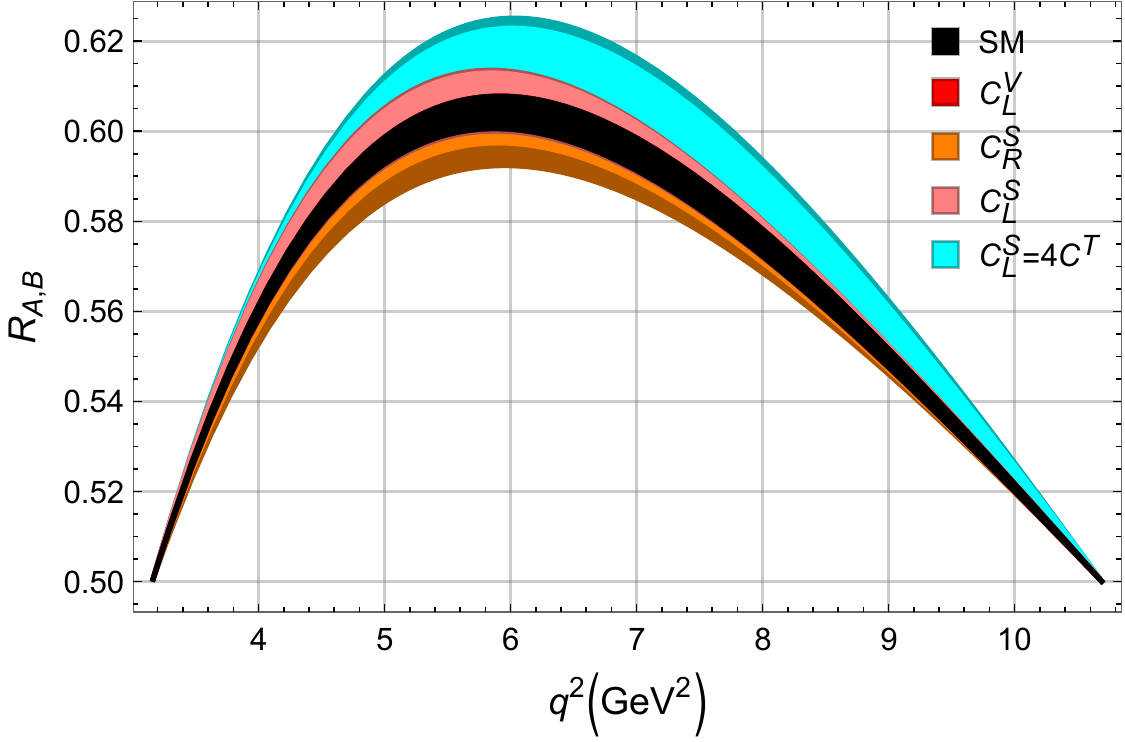}
\caption{}
\end{subfigure}
\begin{subfigure}[b]{0.48\textwidth}
\centering 
\includegraphics[width=7cm, height=4cm]{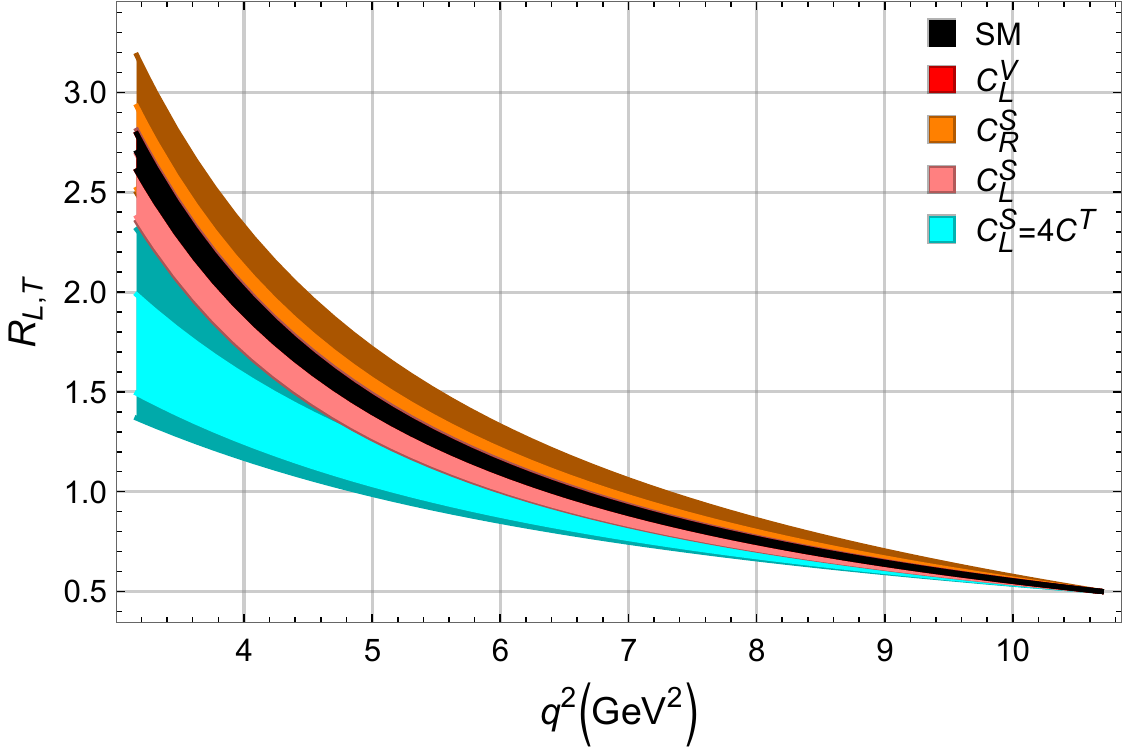} 
\caption{}
\end{subfigure}
\begin{subfigure}[b]{0.48\textwidth}
\centering 
\includegraphics[width=7cm, height=4cm]{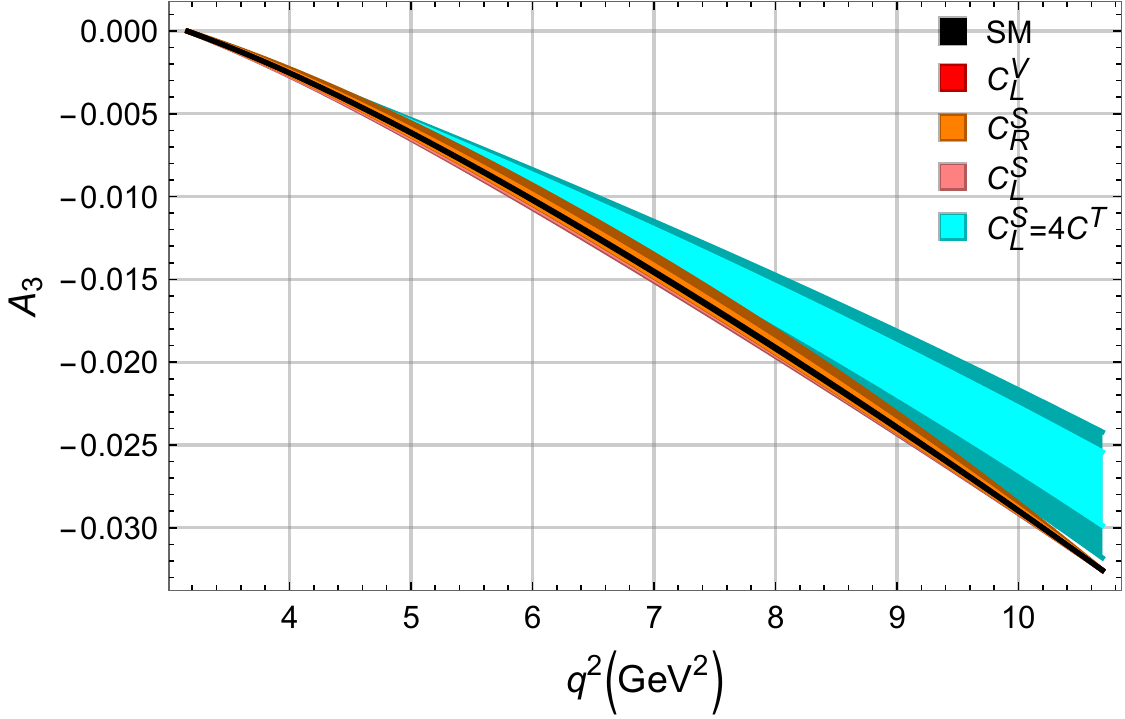}
\caption{}
\end{subfigure}
\begin{subfigure}[b]{0.48\textwidth}
\centering 
\includegraphics[width=7cm, height=4cm]{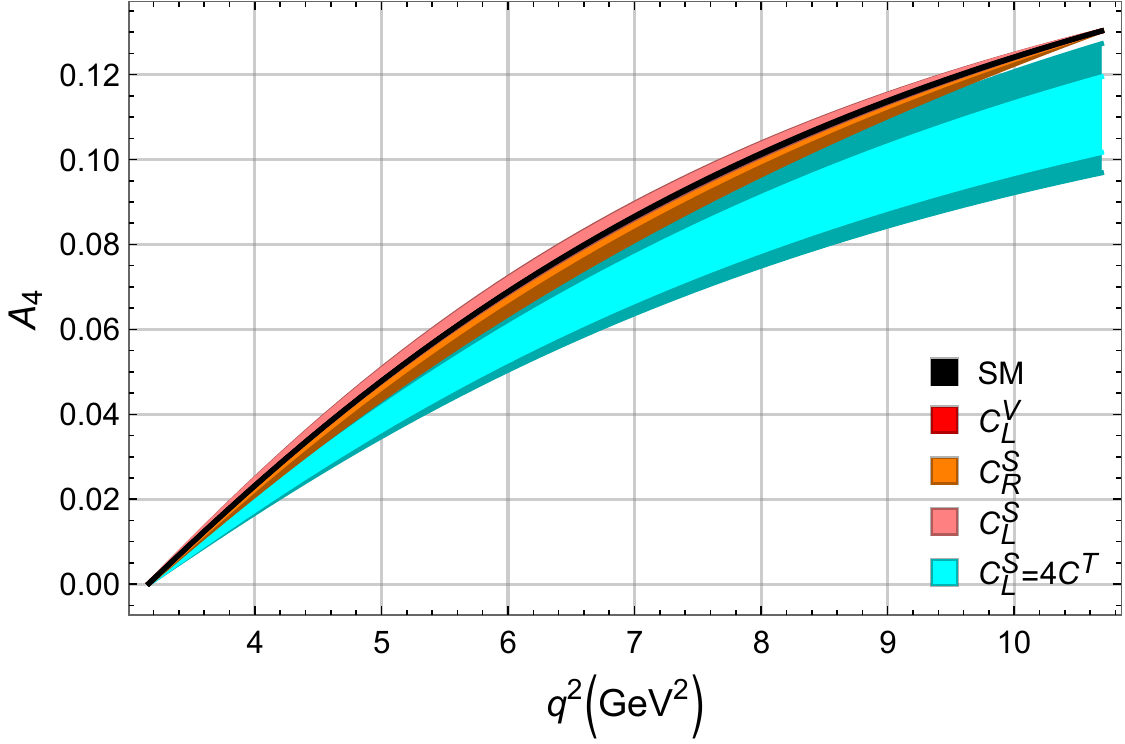} 
\caption{}
\end{subfigure}
\begin{subfigure}[b]{0.48\textwidth}
\centering 
\includegraphics[width=7cm, height=4cm]{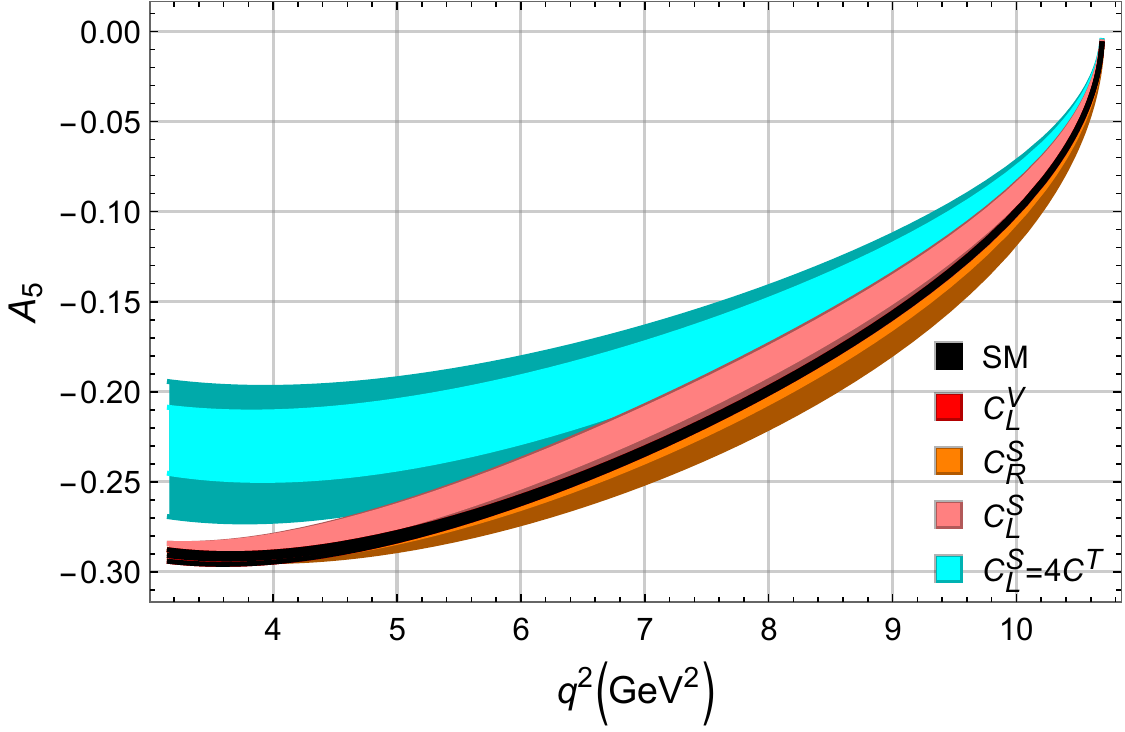}
\caption{}
\end{subfigure}
\begin{subfigure}[b]{0.48\textwidth}
\centering 
\includegraphics[width=7cm, height=4cm]{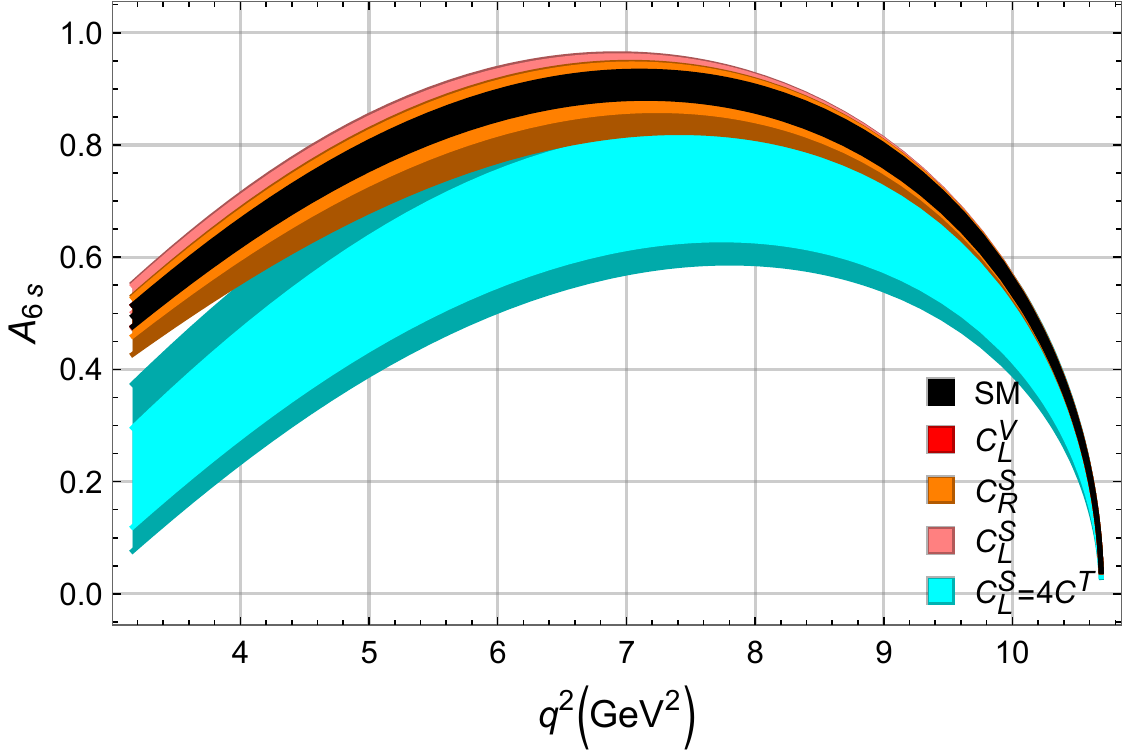}
\caption{}
\end{subfigure}
\caption{\label{Angular CPe LH}The $CP$-even angular observables exhibited for
various NP coupling as a function of $q^{2}$. The width of each curve
comes from the theoretical uncertainties in hadronic form factors
and quark masses. The $1\sigma$ ($2\sigma$) intervals
(dark color) of the complex NP WCs $C_{L}^{V}$, $C_{R}^{S}$,
$C_{L}^{S}$, $C_{L}^{S}=4C^{T}$, for the set of observables $\mathcal{S}_1$
are drawn with red, orange, pink, and cyan colors, respectively.}
\end{figure}

In this section, we will discuss the impact of the constraints of the NP, vector, scalar, and tensor WCs calculated in Sec. III on the above-mentioned physical observables, and check their NP discriminatory power. Fig. \ref{Angular CPe LH}
shows the predictions for $A_{FB}$, $R_{A,B}$, $R_{L,T}$, $A_{3}$,
$A_{4}$, $A_{5}$ and $A_{6s}$ as a function of the square of momentum transfer $\left(q^{2}\right)$. The band in each curve shows the theoretical
uncertainties coming through FFs and other input parameters. The SM value is shown by the black band, whereas the NP couplings $C_{L}^{V}$, $C_{R}^{S}$,
$C_{L}^{S}$ and $C^{T}$ are drawn in red, orange, pink, and cyan bands,
respectively. The corresponding
light and dark shades represent $1\sigma$ and $2\sigma$ intervals. The corresponding numerical values in different $q^2$ bins are given in Table \ref{TableIII}. We can see that in the case of $C_{L}^{S}=4C^{T}$, a significant deviation from the SM value is found
in almost all $CP$-even angular observables, except for the $A_{FB}$. Particularly, its effects are constructive (enhancing the SM value) for $R_{AB},\; A_3,\; A_5$ and destructive (decreasing the SM value) for the remaining angular observables, where the maximum fall-off is for the $A_{6s}$. These outcomes make $C_{L}^{S}=4C^{T}$ a promising
candidate for investigating physics beyond the SM. 
Other than this, the WC $C_{L}^{S}$ shows the second most substantial
deviation, particularly, in the case of $A_{FB}$,
$R_{A,B}$, and $A_{5}$, whereas, the WC $C_{R}^{S}$ does the same for the observable $R_{L,T}$ and $A_{6s}$. 
However, the WC $C_{L}^{V}$ exhibits minimal deviation
for all angular observables, because it has the same SM Lorentz structure and has a small real component. Quantitatively, these findings are given in Table \ref{TableIII}. 
We hope that the experimental measurement of these physical observables will aid in finding particular NP phenomena.

Contrary to these physical observables the values of the $CP$ odd observables $A_{7}$ is found to be too small, to measure experimentally. In Eq. (\ref{IntIs}), we can see that the expressions of $I_{8,\; 9}$ are proportional to the imaginary part of the combination of WCs that comes out to real, and hence the asymmetries $A_{8,\; 9}$ are zero. If one consider the right-handed neutrino operators in the WEH (\ref{eq1}), which are not considered in the current study, their interference with the corresponding left-handed operators lead to the non-zero values of $I_{8,\; 9}$.

\begin{figure}[H]
\centering 
\begin{subfigure}[b]{0.48\textwidth}
\centering 
\includegraphics[width=8cm, height=5cm]{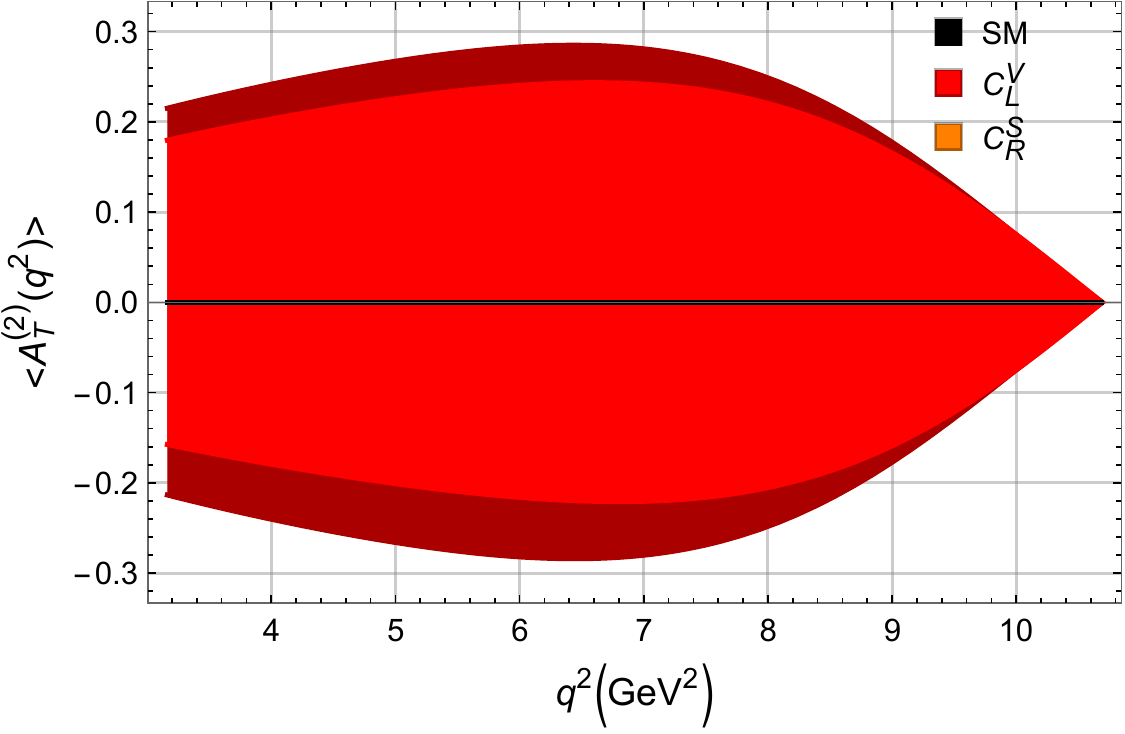} 
\caption{}
\end{subfigure}
\begin{subfigure}[b]{0.48\textwidth}
\centering 
\includegraphics[width=8cm, height=5cm]{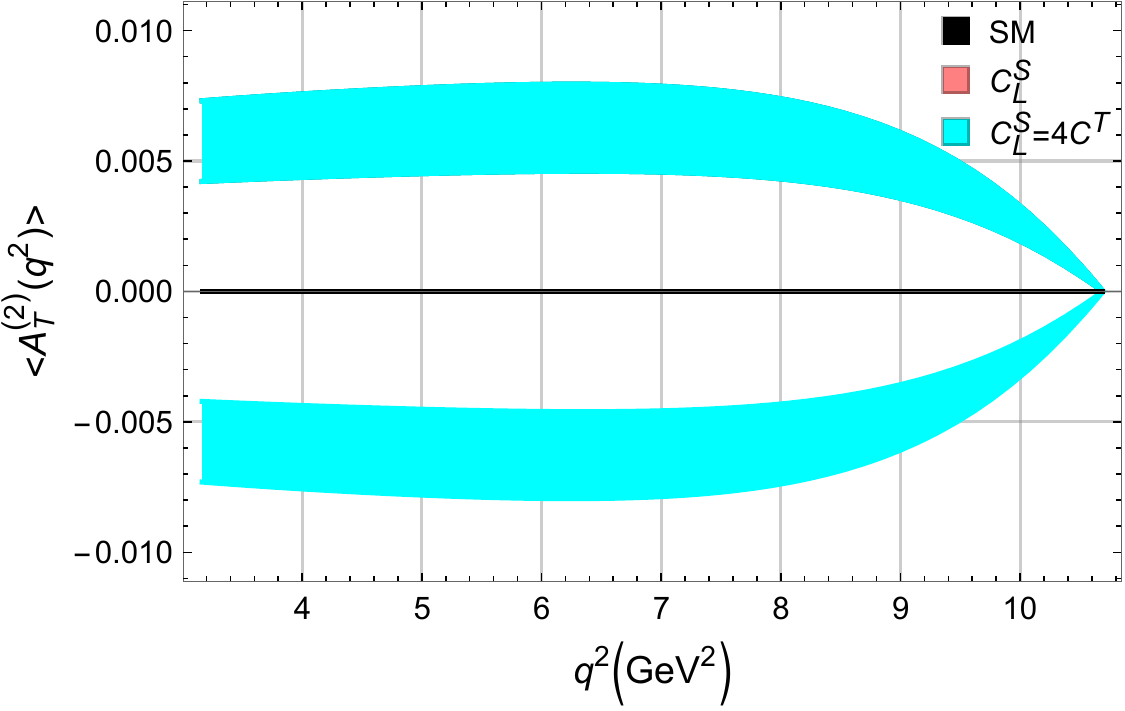}
\caption{}
\end{subfigure}
\caption{\label{CP-3}The $CP$-violating TPAs exhibited for various NP coupling
as a function of $q^{2}$. The width of each curve comes from the theoretical
uncertainties in hadronic form factors and quark masses. The $1\sigma$
sigma ($2\sigma$) intervals in color (dark color) of the complex
coupling WCs $C_{L}^{V}$, $C_{R}^{S}$, $C_{L}^{S}$, $C_{L}^{S}=4C^{T}$,
for the set of observables $\mathcal{S}_1$ are in red, orange, pink, and cyan
colors, respectively.}
\end{figure}

Fig. \ref{CP-3} shows the predictions for
$\left\langle A_{T}^{\left(2\right)}\left(q^{2}\right)\right\rangle $
as a function of $q^{2}$. The description of the various bands and colors is the same as in the case of the above angular asymmetries. In SM, the values of these triple product
asymmetries (TPAs) are zero; hence, any non-zero value will be attributed to the presence
of NP. Specifically, from (\ref{CP-2}) we can see that $V_{4}^{0T}$ and $V_{5}^{T}$ are zero, giving us $\left\langle A_{T}^{\left(1\right)}\left(q^{2}\right)\right\rangle$ and $\left\langle A_{T}^{\left(3\right)}\left(q^{2}\right)\right\rangle$ zero for any NP WCs. The only non-zero contribution comes for the $CP$ violating triple product asymmetry $\left\langle A_{T}^{\left(2\right)}\left(q^{2}\right)\right\rangle$. In this context, for the NP WC $C_{L}^{V}$, 
the largest non-zero value or maximum deviation from the SM reaches
$0.28$ at $q^{2}$$=6.5 \text{GeV}^2$ when $C_{L}^{V}$$=-1.5 \pm 0.9 i$. This
data point falls within the $2\sigma$ interval. 
In a $1\sigma$
interval, the maximum deviation is $0.23$, occurring at $q^{2}=6.5\text{GeV}^2$
for $C_{L}^{V}$ equals to $-1 - 1.05 i$ and $-1.17 + 1.06i$. 
The other non-zero deviation from the SM result is coming for $C_{L}^{S}=4C^{T}$; however, the maximum value in this case is $0.008$ at $q^{2}=6.5\text{GeV}^2$, corresponding to WC $C_{L}^{S}$ equals to $-0.08+0.33i$
and $-0.11+0.34i$ - but this is too small to be measured experimentally. Similarly, for the WCs $C_{R}^{S}$ and
$C_{L}^{S}$ this asymmetry remains zero in the whole $q^2$ range. Therefore, we can say that experimental observation of $\left\langle A_{T}^{\left(2\right)}\left(q^{2}\right)\right\rangle$ will make the new vector type operators an important candidate to hunt for the NP in flavor sector.

\renewcommand{\arraystretch}{1.8}
\begin{table}

\begin{tabular}{|c|l|r|r|r|r|r|}
\hline 
Observables & $\text{Bins (GeV$^2$)}$ & SM & $C_{L}^{V}$ & $C_{R}^{S}$ & $C_{L}^{S}$ & $C_{L}^{S}=4C^{T}$\tabularnewline
\hline 
\hline 
\multirow{4}{*}{$A_{FB}$} & $\left[3.16,5\right]$ & $0.144\pm0.009$ & $0.144\pm0.009$ & $0.135\pm0.054$ & $0.108\pm0.045$ & $0.160\pm0.015$\tabularnewline
\cline{2-7} \cline{3-7} \cline{4-7} \cline{5-7} \cline{6-7} \cline{7-7} 
 & $\left[5,7\right]$ & $-0.015\pm0.008$ & $-0.015\pm0.008$ & $-0.027\pm0.068$ & $-0.055\pm0.049$ & $0.007\pm0.027$\tabularnewline
\cline{2-7} \cline{3-7} \cline{4-7} \cline{5-7} \cline{6-7} \cline{7-7} 
 & $\left[7,9\right]$ & $-0.095\pm0.007$ & $-0.095\pm0.007$ & $-0.104\pm0.057$ & $-0.125\pm0.038$ & $-0.075\pm0.028$\tabularnewline
\cline{2-7} \cline{3-7} \cline{4-7} \cline{5-7} \cline{6-7} \cline{7-7} 
 & $\left[9,10.69\right]$ & $-0.082\pm0.004$ & $-0.082\pm0.004$ & $-0.087\pm0.029$ & $-0.097\pm0.019$ & $-0.072\pm0.018$\tabularnewline
\hline 
\multirow{4}{*}{$R_{A,B}$} & $\left[3.16,5\right]$ & $0.559\pm0.002$ & $0.559\pm0.002$ & $0.560\pm0.008$ & $0.562\pm0.005$ & $0.564\pm0.005$\tabularnewline
\cline{2-7} \cline{3-7} \cline{4-7} \cline{5-7} \cline{6-7} \cline{7-7} 
 & $\left[5,7\right]$ & $0.601\pm0.004$ & $0.601\pm0.004$ & $0.603\pm0.013$ & $0.605\pm0.008$ & $0.612\pm0.012$\tabularnewline
\cline{2-7} \cline{3-7} \cline{4-7} \cline{5-7} \cline{6-7} \cline{7-7} 
 & $\left[7,9\right]$ & $0.574\pm0.003$ & $0.574\pm0.003$ & $0.575\pm0.007$ & $0.576\pm0.005$ & $0.584\pm0.008$\tabularnewline
\cline{2-7} \cline{3-7} \cline{4-7} \cline{5-7} \cline{6-7} \cline{7-7} 
 & $\left[9,10.69\right]$ & $0.525\pm0.001$ & $0.525\pm0.001$ & $0.525\pm0.002$ & $0.525\pm0.001$ & $0.529\pm0.004$\tabularnewline
\hline 
\multirow{4}{*}{$R_{L,T}$} & $\left[3.16,5\right]$ & $1.953\pm0.061$ & $1.953\pm0.061$ & $1.896\pm0.430$ & $1.794\pm0.236$ & $1.377\pm0.382$\tabularnewline
\cline{2-7} \cline{3-7} \cline{4-7} \cline{5-7} \cline{6-7} \cline{7-7} 
 & $\left[5,7\right]$ & $1.136\pm0.026$ & $1.136\pm0.026$ & $1.105\pm0.244$ & $1.060\pm0.114$ & $0.951\pm0.140$\tabularnewline
\cline{2-7} \cline{3-7} \cline{4-7} \cline{5-7} \cline{6-7} \cline{7-7} 
 & $\left[7,9\right]$ & $0.762\pm0.011$ & $0.762\pm0.011$ & $0.747\pm0.120$ & $0.731\pm0.048$ & $0.702\pm0.049$\tabularnewline
\cline{2-7} \cline{3-7} \cline{4-7} \cline{5-7} \cline{6-7} \cline{7-7} 
 & $\left[9,10.69\right]$ & $0.567\pm0.003$ & $0.567\pm0.003$ & $0.563\pm0.034$ & $0.559\pm0.012$ & $0.554\pm0.011$\tabularnewline
\hline 
\multirow{4}{*}{$A_{3}$} & $\left[3.16,5\right]$ & $-0.003\pm0.000$ & $-0.003\pm0.000$ & $-0.003\pm0.000$ & $-0.003\pm0.000$ & $-0.003\pm0.000$\tabularnewline
\cline{2-7} \cline{3-7} \cline{4-7} \cline{5-7} \cline{6-7} \cline{7-7} 
 & $\left[5,7\right]$ & $-0.010\pm0.000$ & $-0.010\pm0.000$ & $-0.010\pm0.001$ & $-0.011\pm0.000$ & $-0.009\pm0.001$\tabularnewline
\cline{2-7} \cline{3-7} \cline{4-7} \cline{5-7} \cline{6-7} \cline{7-7} 
 & $\left[7,9\right]$ & $-0.019\pm0.000$ & $-0.019\pm0.000$ & $-0.019\pm0.001$ & $-0.020\pm0.000$ & $-0.016\pm0.002$\tabularnewline
\cline{2-7} \cline{3-7} \cline{4-7} \cline{5-7} \cline{6-7} \cline{7-7} 
 & $\left[9,10.69\right]$ & $-0.028\pm0.000$ & $-0.028\pm0.000$ & $-0.028\pm0.001$ & $-0.028\pm0.000$ & $-0.024\pm0.004$\tabularnewline
\hline 
\multirow{4}{*}{$A_{4}$} & $\left[3.16,5\right]$ & $0.025\pm0.000$ & $0.025\pm0.000$ & $0.025\pm0.003$ & $0.026\pm0.001$ & $0.020\pm0.003$\tabularnewline
\cline{2-7} \cline{3-7} \cline{4-7} \cline{5-7} \cline{6-7} \cline{7-7} 
 & $\left[5,7\right]$ & $0.068\pm0.000$ & $0.068\pm0.000$ & $0.069\pm0.006$ & $0.071\pm0.003$ & $0.056\pm0.008$\tabularnewline
\cline{2-7} \cline{3-7} \cline{4-7} \cline{5-7} \cline{6-7} \cline{7-7} 
 & $\left[7,9\right]$ & $0.101\pm0.000$ & $0.101\pm0.000$ & $0.102\pm0.006$ & $0.103\pm0.002$ & $0.084\pm0.013$\tabularnewline
\cline{2-7} \cline{3-7} \cline{4-7} \cline{5-7} \cline{6-7} \cline{7-7} 
 & $\left[9,10.69\right]$ & $0.122\pm0.000$ & $0.122\pm0.000$ & $0.122\pm0.002$ & $0.123\pm0.001$ & $0.103\pm0.016$\tabularnewline
\hline 
\multirow{4}{*}{$A_{5}$} & $\left[3.16,5\right]$ & $-0.289\pm0.003$ & $-0.289\pm0.003$ & $-0.288\pm0.004$ & $-0.280\pm0.011$ & $-0.226\pm0.044$\tabularnewline
\cline{2-7} \cline{3-7} \cline{4-7} \cline{5-7} \cline{6-7} \cline{7-7} 
 & $\left[5,7\right]$ & $-0.259\pm0.002$ & $-0.259\pm0.002$ & $-0.255\pm0.017$ & $-0.241\pm0.019$ & $-0.207\pm0.038$\tabularnewline
\cline{2-7} \cline{3-7} \cline{4-7} \cline{5-7} \cline{6-7} \cline{7-7} 
 & $\left[7,9\right]$ & $-0.198\pm0.002$ & $-0.198\pm0.002$ & $-0.193\pm0.025$ & $-0.178\pm0.022$ & $-0.161\pm0.028$\tabularnewline
\cline{2-7} \cline{3-7} \cline{4-7} \cline{5-7} \cline{6-7} \cline{7-7} 
 & $\left[9,10.69\right]$ & $-0.105\pm0.001$ & $-0.105\pm0.001$ & $-0.101\pm0.020$ & $-0.092\pm0.014$ & $-0.086\pm0.015$\tabularnewline
\hline 
\multirow{4}{*}{$A_{6s}$} & $\left[3.16,5\right]$ & $0.650\pm0.023$ & $0.650\pm0.023$ & $0.662\pm0.100$ & $0.686\pm0.062$ & $0.371\pm0.186$\tabularnewline
\cline{2-7} \cline{3-7} \cline{4-7} \cline{5-7} \cline{6-7} \cline{7-7} 
 & $\left[5,7\right]$ & $0.862\pm0.025$ & $0.862\pm0.025$ & $0.874\pm0.110$ & $0.893\pm0.061$ & $0.632\pm0.188$\tabularnewline
\cline{2-7} \cline{3-7} \cline{4-7} \cline{5-7} \cline{6-7} \cline{7-7} 
 & $\left[7,9\right]$ & $0.870\pm0.022$ & $0.870\pm0.022$ & $0.877\pm0.077$ & $0.885\pm0.041$ & $0.703\pm0.165$\tabularnewline
\cline{2-7} \cline{3-7} \cline{4-7} \cline{5-7} \cline{6-7} \cline{7-7} 
 & $\left[9,10.69\right]$ & $0.557\pm0.013$ & $0.557\pm0.$$013$ & $0.559\pm0.027$ & $0.560\pm0.017$ & $0.472\pm0.100$\tabularnewline
\hline 
\end{tabular}
\caption{The numerical values of physical variables which are mentioned above at the best-fit-points of various NP coupling
in the various $q^{2}$ bins. The uncertainties come from hadronic form factors and the other input parameters. For distinct WCs $C_{L}^{V}$, $C_{R}^{S}$, $C_{L}^{S}$, $C_{L}^{S}=4C^{T}$, incorporated the maximum deviations within $2\sigma$ intervals in the uncertainties.
}\label{TableIII}

\end{table}

\section{Conclusion}\label{sec6}

The Standard Model (SM) elegantly explains most of the experimentally observed phenomena, and no new particles beyond those present in the SM have been discovered so far. However, there are some deviations from the SM predictions observed in flavor-changing-charged-current (FCCC) transitions involving the third generation of leptons i.e., $b\to c\tau \bar{\nu}_\tau$. In this situation, it is important to explain how to study these indirect signatures of the new physics (NP).

In this work, we have analyzed anomalies observed in $b\to c\tau \bar{\nu}_\tau$ decays by including new vector, scalar, and tensor operators in the SM effective Hamiltonian. We assumed the corresponding Wilson coefficients (WCs) to be complex and also we took the neutrinos to be left-handed. Our study is composed of two sets of observables. In the first set $\left(\mathcal{S}_1\right)$ we have included $R_{\tau/{\mu,e}}\left(D\right), R_{\tau/{\mu,e}}\left(D^*\right), F_{L}\left(D^*\right)$ and $P_{\tau}\left(D^*\right)$, whereas in the second one $\left(\mathcal{S}_2\right)$, we included two more experimentally measured observables $R_{\tau/\mu}(J/\psi)$ and $R_{\tau/\ell}(X_c)$. For these two sets, we have estimated the best-fit points, $p-$value, $\chi_{\text{SM}}^{2}$, $\text{pull}_{\text{SM}}$ and $1,2\sigma$ deviations for these complex WCs. As the branching ratio of $B_c\to \tau \bar{\nu}_\tau$ is not yet measured, therefore, we have shown our results for the $60\%,\; 30\%$ and $10\%$ limits on it. We found that the NP WCs $C^{V}_{L}$ and $C^S_{R}$ are not sensitive to these bounds, whereas the for $C^S_{L}$ the $p-$ value is maximum for $60\%$ limits. The value of BFPs and the $p-$value significantly change for the $\mathcal{S}_2$. This is attributed to the large error in the experimental measurements of $R_{\tau/\mu}\left(J/\psi\right)$ and $R_{\tau/\ell}\left(X_c\right)$. We hope that in the future when these measurements are refined, we will have better control over the parametric space of these new WCs.

Due to the same quark level transition, there is a strong theoretical correlation between $R_{\tau/{\mu,e}}\left(D^{(*)}\right)$ and $R_{\tau/{\mu,e}}\left(\Lambda_c\right)$. Using the updated values of different input parameters and the experimental measurements for $R_{\tau/{\mu,e}}\left(D^{(*)}\right)$, the updated sum-rule was derived in \cite{Fedele:2022iib}. We validated the sum rule in our case and found that the remainder in our case is $<10^{-3}$. On the same lines, we derived the similar sum rules for $R_{\tau/\mu}\left(J/\psi\right)$ and $R_{\tau/\ell}\left(X_c\right)$ relating them with $R_{\tau/{\mu,e}}\left(D^{(*)}\right)$. We have seen that in  $R_{\tau/\mu}\left(J/\psi\right)$ the maximum contribution comes from the LFU ratio of $D^*$, and hence large deviations in the $R_{\tau/{\mu,e}}\left(D^{*}\right)$ will strongly impact this ratio. Using the best-fit values of our complex WCs, and the latest measurements of the  $R_{\tau/{\mu,e}}\left(D^{(*)}\right)$, we calculated $R_{\tau/\mu}\left(J/\psi\right) = 0.289$ and $R_{\tau/\ell}\left(X_c\right)=0.248$. In addition to this, we have plotted the correlation of $R_{D}$ and $R_{D^*}$ with other observables and hope that the correlation of $R_{\tau/{\mu,e}}\left(D^{(*)}\right)$ with other physical observables will help us to discriminate between different NP scenarios. 

Furthermore, we also investigated the effects of these NP WCs on different angular asymmetries and the $CP$-violation triple product in $B\to D^*\tau \bar{\nu}_\tau$ decays. We found that the most promising effects on the different angular asymmetries are attributed to the NP WC $C_{L}^{S}=4C^{T}$. We know that the $CP$-triple product is zero in the SM, and any non-zero value would hint towards the NP. We found that for the vector-like new operators, the maximum value of $\left\langle A_{T}^{\left(2\right)}\left(q^{2}\right)\right\rangle = 0.28$.  We hope our results can be tested at the LHCb and future high-energy experiments.

\section*{Acknowledgements}

The authors would like to thank Prof. Monika Blanke and Dr. Teppei Kitahara for helping us understand their work and suggesting improvements in our code, and appreciate this kind gesture. MJA and SS would like to thank Wang Yu-Ming for reading the manuscript and for suggesting us some important references.  

\section*{Data Availability Statement}

No Data associated in the manuscript.

\appendix
\numberwithin{equation}{section}
\section{Goodness of fit}\label{GoF}

We accomplish $\chi^2$ to test the hypothesis about the distribution
of observables in distinct effective operators. This helps us to quantify the discrepancy between the
theoretical and experimental data used to fit. The $\chi^{2}$ expression is coded
as \cite{ParticleDataGroup:2020ssz,Workman:2022ynf}:
\begin{equation}
\chi^{2}\left(C_{M}^{X}\right)=\sum_{i,j}^{N_{obs}}\left[O_{i}^{\text{exp.}}-O_{i}^{\text{th}}\left(C_{M}^{X}\right)\right]C_{ij}^{-1}\left[O_{j}^{\text{exp.}}-O_{j}^{\text{th}}\left(C_{M}^{X}\right)\right],
\end{equation}
where $N_{obs}$ is the number of observables, $O_{i}^{exp.}$ are
the data from experiments and $O_{i}^{th}$ are the observables theoretical
parameters, which are complex functions of scalar and vector WCs $C_{M}^{X}$
$(X=S,V)$ and $(M=L,R)$. The covariance matrix is the sum of
theoretical and experimental uncertainties and let in the experimental
correlation between $R_{\tau/{\mu,e}}(D)$ and $R_{\tau/{\mu,e}}(D^{*})$ in terms of $pull$ calculated
by the formulas 
\begin{equation}
\chi_{R_{\tau/{\mu,e}}\left(D\right)-R_{\tau/{\mu,e}}\left(D^{*}\right)}^{2}=\frac{\chi_{R_{\tau/{\mu,e}}\left(D\right)}^{2}+\chi_{R_{\tau/{\mu,e}}\left(D^{*}\right)}^{2}-2*\rho*\text{pull}_{R_{\tau/{\mu,e}}\left(D\right)}*\text{pull}_{R_{\tau/{\mu,e}}\left(D^{*}\right)}}{1-\rho^{2}},
\end{equation}
where, the correlation value $\rho$ taking from the HFLAV is $-0.37$. Firstly, we work out how many degrees of freedom 
($dof$) we have, which is equal to $N_{dof}=N_{obs}-N_{par}$, where
$N_{par}$ is the number of parameters to be fitted, for each parameter. For the complex WCs, as we have in our case, we have  $N_{par}=2$ and
$N_{obs}=4$ and $6$ for the sets of observables $\mathcal{S}_1$ and $\mathcal{S}_2$. As a second step, we obtain the minimum value of $\chi^{2}$ for each parameter
to acquire Best fit points (BFP). Thirdly, we utilized the value of
$\chi^{2}$ to get $p-\text{value}$. The $p-\text{value}$ for the hypothesis can
be calculated as \cite{ParticleDataGroup:2020ssz,Workman:2022ynf}
\begin{equation}
p=\intop_{\chi^{2}}^{\infty}f\left(z;n_{d}\right)dz,
\end{equation}
where $f(z;n_{d})$ is the $\chi^{2}$ probability distribution function
and $n_{d}$ is the number of degrees of freedom. The $p-\text{value}$ quantify
the consistency between data and the hypothesis of the NP scenario. Finally,
we estimate the value of $\text{pull}$ from the SM in units of standard deviation
($\sigma$) determined by 
\begin{equation}
\text{pull}_{SM}=\sqrt{\chi_{\text{SM}}^{2}-\chi_{\text{min}}^{2}},
\end{equation}
where $\chi_{\text{\text{SM}}}^{2}=\chi^{2}(0)$.

\section{ Angular Observables}\label{AO}
 After putting the form factors and other input parameters, the integration over $q^2$ leads to the expressions of the angular coefficients $\left(\times 10^{-16}\right)$ in terms of our complex NP WCs, i.e.,
\begin{eqnarray}
I_{1}^{c} & = & 4.36\left\{ \left|1+C_{L}^{V}\right|^{2}+0.002\left|1+C_{L}^{V}-1.1\left(C_{R}^{S}-C_{L}^{S}\right)\right|^{2}+12.67\left|C^{T}\right|^{2}-12.67\Re\left[\left(1+C_{L}^{V}\right)\left(C^{T}\right)^{*}\right]\right\} ,\nonumber \\
I_{1}^{s} & = & 5.55\left\{ \left|1+C_{L}^{V}\right|^{2}+7.26\left|C^{T}\right|^{2}-4.69\Re\left[\left(1+C_{L}^{V}\right)\left(C^{T}\right)^{*}\right]\right\} ,\nonumber \\
I_{2}^{c} & = & 2.37\left\{ -\left|1+C_{L}^{V}\right|^{2}+12.67\left|C^{T}\right|^{2}\right\} ,\nonumber \\
I_{2}^{s} & = & 1.19\left\{ \left|1+C_{L}^{V}\right|^{2}+12.68\left|C^{T}\right|^{2}\right\} ,\nonumber \\
I_{3} & = & 2.37\left\{ -\left|1+C_{L}^{V}\right|^{2}+12.67\left|C^{T}\right|^{2}\right\} ,\nonumber \\
I_{4} & = & 2.37\left\{ -\left|1+C_{L}^{V}\right|^{2}+12.67\left|C^{T}\right|^{2}\right\} ,\nonumber \\
I_{5} & = & 0.069\left\{ \Re\left[\left(1+C_{L}^{V}-1.93C^{T}\right)\left(1+C_{L}^{V}+3.62C^{T}\right)^{*}\right]\right.\nonumber \\
 &  & \left.+0.53\Re\left[\left(1+C_{L}^{V}-6.55C^{T}\right)\left(1+C_{L}^{V}-1.1\left(C_{R}^{S}-C_{L}^{S}\right)\right)^{*}\right]\right\} ,\nonumber \\
I_{6}^{c} & = & 0.074\left\{ \Re\left[\left(1+C_{L}^{V}-6.55C^{T}\right)\left(1+C_{L}^{V}-1.1\left(C_{R}^{S}-C_{L}^{S}\right)\right)^{*}\right]\right\} ,\nonumber \\
I_{6}^{s} & = & 0.137\left\{ -\Re\left[\left(1+C_{L}^{V}-1.93C^{T}\right)\left(1+C_{L}^{V}+3.62C^{T}\right)^{*}\right]\right\} ,\nonumber \\
I_{7} & = & 6.73\left\{ \Im\left[\left(1+C_{L}^{V}-1.93C^{T}\right)\left(1+C_{L}^{V}-1.94C^{T}\right)^{*}\right]\right.\nonumber \\
 &  & \left.+0.00006\Im\left[\left(1+C_{L}^{V}+12.27C^{T}\right)\left(1+C_{L}^{V}-1.1\left(C_{R}^{S}-C_{L}^{S}\right)\right)^{*}\right]\right\} ,\nonumber \\
I_{8} & = & 0.024\left\{ \Im\left[\left|1+C_{L}^{V}\right|^{2}+23.76\left|C^{T}\right|^2\right]\right\}=0 ,\nonumber \\
I_{9} & = & 0.048\left\{ \Im\left[-\left|1+C_{L}^{V}\right|^{2}-23.74\left|C^{T}\right|^2\right]\right\}=0 .\label{IntIs}
\end{eqnarray}

\section{$CP$-violating triple product}\label{CPTP}

Defining $\theta_{D}$ and $\theta_{\tau}$ as angles between the $D$ meson (that originate from the cascade decay of $D^*\to D\pi$ meson) and that of $\tau$ with the decaying $B-$meson, respectively. Also, the angle $\varphi$, gives of the angular difference between
the decay plane of $D^{\ast}$ and the plane defined by the momenta
of the $\tau$ lepton and the corresponding $\left(\nu_\tau\right)$ neutrino.

The relevant triple products are computed through the integration
of the complete decay distribution across distinct ranges of the polar
angles $\theta_{D}$ and $\theta_{\tau}$ . This process yields the
following triple products as the outcomes of interest. \cite{Kumbhakar:2020jdz, Duraisamy:2014sna,Duraisamy:2013pia,Alok:2011gv,Bhattacharya:2019olg,Bhattacharya:2020lfm}
\begin{align*}
\frac{d^{2}\Gamma^{\left(1\right)}}{dq^{2}d\phi} & =\intop_{-1}^{1}\intop_{-1}^{1}\frac{d^{4}\Gamma}{dq^{2}d\cos\theta_{\tau}d\cos\theta_{D}d\phi}d\cos\theta_{\tau}d\cos\theta_{D}\\
 & =\frac{1}{2\pi}\frac{d\Gamma}{dq^{2}}\left(1+A_{C}^{\left(1\right)}\cos2\phi+A_{T}^{\left(1\right)}\sin2\phi\right),
\end{align*}

\begin{align*}
\frac{d^{2}\Gamma^{\left(2\right)}}{dq^{2}d\phi} & =\intop_{-1}^{1}dcos\theta_{\tau}\left[\intop_{0}^{1}-\intop_{-1}^{0}\right]\frac{d^{4}\Gamma}{dq^{2}d\cos\theta_{\tau}d\cos\theta_{D}d\phi}d\cos\theta_{D}\\
 & =\frac{1}{4}\frac{d\Gamma}{dq^{2}}\left(A_{C}^{\left(2\right)}\cos\phi+A_{T}^{\left(2\right)}\sin\phi\right),
\end{align*}
and
\begin{align*}
\frac{d^{2}\Gamma^{\left(3\right)}}{dq^{2}d\phi} & =\left[\intop_{0}^{1}-\intop_{-1}^{0}\right]d\cos\theta_{\tau}\left[\intop_{0}^{1}-\intop_{-1}^{0}\right]\frac{d^{4}\Gamma}{dq^{2}d\cos\theta_{\tau}d\cos\theta_{D}d\phi}d\cos\theta_{D}\\
 & =\frac{2}{3\pi}\frac{d\Gamma}{dq^{2}}\left(A_{C}^{\left(3\right)}\cos\phi+A_{T}^{\left(3\right)}\sin\phi\right).
\end{align*}

The coefficients $A_{C}^{\left(i\right)}$of $\cos\phi$ and $\cos2\phi$
are $CP-$even, and the coefficients
$A_{C}^{\left(i\right)}$of $\sin\phi$ and $\sin2\phi$ are CP-odd,  under the $CP$ transformation. 
The expressions of $V$'s in terms of the NP WCs are:
\begin{eqnarray}
V_{1}^{0} & = & 1787.12\left\{ \left|1+C_{L}^{V}\right|^{2}+0.71\left|1+C_{L}^{V}+0.34\left(C_{R}^{S}-C_{L}^{S}\right)\right|^{2}+4.73\left|1-0.9C^{T}\right|^{2}\right.\notag\\
& & \left.+4.35\Re\left[\left(1-0.9C^{T}\right)\left(1+C_{L}^{V}\right)^{*}\right]\right\}, \notag\\
V_{2}^{0} & = & 195\left\{ \left|1+C_{L}^{V}\right|^{2}+3.06\left|C^{T}\right|^{2}\right\},\notag\\
V_{1}^{T} & = & 455.81\left\{ 1+\left|C_{L}^{V}\right|^{2}+2\Re\left[C_{L}^{V}\right]+1.76\left|C^{T}\right|^{2}+4.1\Re\left[C^{T}\right]-2.05\left(C_{L}^{V}C^{T\;*}+C^{T}C_{L}^{V\;*}\right)\right\} ,\notag \\
V_{2}^{T} & = &97.47\left\{ 1+\left|C_{L}^{V}\right|^{2}+2\Re\left[C_{L}^{V}\right]-3.07\left|C^{T}\right|^{2}\right\} ,\nonumber \\
V_{3}^{0T} & = & 0.03\Im\left[\left(1+C_{L}^{V}-12.27C^{T}\right)\left(-C_{R}^{S}+C_{L}^{S}\right)^{*}+14.37C_{L}^{V}+7.18\left(1+C^{T}\right)C_{L}^{V\;*}\right],\notag \\
V_{5}^{T} & = & 0,\quad\quad\quad\quad V_{4}^{0T}  = 0 .\label{CP-2}
\end{eqnarray}
We can see from ref. \cite{Duraisamy:2013pia} that $V_{4}^{0T} = \sqrt{2}\left(1-\frac{m_{\tau}^2}{q^2} \right)\Im\left[\mathcal{A}_{\perp}^*\mathcal{A}_{0}\right]$, which after integrating over  $q^2$ is proportional to $\Im\left|1+C_{L}^{V}\right|^2$, and hence it is zero.

\end{document}